%% file: main.tex
\documentclass[acmtog, nonacm]{acmart}
\usepackage{booktabs} %

\usepackage{hhline}
\usepackage[ruled]{algorithm2e} %
\usepackage{subcaption}
\usepackage{pifont}
\usepackage{arydshln}
\usepackage{multirow}
\usepackage{multicol}
\usepackage{xcolor}
\usepackage[export]{adjustbox}
\usepackage[normalem]{ulem}
\usepackage{cleveref}
\usepackage{xspace}
\usepackage{rotating}
\usepackage{xhfill}
\usepackage{tikz}
\usetikzlibrary {spy}
\usepackage{xcolor}
\usepackage{siunitx}
\definecolor{myspycolor}{RGB}{120,120,255}

\acmDOI{}

\makeatletter

\setcopyright{none}
\acmJournal{TOG}
\copyrightyear{2025}
\acmYear{2025}
\acmSubmissionID{1862}

\input{macros}

\title{EditP23: 3D Editing via Propagation of Image Prompts to Multi-View}

\begin{document}

\begin{abstract}
\input{sections/1_abstract}

\end{abstract}
\author{Roi Bar-On}
\affiliation{
  \institution{Tel-Aviv University}
  \country{Israel}
}
\author{Dana Cohen-Bar}
\affiliation{
  \institution{Tel-Aviv University}
  \country{Israel}
}
\author{Daniel Cohen-Or}
\affiliation{
  \institution{Tel-Aviv University}
  \country{Israel}
}

\input{figures/teaser}

\maketitle
\begingroup
\renewcommand\thefootnote{}
\footnotetext{Project page: \href{https://editp23.github.io/}{\texttt{https://editp23.github.io/}}}
\endgroup

\input{sections/2_intro}

\input{sections/3_related}

\input{sections/4_method}

\input{sections/5_experiments}
\input{sections/6_conclusion}

\begin{acks}
We thank Daniel Garibi, Rinon Gal, Or Patashnik, Amir Barda, Nir Goren, Gal Metzer, and our colleagues at Tel Aviv University for their valuable feedback and support.
\end{acks}
\clearpage
\bibliographystyle{ACM-Reference-Format}
\bibliography{main}

\input{sections/7_additional_res}

\clearpage
\end{document}

%% file: macros.tex
\crefname{section}{Sec.}{Secs.}
\Crefname{section}{Section}{Sections}
\Crefname{table}{Table}{Tables}
\crefname{table}{Tab.}{Tabs.}
\usepackage{xcolor}
\usepackage{longtable}
\usepackage{array}
\usepackage{adjustbox}

\definecolor{mydarkblue}{rgb}{0,0.08,1}
\definecolor{mydarkgreen}{rgb}{0.02,0.6,0.02}
\definecolor{mydarkorange}{rgb}{0.40,0.2,0.02}

\makeatletter
\DeclareRobustCommand\onedot{\futurelet\@let@token\@onedot}
\def\@onedot{\ifx\@let@token.\else.\null\fi\xspace}

\def\eg{\emph{e.g}\onedot}

\def\naive{na\"{\i}ve\xspace}
\def\Naive{Na\"{\i}ve\xspace}

\newif\ifdraft
\drafttrue

\definecolor{mygreen}{RGB}{112,173,71}
\definecolor{myblue}{RGB}{91,155,213}
\definecolor{myorange}{RGB}{237,125,49}
\definecolor{myred}{RGB}{255,22,67}
\definecolor{honey}{RGB}{201,89,95}
\definecolor{amber}{RGB}{180,148,41}
\definecolor{darkblue}{RGB}{84,96,126}

\newcommand\ourmethod{EditP23}

%% file: sections/1_abstract.tex
We present EditP23, a method for mask-free 3D editing that propagates 2D image edits to multi-view representations in a 3D-consistent manner. In contrast to traditional approaches that rely on text-based prompting or explicit spatial masks, EditP23 enables intuitive edits by conditioning on a pair of images: an original view and its user-edited counterpart. These image prompts are used to guide an edit-aware flow in the latent space of a pre-trained multi-view diffusion model, allowing the edit to be coherently propagated across views. Our method operates in a feed-forward manner, without optimization, and preserves the identity of the original object, in both structure and appearance. We demonstrate its effectiveness across a range of object categories and editing scenarios, achieving high fidelity to the source while requiring no manual masks.

%% file: figures/teaser.tex
\begin{teaserfigure}
    \centering
    \includegraphics[width=0.95\linewidth,trim={0cm 10cm 0cm 1cm}, clip]{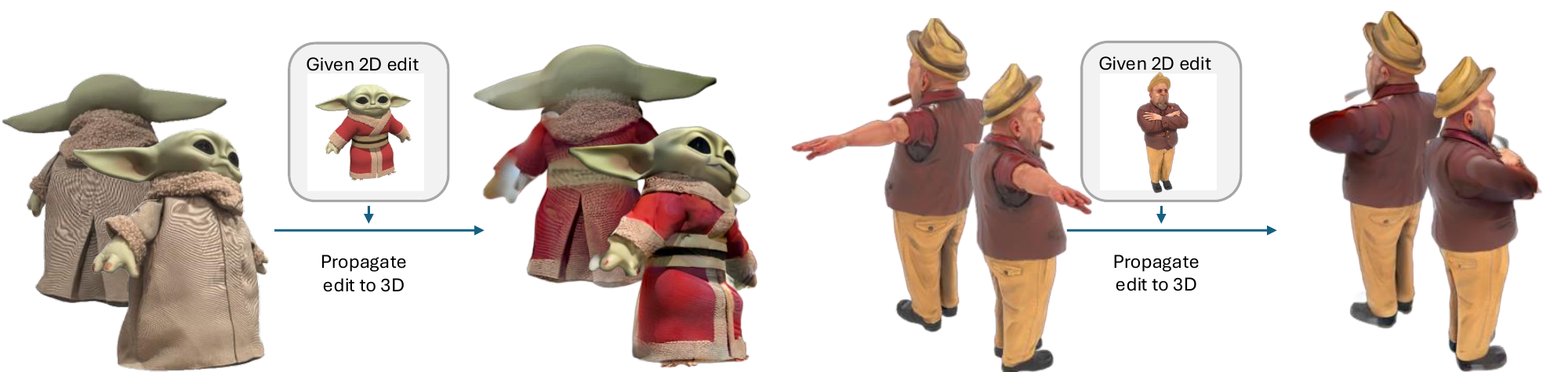}
    \caption{
    Our method enables fast, mask-free 3D object editing by propagating a user-provided 2D modification to the full 3D shape. The figure illustrates two use cases: appearance editing (left) and geometry editing (right). Given a single edited view (inset), the change is consistently applied to the entire subject in just a few seconds.}
    \label{fig:teaser}
\end{teaserfigure}

%% file: sections/2_intro.tex
\section{Introduction}

3D editing plays a central role in numerous domains, including entertainment, design, e-commerce, and manufacturing. While recent advances in generative models have significantly improved the quality of 3D object generation, editing remains significantly more challenging. In contrast to generation, editing involves the core challenge of balancing user-intended modifications with source fidelity, a necessity for ensuring spatial and semantic consistency in structured visual domains.

3D editing is fundamentally more difficult than 2D manipulation because 3D training data is scarce and hard to annotate. To overcome this, many existing 3D editing approaches require extra user input to guide the process. Specifically, they often rely on manually defined masks to explicitly constrain and localize modifications, which in turn limits their practicality and user accessibility.

Recent 3D generation pipelines follow a two-stage strategy: (i) a diffusion model synthesizes a multi-view image grid, and (ii) a reconstruction method lifts those views to geometry. A natural approach to 3D editing would be to edit a single view and use the multi-view model to complete the remaining viewpoints. However, this \naive strategy often fails to preserve the original object's identity because the model lacks conditioning on object-specific features that are absent from the edited view, leading to hallucinated geometry and appearance that diverge from the source object when viewed from other angles (see~\cref{fig:baseline}).

Building on the principles of \emph{Edit-Aware Denoising}~\cite{deltadenoisingscore, flowedit}, we propose \ourmethod{}, a novel method that adapts this concept to multi-view images for 3D-consistent editing. Unlike methods that rely on text prompts, our approach uses a pair of image prompts: an original view and its user-edited counterpart to guide the edit. These images define an edit-aware flow in the latent space of a pre-trained multi-view generator, enabling the edit to propagate consistently across all views. By conditioning on both images, our method preserves the object's overall identity, including structure and appearance, while ensuring the user's modifications are faithfully and globally applied to the underlying 3D representation.

To summarize, we present a method with the following key contributions:

\begin{itemize}
    \item \textbf{Mask-Free Editing.} Our method requires no manual 3D annotations or 2D segmentation masks, overcoming a key limitation of prior mask-assisted approaches and improving user accessibility. The user only needs to edit a single 2D view to guide the 3D edit.

    \item \textbf{Flexible Image Prompts.} The edit is guided by an image pair, allowing users to leverage any preferred 2D editing tool, from manual painting to generative pipelines. This provides greater flexibility and intuitive control compared to text-prompt-driven systems.

    \item \textbf{Training-Free Framework.} Our method leverages a frozen, pre-trained multi-view diffusion backbone, requiring no new training or fine-tuning. This approach avoids a data- and compute-intensive training process  and preserves the original richness of the generative model by harnessing its full capacity. Furthermore, this design allows our editing technique to be easily extended to future backbone models.

    \item \textbf{Fast, Feed-Forward Inference.} Our approach operates in a feed-forward manner without lengthy gradient-based optimization. Edit propagation is efficient, requiring fewer denoising steps than full generation and completing 3D updates in seconds on a single GPU.
\end{itemize}

Together, these properties make our pipeline both flexible and efficient, offering intuitive control over a wide range of editing tasks, including pose changes, object additions, and global modifications.

%% file: sections/3_related.tex
\section{Related Work}

\subsection{Image Editing with Diffusion Models}
Diffusion models have enabled powerful 2D image editing methods, including global edits, local changes, and style transfers. These range from \textit{training-based} approaches like InstructPix2Pix~\cite{brooks2023instructpix2pix}, Imagic~\cite{kawar2023imagic}, Pix2Pix-Zero~\cite{pix2pixzero2023}, and Glide~\cite{glide} to \textit{inversion-based} methods such as Null-Text Inversion~\cite{nulltext}, Plug-and-Play~\cite{tumanyan2023plug}, Edit Friendly DDPM Inversion~\cite{editfriendlyddpm},  LEDITS++~\cite{ledits++} and MasaCtrl~\cite{masactrl}. In contrast, \textit{non-inversion} methods steer the denoising process directly. This family includes foundational techniques like SDEdit~\cite{sdedit} and Blended Diffusion~\cite{avrahami2022blended}, as well as  approaches like Delta Denoising Score (DDS)~\cite{deltadenoisingscore} and FlowEdit~\cite{flowedit} that introduce an ``edit-aware'' denoising. Building upon the strong foundation of 2D diffusion editing, and inspired specifically by the edit-aware principles in DDS and FlowEdit, our work extends these techniques to diffusion-based editing of multi-view image grids.

\subsection{3D Generation with Multi-View Diffusion}
Multi-view diffusion models have emerged as a powerful paradigm for 3D content generation, often operating via a two-stage process. First, a diffusion model synthesizes a set of consistent 2D viewpoints of an object \eg, Zero123++~\cite{zero123++}, MVDream~\cite{mvdream}, SyncDreamer~\cite{syncdreamer} and Wonder3D~\cite{wonder3d}. Second, these views are lifted to a 3D representation using reconstruction algorithms \eg, InstantMesh~\cite{instantmesh}, LGM ~\cite{lgm}, CRM~\cite{crm}. This approach leverages strong 2D priors, offering advantages over direct 3D generation. While multi-view diffusion is widely used for generation, its potential for editing remains underexplored. Our work leverages the multi-view representation of these models to extend well-established 2D image editing techniques, while maintaining 3D consistency across views.

\subsection{3D Editing}
Current 3D editing methods present a fundamental trade-off between precision and scope: \textit{mask-assisted} approaches offer precise local control but are ill-suited for global transformations, while \textit{mask-free} methods provide global flexibility but often struggle to preserve details in unchanged regions.

\paragraph{Mask-assisted 3D editing.}
These methods achieve precise local control by using spatial constraints to define the editing region. For instance, some approaches perform inpainting within multi-view masks, such as Instant3Dit~\cite{barda2024instant3dit} and NeRFiller~\cite{nerfiller}, while PrEditor3D~\cite{erkoc2024preditor3d} uses a segmentation module to localize the edits. TRELLIS~\cite{trellis} defines a bounding box in its latent representation to constrain the modification. Sked~\cite{mikaeili2023sked} offers an alternative by using a sketch-based constraint, though it remains limited to local changes. While effective for targeted control, this reliance on pre-defined spatial constraints makes these methods impractical for global transformations and limits their overall flexibility.

\paragraph{Mask-free 3D editing.}
Recent mask-free approaches guide edits using prompts or view-level modifications, offering greater flexibility by avoiding explicit spatial annotations. A prominent family of these methods relies on iterative, per-edit optimization. One such category uses Score Distillation Sampling (SDS)~\cite{dreamfusion} to align a 3D representation with text prompts, with methods like Vox-E~\cite{voxe}, DreamEditor~\cite{zhuang2023dreameditortextdriven3dscene}, and TIP-Editor~\cite{zhuang2024tipeditoraccurate3deditor} building on this framework. Another optimization-based category involves modifying individual 2D views and then consolidating them into a consistent 3D representation; these methods often leverage structures like a NeRF~\cite{nerf} or Gaussian Splatting~\cite{gs} to harmonize the edits, as seen in Instruct-NeRF2NeRF~\cite{instructnerf2nerf}, QNeRF~\cite{qnerf} and DGE~\cite{dge}. While powerful, the primary drawback for all these optimization-based approaches is a lengthy and computationally intensive per-edit process.

To accelerate editing, other works propose faster, feed-forward solutions. MVEdit~\cite{chen2024mvedit}, for instance, introduces a training-free adapter for multi-view diffusion, but this speed can come at the cost of preserving fine details and structural fidelity. Another fast approach operates in the compact latent space of generative 3D models like Shape-E~\cite{jun2023shapegeneratingconditional3d}. However, methods in this category like Sharp-It~\cite{edelstein2024sharpitmultiviewmultiviewdiffusion} and SHAP-Editor~\cite{shapeditor} can struggle to reconstruct intricate geometry from the original object due to information loss during the encoding step.

\ourmethod{} introduces a novel editing paradigm that avoids the core limitations of previous methods. Our approach is mask-free, training-free, and operates via fast, feed-forward updates directly on the multi-view grid, unlike techniques that require lengthy optimization or operate in detail-sparse latent spaces.  \ourmethod{} efficiently handles both global and local edits in a unified framework, preserving the structure and fine details of the source object.

%% file: sections/4_method.tex
\section{Method}
\input{figures/method}
In this section, we introduce our approach for 3D object editing by propagating a single-view edit into a multi-view grid.
\subsection{Overview}
Our approach adapts the standard two-stage pipeline from recent 3D \emph{generation} methods for the task of 3D \emph{editing}. This pipeline first generates a multi-view grid (mv-grid), then reconstructs a 3D object from it.
\input{figures/baseline}

We work with mv-grids that represent multiple views of an object, concatenated into a single grid image. Given an input 3D object and a single user-edited view, our goal is to generate a new 3D object that integrates the intended edit while maintaining fidelity to the original shape. Our method operates by rendering the original object to obtain a source mv-grid, applying our adapted \textbf{Edit-Aware Denoising} technique to transform the mv-grid based on the edited view, and finally reconstructing the edited 3D object.

Formally, our method takes as input a 3D object, a source view $I_{\text{src}}$, and its user-edited counterpart $I_{\text{tar}}$. We first render the 3D object to obtain a multi-view grid $\mathbf{X}_{\text{src}}=\{x_i\}_{i=1}^{6}$. \Cref{alg:edit_mvgrid} then generates an edited mv-grid $\mathbf{x}_{\text{tar}}$ by using $I_{\text{tar}}$ as a conditioning signal for the diffusion process. This ensures the edit is coherently propagated across all views while preserving the original shape's content.

\subsection{Preliminaries: Multi-View Diffusion Models}
\label{sec:preliminaries}
To propagate edits coherently across multiple views, our method leverages Zero123++~\cite{zero123++}, an image-conditioned multi-view diffusion model. This model takes a single conditioning image $I_{\mathrm{cond}}$ and synthesizes a complete multi-view output structured as a 2D image grid (mv-grid) by concatenating six view tiles. This grid is processed jointly by the underlying UNet, which employs a $v$-prediction formulation~\cite{salimans2022progressive} for iterative refinement from noise to clean multi-view images.

Zero123++ employs a two-pass UNet mechanism at each diffusion step $t$ to incorporate conditioning information: First, the condition image $I_{\mathrm{cond}}$ is noised to match the noise level $\sigma_t$ of the current multi-view grid $\mathbf{Z}^t$, yielding $\tilde{I}_{\mathrm{cond}}^{\,t}$. In the \textbf{reference pass}, the UNet processes $\tilde{I}_{\mathrm{cond}}^{\,t}$ and caches attention keys and values $(K_{\text{ref}}^{t}, V_{\text{ref}}^{t})$ that capture salient conditioning features. During the subsequent \textbf{grid pass}, the noisy multi-view grid $\mathbf{Z}^{t}$ is processed with self-attention layers augmented by the cached $(K_{\text{ref}}^{t}, V_{\text{ref}}^{t})$. This enables each view tile to reference spatial features from the conditioning image, promoting 3D consistency across the output.

Building upon this conditioning mechanism, \ourmethod{} introduces an image-prompted, edit-aware denoising step to guide the multi-view generation process, as detailed in the following section.

\subsection{Edit-Aware Denoising for Multi-View Diffusion}
\label{sec:edit-aware-denoising}
Our method operates by directly guiding the denoising process of the multi-view diffusion model. This approach is inspired by recent inversion-free 2D editing techniques like Delta Denoising Score (DDS)~\cite{deltadenoisingscore} and FlowEdit~\cite{flowedit}. These methods accept a pair of prompts to specify an edit: a source prompt representing the original concept and a target prompt representing the desired, edited concept. They establish an ``edit direction'' in the latent space by using the the difference between the model's prediction guided by the target prompt and its prediction guided by the source prompt.

At each iterative step $t_i$, we compare a \textbf{source configuration} with a \textbf{target configuration} (illustrated in~\cref{fig:method}). The source configuration comprises the original mv-grid $\mathbf{X}_\text{src}$ (with its noised version $\mathbf{Z}_\text{src}^{t_i}$) and the source condition view $I_{\text{src}}$. The target configuration includes the mv-grid currently being edited $\mathbf{X}_\text{edit}^{t_i}$ (with its noised version $\mathbf{Z}_\text{edit}^{t_i}$) and the target condition view $I_{\text{tar}}$.

The core of our editing mechanism is a differential \emph{edit direction}, computed by taking the difference between the model's predictions for the target and source configurations. For a diffusion model $\phi$ parameterized to predict velocity $v_\phi$, this is:
\begin{equation}
\Delta \mathbf{v}_\phi^{t_i} = v_\phi\left(\mathbf{Z}_\text{edit}^{t_i}, I_{\text{tar}}\right) - v_\phi\left(\mathbf{Z}_\text{src}^{t_i}, I_\text{src}\right)
\label{eq:delta_v}
\end{equation}
This $\Delta \mathbf{v}^{t_i}$ vector is designed to isolate the specific transformations required for the edit. By subtracting the source prediction from the target prediction, components related to shared content and common noise artifacts ideally cancel out. The subsequent update step uses this $\Delta \mathbf{v}^{t_i}$ to refine $\mathbf{X}_\text{edit}^{t_i}$, guiding its evolution towards the desired target.

\subsection{Algorithm}
We apply the edit-aware denoising mechanism described above for our multi-view editing task. Our method utilizes the pre-trained Zero123++ model as its backbone and adapts the denoising formulation for image-prompted conditioning. A key aspect of this adaptation is the correlated noising strategy, which is crucial for effectively isolating the edit signal.

Inspired by DDS, we apply an identical Gaussian noise realization $\mathbf{N}_\text{grid}^{t_i}$ to both the current edited grid $\mathbf{X}_\text{edit}^{t_i}$ and the original source grid $\mathbf{X}_\text{src}$ at each timestep $t_i$. This produces their noised counterparts $\mathbf{Z}_\text{edit}^{t_i}$ and $\mathbf{Z}_\text{src}^{t_i}$. Concurrently, another identical Gaussian noise realization $\mathbf{N}_{\text{cond}}^{t_i}$ is used to noise both the target condition image $I_{\text{tar}}$ and the source condition image $I_{\text{src}}$. This ensures that the inputs to the diffusion model $\phi$ are highly correlated, allowing the subtraction in~\cref{eq:delta_v} to effectively capture the edit.

By integrating these elements, our algorithm (\cref{alg:edit_mvgrid}) effectively propagates edits from a single modified view to all views in the multi-view grid, maintaining 3D geometric coherence while preserving the object's original structure and appearance. The overall flow of this differential denoising mechanism at each timestep is depicted in~\cref{fig:method}.

\begin{algorithm}[htb]
\caption{\ourmethod{}: Single-View Edit Propagation}
\label{alg:edit_mvgrid}
\KwIn{source grid $\mathbf{x}_{\text{src}}$, reference view $I_{\text{src}}$, edited view $I_{\text{tar}}$, noise schedule $\{t_i\}_{i=0}^{T}$}
\KwOut{edited grid $\mathbf{x}_{\text{tar}}$}

$\mathbf{x}_{\text{edit}}^{t_T} \leftarrow \mathbf{x}_{\text{src}}$\\
\For{$i = T,\dots,1$}{
Draw $\mathbf{N}_{\text{grid}}^{t_i},\;\mathbf{N}_{\text{cond}}^{t_i}\sim\mathcal{N}(\mathbf{0},\sigma_{t_i}^2\mathbf{I})$\\
 $\mathbf{z}_{\text{src}}^{t_i} \leftarrow\texttt{{add\_noise}}\left( \mathbf{x}_{\text{src}}, \mathbf{N}_{\text{grid}}^{t_i}\right)$\\
 $\mathbf{z}_{\text{edit}}^{t_i}\leftarrow \texttt{{add\_noise}}\left( \mathbf{x}_{\text{edit}}^{t_i}, \mathbf{N}_{\text{grid}}^{t_i}\right)$\\
$\tilde{I}_{\text{src}}^{t_i} \leftarrow \texttt{{add\_noise}}\left(I_{\text{src}}, \mathbf{N}_{\text{cond}}^{t_i}\right)$\\
$\tilde{I}_{\text{tar}}^{t_i} \leftarrow \texttt{{add\_noise}}\left(I_{\text{tar}}, \mathbf{N}_{\text{cond}}^{t_i}\right)$\\

$\mathbf{v}_{\text{src}}^{t_i} \leftarrow v_\phi(\mathbf{z}_{\text{src}}^{t_i},\tilde{I}_{\text{src}}^{t_i})$\\
$\mathbf{v}_{\text{tar}}^{t_i} \leftarrow v_\phi(\mathbf{z}_{\text{edit}}^{t_i},\tilde{I}_{\text{tar}}^{t_i})$\\
$\Delta\mathbf{v}^{t_i} \leftarrow \mathbf{v}_{\text{tar}}^{t_i} - \mathbf{v}_{\text{src}}^{t_i}$\\
$\mathbf{x}_{\text{edit}}^{t_{i-1}} \leftarrow \mathbf{x}_{\text{edit}}^{t_i} + \Delta\mathbf{v}^{t_i}dt$\\

}
\Return{$\mathbf{x}_\mathrm{edit}^{t_0}$}\\
\textbf{Note:}
\texttt{add\_noise} refers to the standard forward diffusion (noising) process.

\end{algorithm}

%% file: figures/method.tex
\begin{figure*}[tb]
    \centering
    \includegraphics[width=0.9\textwidth, trim={10cm 6cm 4cm 4cm}, clip]{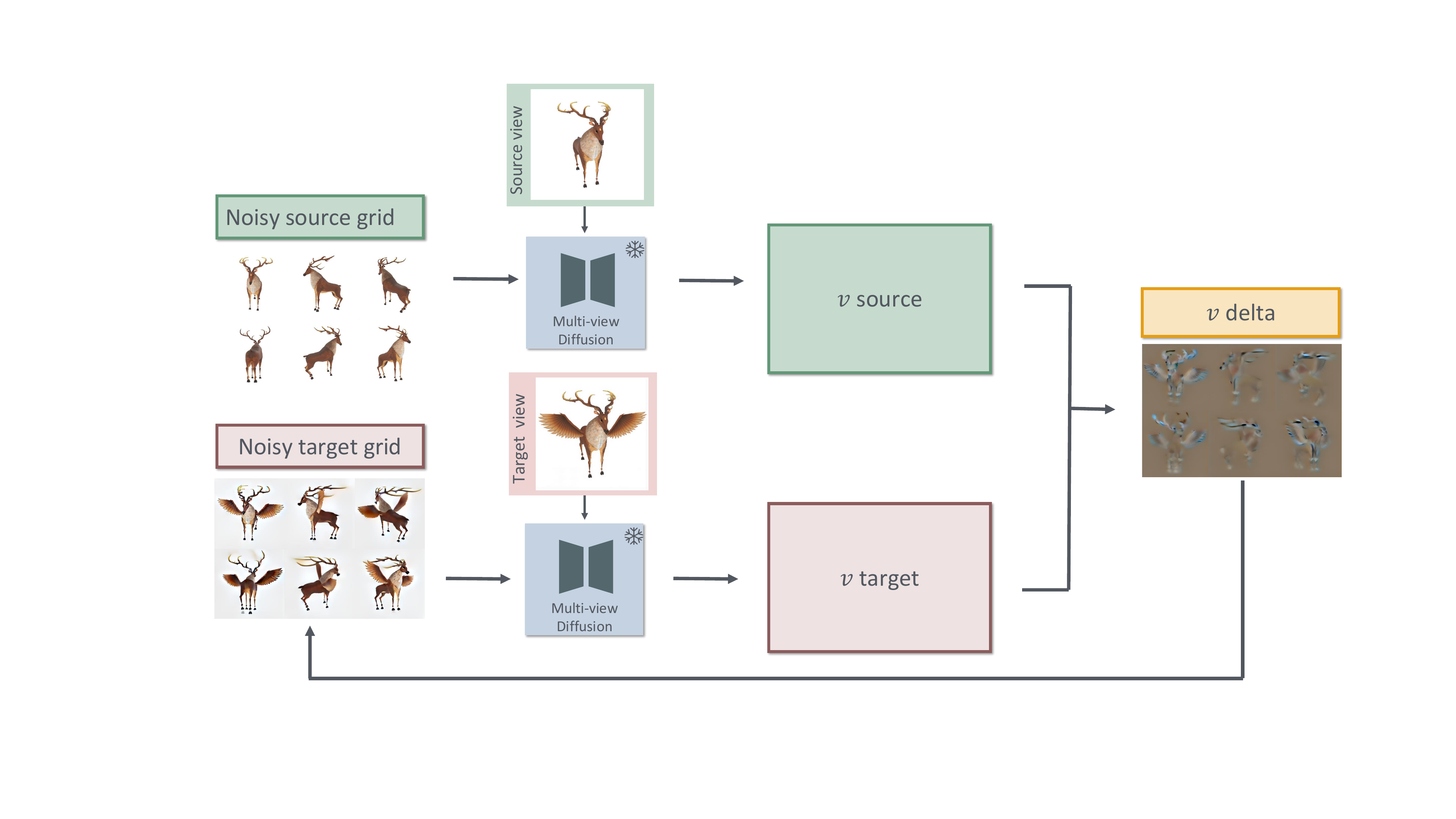}
    \caption{\textbf{Overview of our edit-aware denoising mechanism at a single timestep.} 
Top branch: The original source grid is fed to the multi-view diffusion model along with the  source condition view  to predict the velocity towards the source.
Bottom branch: The current edited grid is conditioned on the target view  to predict the velocity towards the target. 
The resulting delta isolates the edit and guides the subsequent update of the edited grid.}
    \label{fig:method}
\end{figure*}

%% file: figures/baseline.tex
\begin{figure}[htb]
\centering
\begin{tabular}{@{}p{0.23\linewidth}@{}p{0.25\linewidth}@{}p{0.21\linewidth}@{}p{0.26\linewidth}@{}}
Condition & Source & Ours & Baseline \\
\end{tabular}

\begin{subfigure}[b]{0.48\textwidth}
\centering
\hfill

\includegraphics[width=\linewidth]{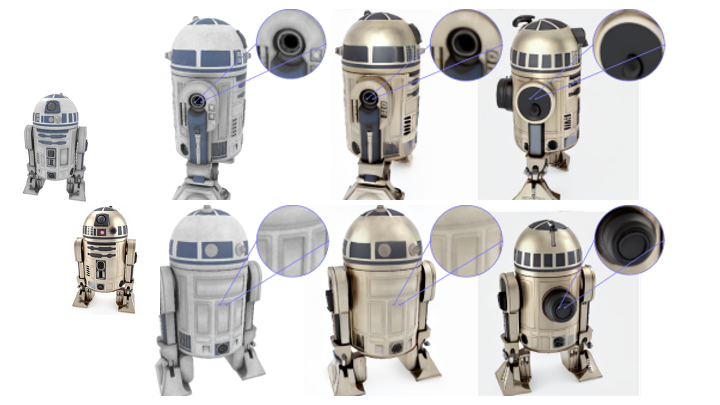}
\end{subfigure}
\hfill
\begin{subfigure}[b]{0.48\textwidth}
\centering
\includegraphics[width=\linewidth]{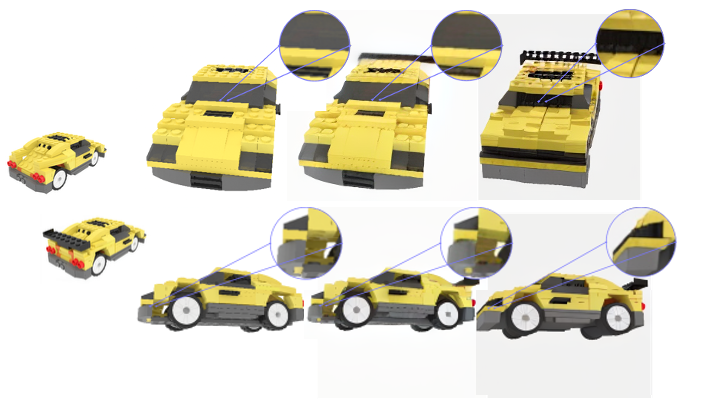}
\end{subfigure}

\caption{\textbf{Comparison  with a \Naive Baseline.} We compare our method with the baseline on two examples: R2D2 (top) and a yellow LEGO car (bottom). The baseline conditions the multi-view diffusion model directly on the edited view. In contrast, our method uses edit-aware denoising to propagate the intended edit consistently across the entire object while preserving structure and appearance. Each example is shown in four columns: editing condition (source and target views), the rendered source object, our result, and the baseline. For each edit, we display two viewpoints. The baseline struggles to retain key semantic features, \eg, hallucinating geometry on R2D2, whereas our method applies the changes coherently and meaningfully, even in the generic LEGO case, without relying on masks or frontal supervision.}

\label{fig:baseline}
\end{figure}

%% file: sections/5_experiments.tex
\input{figures/results_2}

\section{Experiments And Results}
\subsection{Implementation Details}
\label{sec:impl}

\paragraph{Guidance presets.}
We group edits by ``hardness'', from mild texture tweaks to large
geometry additions, and choose between four configurations of $n_{\max}$ and  \(\text{CFG}_{\!\text{tar}}\), keeping all other parameters constant.
Here \(\text{CFG}_{\!\text{tar}}\) is the classifier-free guidance
weight used when predicting the \emph{target} velocity, and
\(n_{\max}\) is the number of scheduler steps we keep for guidance.  
For every example in the paper we select the best of these four presets, this simple sweep was
robust to a wide range of edit types.

\paragraph{Rendering and reconstruction.}
Source meshes are drawn from Objaverse and Objaverse-XL
\cite{objaverse,objaverseXL} and rendered in Blender to produce the reference view and \(V{=}6\) multi-view grid.
For figures that require a final 3D asset 
(\cref{fig:qualitative_comparison,fig:reconstructed_editing_results})
the edited grid is converted to a textured
mesh with Instant Mesh's reconstruction module
~\cite{instantmesh}.  
All other qualitative results are shown directly in multi-view grid form
(before reconstruction) to highlight the efficacy of the propagation
step.
Any 2D editor can supply the image prompts required by \ourmethod{};
in practice, we use
FlowEdit\,\cite{flowedit} for
\emph{global} changes (\eg, overall colour or silhouette) and
FLUX in-painting,~\cite{flux2024} for \emph{local} modifications.
The resulting edited view \(I_{\mathrm{tar}}\) is then fed to our
multi-view propagation pipeline exactly as described.

\input{figures/qualitative}

\subsection{Experiment Details}

\paragraph{Dataset.} 
We evaluate our method on a diverse set of 24 objects spanning various categories including figures, furniture, vehicles, animals, and everyday items. For these objects, we apply 54 different editing prompts that encompass both local and global transformations. Our editing prompts cover a wide range of modification types including pose changes, element additions, global style transformations and more. 

\paragraph{Baseline Methods.}
We compare our method with both mask-free and mask-assisted 3D editing approaches. For mask-free baselines, we conduct both quantitative and qualitative evaluations against MVEdit~\cite{chen2024mvedit} and Vox-E~\cite{voxe}. In contrast, mask-assisted methods inherently suffer from significant limitations: they are unable to perform global edits and require manual creation of 3D masks for local edits, a process that is both labor-intensive and technically challenging. Given these constraints, we include only qualitative comparisons with Instant3Dit~\cite{barda2024instant3dit}. \Cref{fig:qualitative_comparison} shows qualitative results from this comparison.

\paragraph{Evaluation Metrics.}
Following Vox-E, we evaluate our 3D editing results using two CLIP-based metrics that assess different aspects of edit quality:
\noindent \textbf{CLIP Similarity} ($\text{CLIP}_\text{Sim}$) measures the semantic alignment between our edited 3D objects and the target text prompts. We compute this metric by extracting CLIP embeddings from both the target editing instruction and rendered images of our generated 3D outputs, then calculating the cosine similarity between these text and image embeddings.
\noindent \textbf{CLIP Direction Similarity} ($\text{CLIP}_\text{Dir}$) evaluates the consistency of our editing transformations by measuring directional changes in CLIP embedding space, following the approach introduced by ~\cite{gal2021stylegannadaclipguideddomainadaptation}. This metric computes the cosine similarity between the direction vector from original to edited object in image embedding space and the direction vector from source to target description in text embedding space, ensuring that our edits follow semantically meaningful transformations.

\paragraph{User Study.} 
\input{tables/table_comp}

\input{figures/user_study}

We conducted a user study to complement our quantitative evaluation. Using a 2-alternative forced choice setup, participants were shown the original 3D object along with the editing prompt, followed by results from our method and a baseline method in random order. Participants were asked to select which result better aligned with the edit prompt while avoiding introducing unintended changes to the original shape. We collected responses from 48 different participants who answered \num{1248} questions across our benchmark. The results, presented in~\cref{fig:human_study}, demonstrate that our method is strongly preferred.

\subsection{Results}

As demonstrated in~\cref{tab:performance_comparison}, our approach consistently outperforms all baseline methods on both alignment and preservation metrics. The superior performance on alignment metrics indicates that our method produces edited 3D objects that better correspond to the provided edit prompts, while the higher preservation scores demonstrate that our approach maintains fidelity to the original object geometry in unedited regions.

The quantitative improvements are further supported by our qualitative evaluation presented in~\cref{fig:qualitative_comparison}. Our method demonstrates robust performance across diverse editing scenarios, including local geometric modifications and global texture changes as shown in the examples in \cref{fig:results}. In contrast, other methods diverge significantly from the original objects and yield less realistic results.

The user study results strongly corroborate our quantitative findings. As shown in~\cref{fig:human_study}, participants preferred our method in 81\% of comparisons against MVEdit and 93\% against Vox-E, demonstrating significant user preference across our diverse set of editing scenarios.

\subsection{Ablation Studies}
\input{figures/ablation}

To understand the impact of each component of our proposed specific edit-aware denoising mechanism, we conduct ablation studies. The qualitative results of these studies are presented in~\cref{fig:ablation}.
Our analysis focuses on two key variants, which we term ``SDEdit'' and ``FlowEdit'' to demonstrate why our chosen approach is optimal.
The first variant, \textbf{SDEdit}: This variant ablates the source conditioned prediction term ($v_\text{src}$) from the update step. This simplifies our method to a process akin to a standard SDEdit step, where guidance comes only from the target condition (in our case---a single view edit). As observed in~\cref{fig:ablation}, this simplification leads to several issues: a noticeable degradation in detail, a failure to apply geometric manipulations, and less coherent integration of the intended changes.
The second variant, \textbf{FlowEdit}: This variant modifies our DDS-inspired noise application strategy. Instead of adding the same noise realization $N_\text{grid}$ to both $X_\text{src}$ and $X_\text{edit}$, we perturb $X_\text{src}$ with noise to form $Z_\text{src}$ and apply the corresponding displacement $Z_\text{src}-X_\text{src}$ to $X_\text{edit}$ to form $Z_\text{edit}$. This approach can introduce artifacts such as blurry remainders of the source object (\eg, subtle traces of Superman's original arm positions). It can also struggle to precisely maintain the original object's proportions

 Our full method achieves significantly better results across all metrics, demonstrating superior edit quality, while retaining details unrelated to the editing intent, and improved alignment with image prompts.

%% file: figures/results_2.tex
\setlength{\tabcolsep}{0pt}
\renewcommand{\arraystretch}{0.6} %
\begin{figure*}[t]
\centering
\begin{tabular}{
  >{\centering\arraybackslash}m{0.03\linewidth} %
  >{\centering\arraybackslash}m{0.13\linewidth} 
  >{\centering\arraybackslash}m{0.17\linewidth} 
  >{\centering\arraybackslash}m{0.17\linewidth} 
  @{\hspace{1pt}} %
  >{\centering\arraybackslash}m{0.13\linewidth} 
  >{\centering\arraybackslash}m{0.17\linewidth} 
  >{\centering\arraybackslash}m{0.17\linewidth}
}
& \textbf{Cond. View} & \textbf{View 1} & \textbf{View 2} & \textbf{Cond. View} & \textbf{View 1} & \textbf{View 2} \\

\rotatebox[origin=c]{90}{Original} &
\adjustbox{valign=m}{\includegraphics[width=\linewidth, trim={0.5cm 0.5cm 0.5cm 0.5cm}, clip]{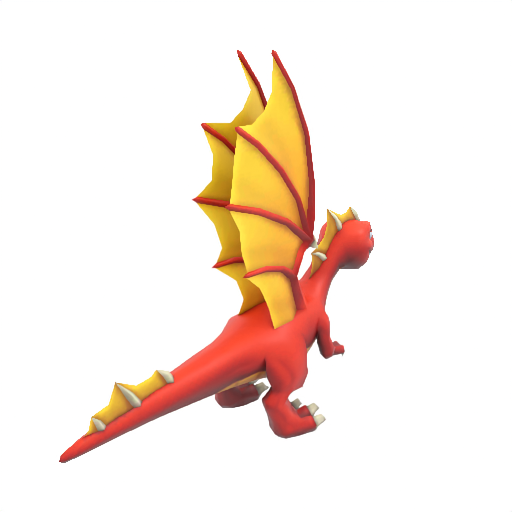}} &
\adjustbox{valign=m}{%
    \begin{tikzpicture}[spy using outlines={circle, myspycolor, magnification=2, size=0.6cm, connect spies}]
    \node {\includegraphics[width=\linewidth, trim={0.5cm 0.5cm 0.5cm 0.5cm}, clip]{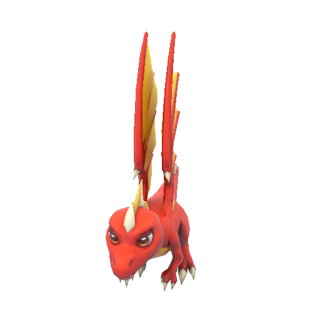}};
    \spy on (0.31,0) in node [left] at (1.4,1); %
    \end{tikzpicture}%
} &
\adjustbox{valign=m}{\includegraphics[width=\linewidth, trim={0cm 0.1cm 0.5cm 0.5cm}, clip]{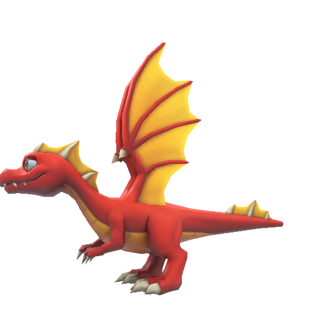}} &
\adjustbox{valign=m}{\includegraphics[width=\linewidth, trim={0.5cm 0.5cm 0.5cm 0.5cm}, clip]{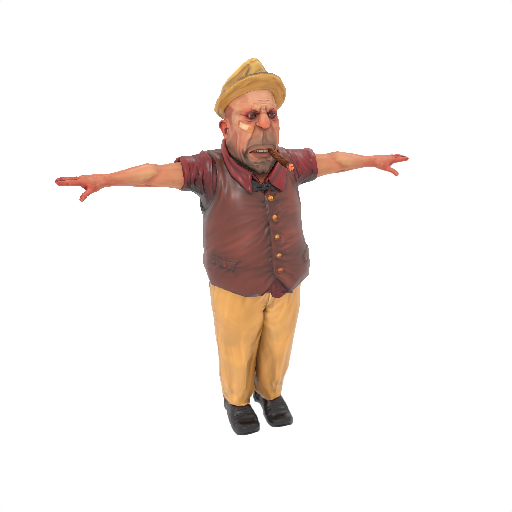}} &
\adjustbox{valign=m}{\includegraphics[width=\linewidth, trim={0.5cm 0.5cm 0.5cm 0.5cm}, clip]{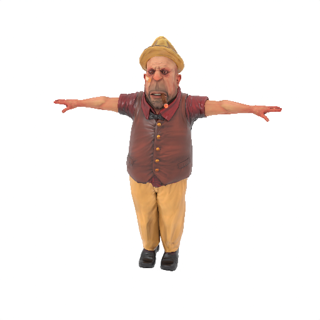}} &
\adjustbox{valign=m}{%
    \begin{tikzpicture}[spy using outlines={circle, myspycolor, magnification=2, size=0.6cm, connect spies}]
    \node {\includegraphics[width=\linewidth, trim={0.5cm 0.5cm 0.5cm 0.5cm}, clip]{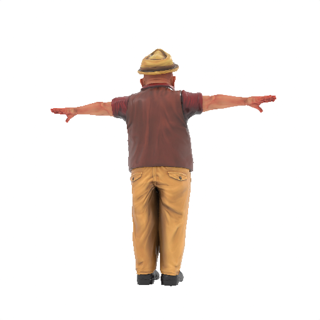}};
    \spy on (0.2,-0.2) in node [left] at (1.4,1); %
    \end{tikzpicture}%
} \\

\rotatebox[origin=c]{90}{Edited} &
\adjustbox{valign=m}{\includegraphics[width=\linewidth, trim={0.5cm 0.5cm 0.5cm 0.5cm}, clip]{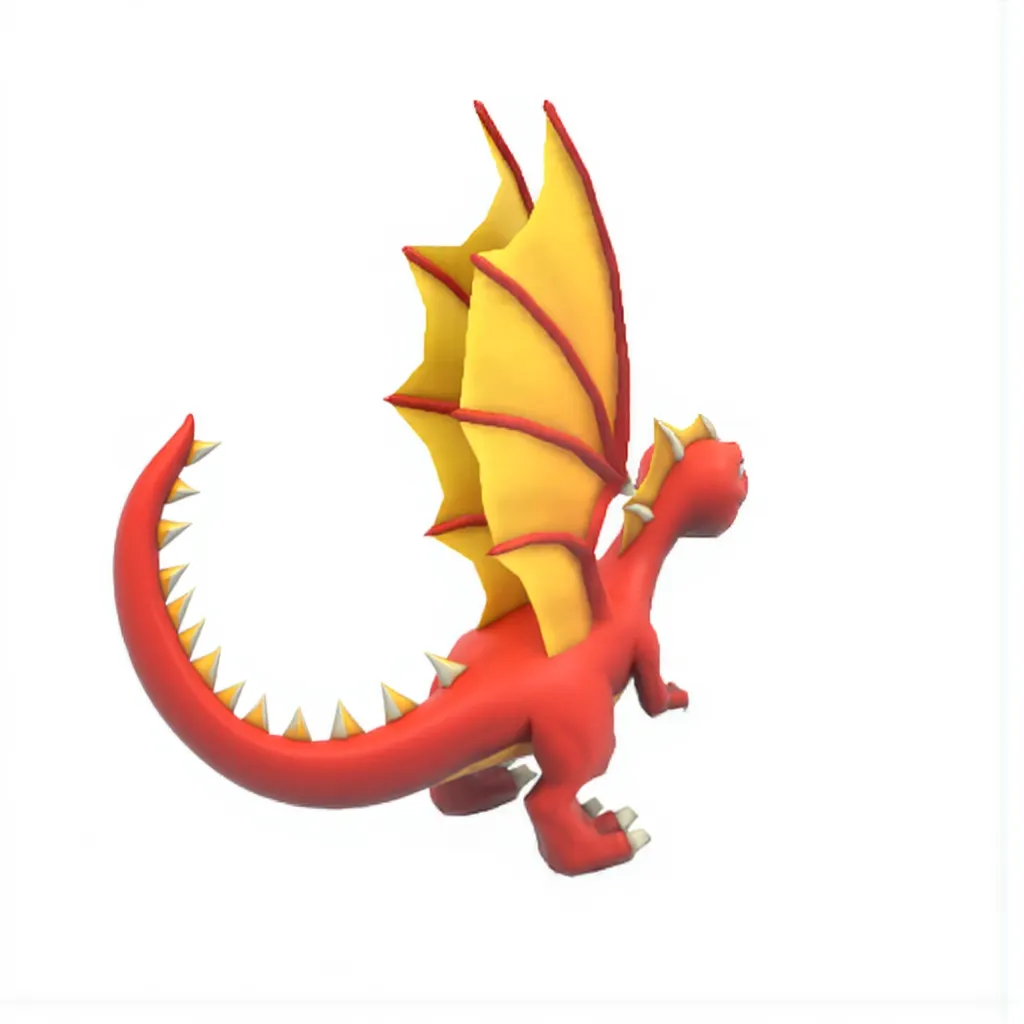}} &
\adjustbox{valign=m}{%
    \begin{tikzpicture}[spy using outlines={circle, myspycolor, magnification=2, size=0.6cm, connect spies}]
    \node {\includegraphics[width=\linewidth, trim={0.5cm 0.5cm 0.5cm 0.5cm}, clip]{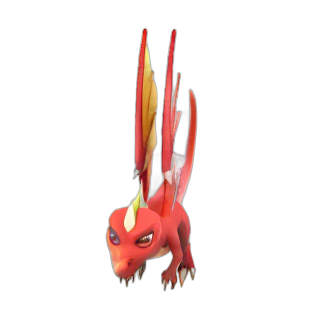}};
    \spy on (0.31,0) in node [left] at (1.45,1); %
    \end{tikzpicture}%
} &
\adjustbox{valign=m}{\includegraphics[width=\linewidth, trim={0.1cm 0.1cm 0.5cm 0.5cm}, clip]{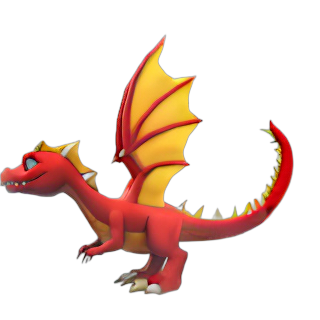}} &
\adjustbox{valign=m}{\includegraphics[width=\linewidth, trim={0.5cm 0.5cm 0.5cm 0.5cm}, clip]{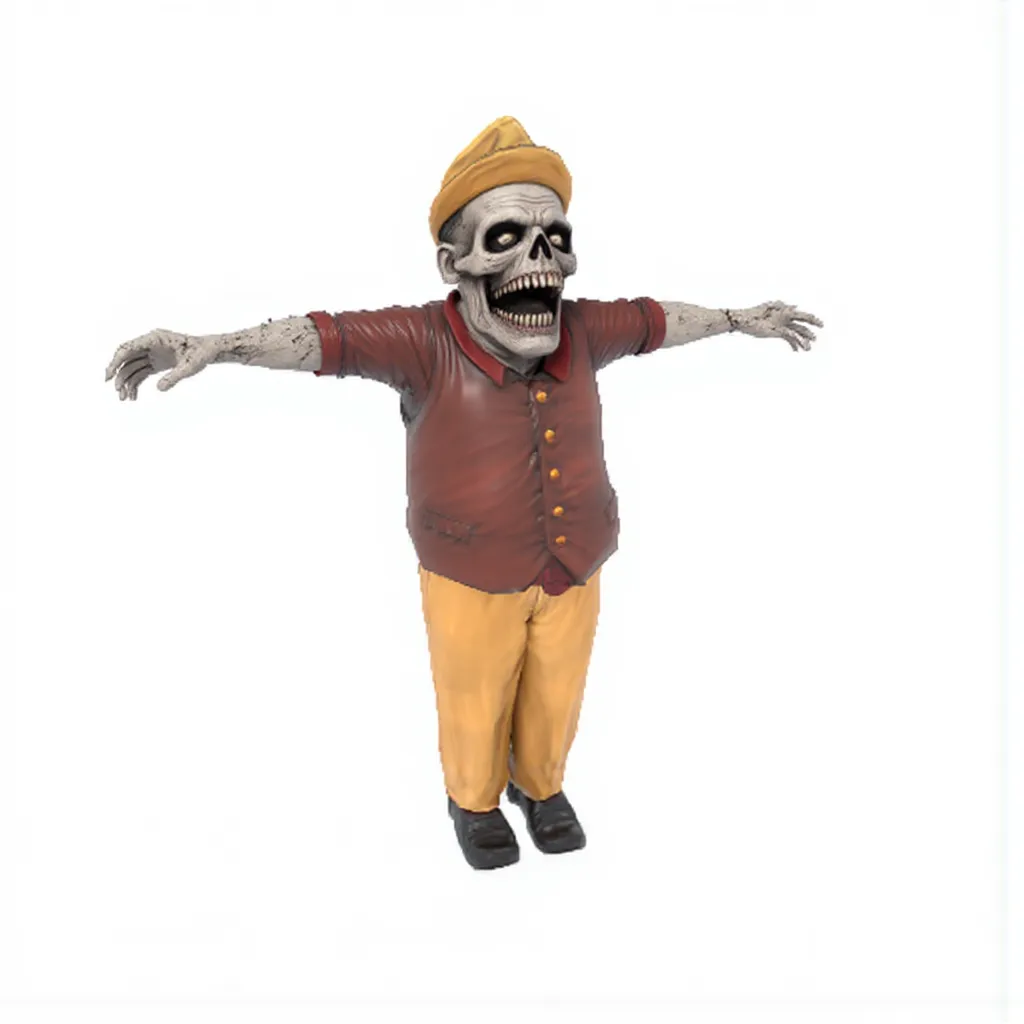}} &
\adjustbox{valign=m}{\includegraphics[width=\linewidth, trim={0.5cm 0.5cm 0.5cm 0.5cm}, clip]{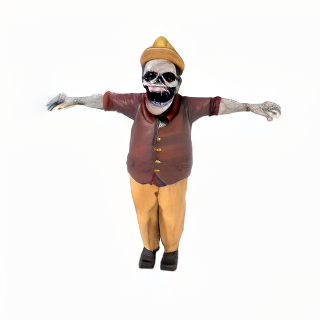}} &
\adjustbox{valign=m}{%
    \begin{tikzpicture}[spy using outlines={circle, myspycolor, magnification=2, size=0.6cm, connect spies}]
    \node {\includegraphics[width=\linewidth, trim={0.5cm 0.5cm 0.5cm 0.5cm}, clip]{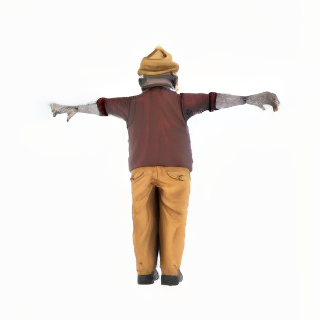}};
    \spy on (0.2,-0.2) in node [left] at (1.45,1); %
    \end{tikzpicture}%
} \\

\rotatebox[origin=c]{90}{Original} &
\adjustbox{valign=m}{\includegraphics[width=\linewidth, trim={0.5cm 0.5cm 0.5cm 0.5cm}, clip]{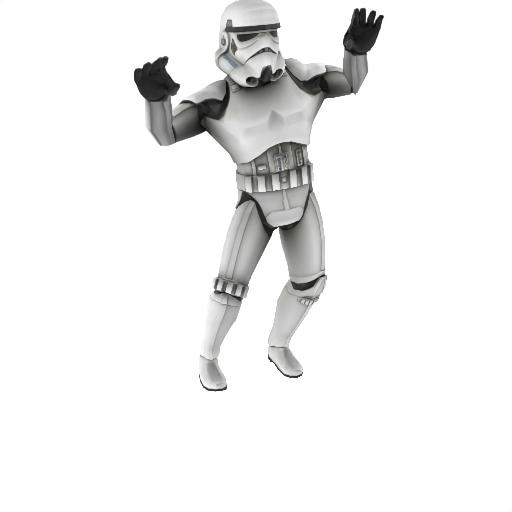}} &
\adjustbox{valign=m}{\includegraphics[width=\linewidth, trim={0.5cm 0.5cm 0.5cm 0.5cm}, clip]{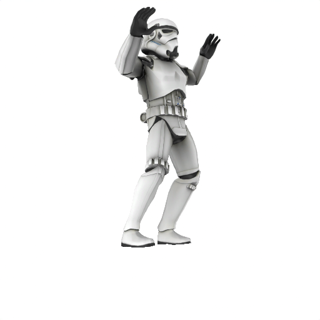}} &
\adjustbox{valign=m}{\includegraphics[width=\linewidth, trim={0.5cm 0.5cm 0.5cm 0.5cm}, clip]{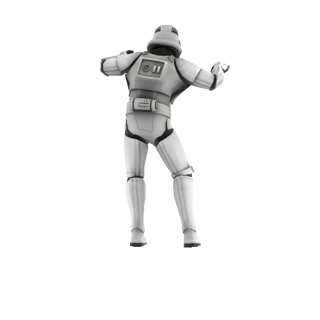}} &
\adjustbox{valign=m}{\includegraphics[width=\linewidth, trim={0.5cm 0cm 0.5cm 2.5cm}, clip]{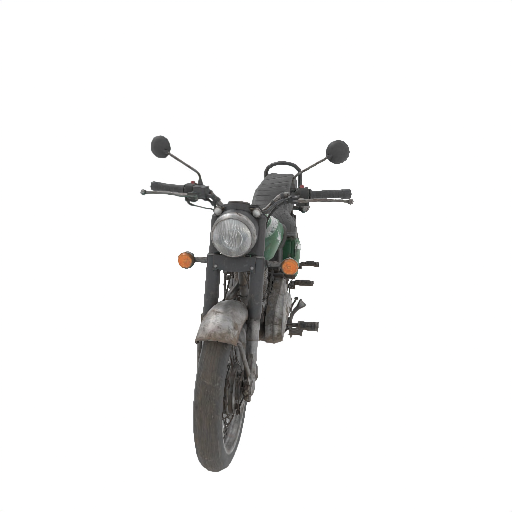}} &
\adjustbox{valign=m}{\includegraphics[width=\linewidth, trim={0.5cm 0.5cm 0.5cm 2.5cm}, clip]{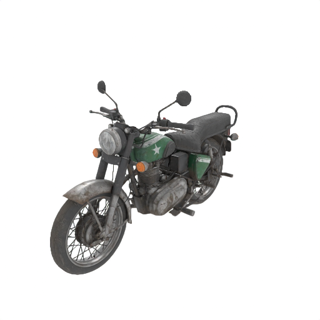}} &
\adjustbox{valign=m}{\includegraphics[width=\linewidth, trim={0.5cm 0.5cm 0.5cm 2.5cm}, clip]{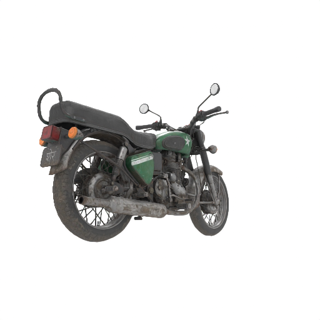}} \\

\rotatebox[origin=c]{90}{Edited} &
\adjustbox{valign=m}{\includegraphics[width=\linewidth, trim={0.5cm 0.5cm 0.5cm 0.5cm}, clip]{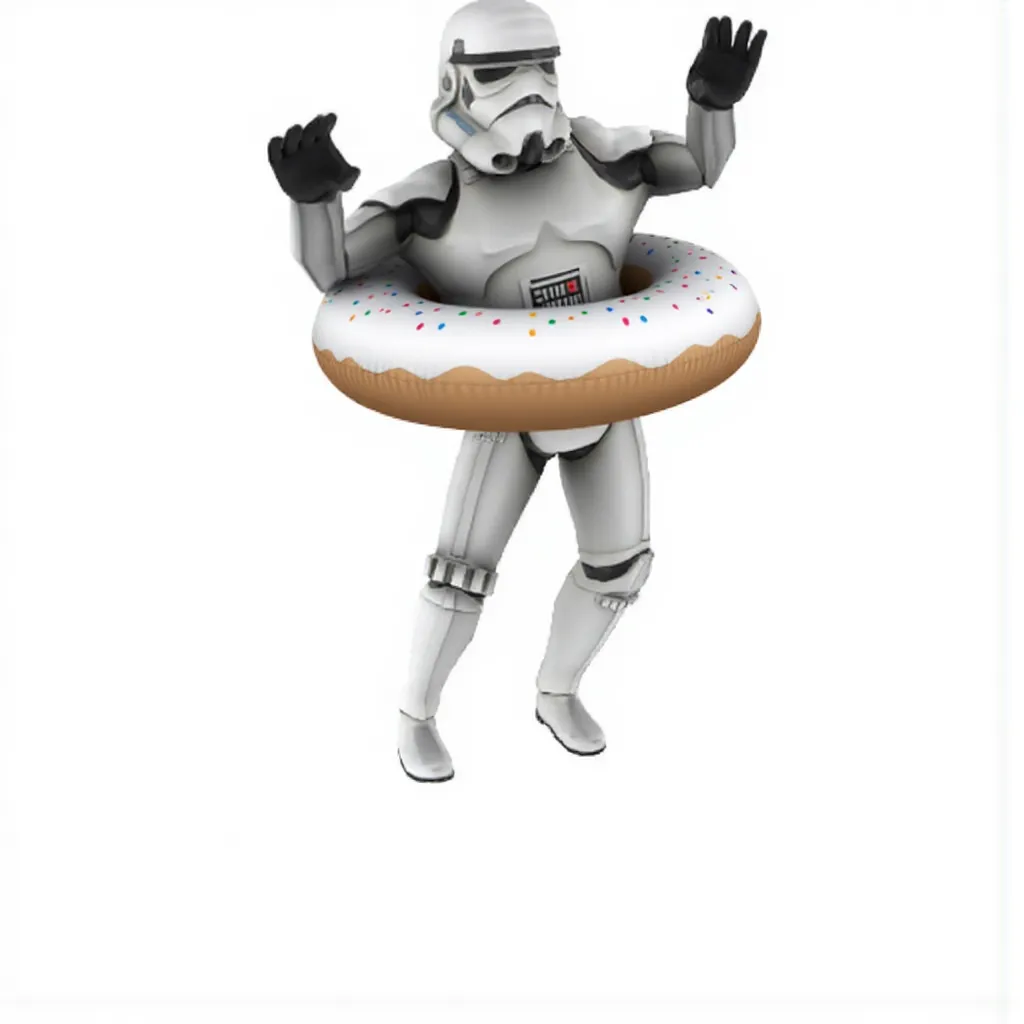}} &
\adjustbox{valign=m}{\includegraphics[width=\linewidth, trim={0.5cm 0.5cm 0.5cm 0.5cm}, clip]{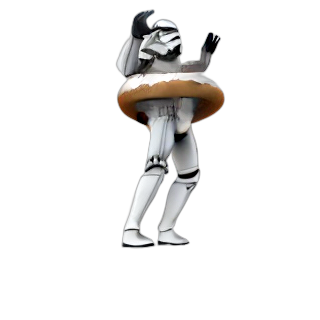}} &
\adjustbox{valign=m}{\includegraphics[width=\linewidth, trim={0.5cm 0.5cm 0.5cm 0.5cm}, clip]{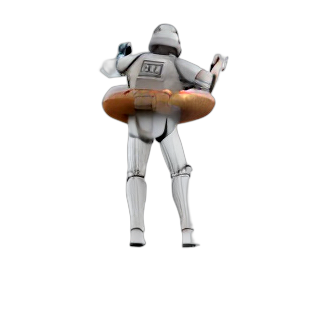}} &
\raisebox{5mm}{\adjustbox{valign=m}{\includegraphics[width=\linewidth, trim={0.5cm 0cm 0.5cm 0.5cm}, clip]{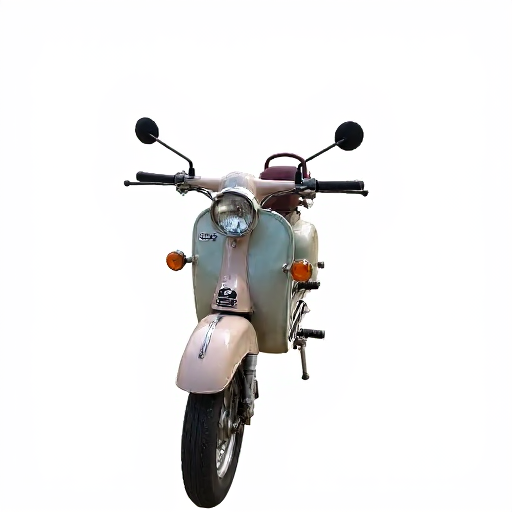}}} &
\adjustbox{valign=m}{\includegraphics[width=\linewidth, trim={0.5cm 0.5cm 0.5cm 2.5cm}, clip]{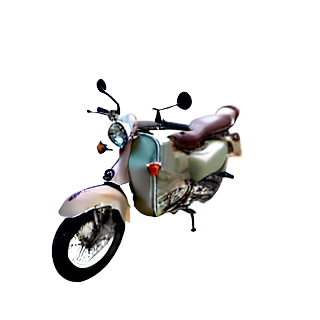}} &
\adjustbox{valign=m}{\includegraphics[width=\linewidth, trim={0.5cm 0.5cm 0.5cm 2.5cm}, clip]{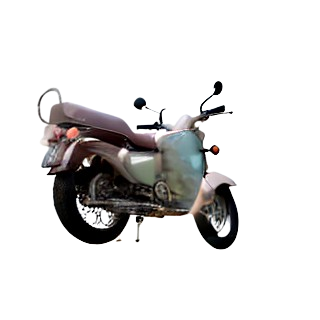}} \\

\end{tabular}
\caption{\textbf{Qualitative Results of EditP23.} This figure showcases results  across diverse object categories. Each block compares a source object (top) with its edited version (bottom). The leftmost column displays the conditioning views (source and target) used to prompt the edit, while the remaining columns show novel views of the result. Our approach consistently applies the desired edit while preserving the object's structure and identity across all viewpoints.
}
\label{fig:results}
\end{figure*}

%% file: figures/qualitative.tex
\begin{figure}[hbtp]
    \centering
    \setlength{\tabcolsep}{1pt}
    \renewcommand{\arraystretch}{1.2}
\begin{tabular}{@{}c@{}c@{}c@{}c@{}c@{}c@{}}
    & Input & Ours & MVEdit & Instant3Dit & Vox-E \\

    \multirow{2}{*}{\rotatebox{90}{\small ``with headphones''}} &
    \includegraphics[width=0.18\linewidth]{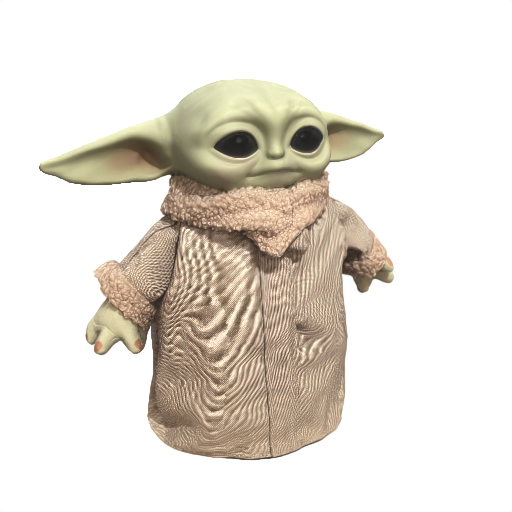} &
    \includegraphics[width=0.18\linewidth]{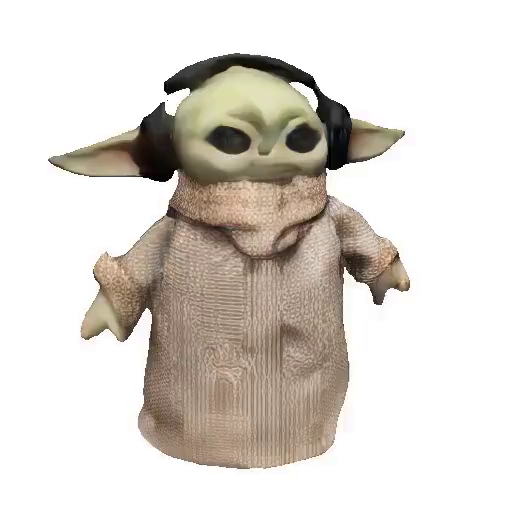} &
    \includegraphics[width=0.18\linewidth]{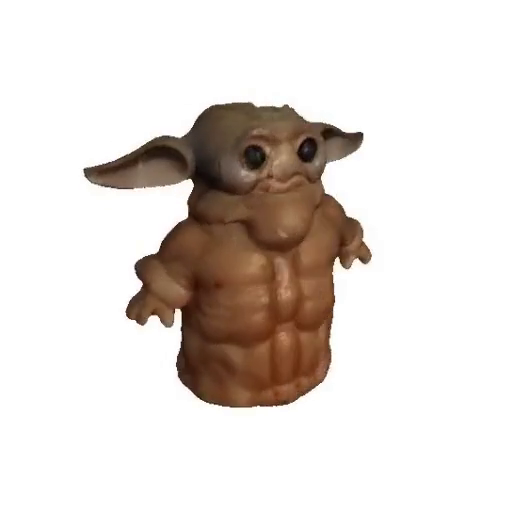} &
    \includegraphics[width=0.18\linewidth, trim={10cm 0cm 10cm 10cm}, clip]{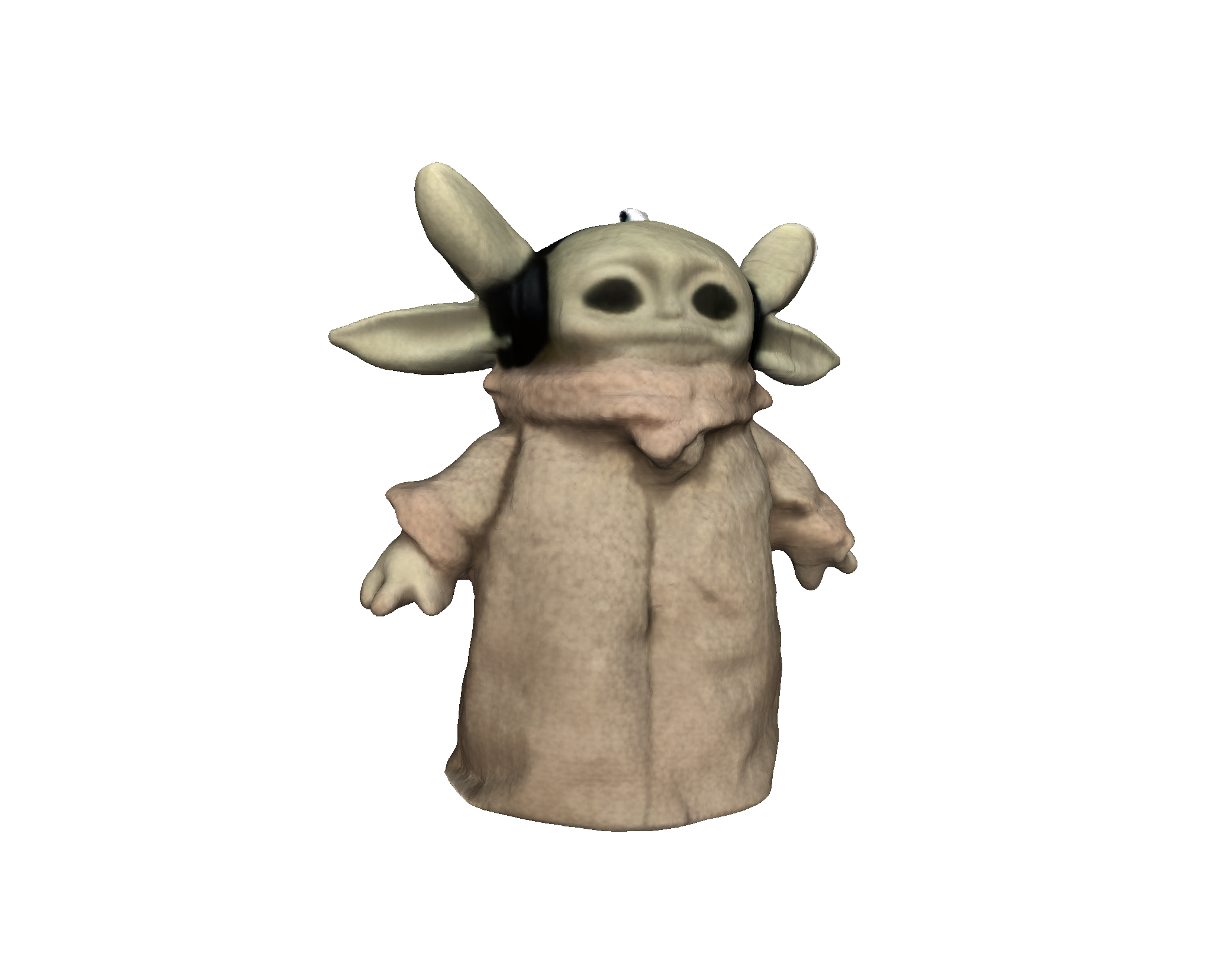} &
    \includegraphics[width=0.18\linewidth]{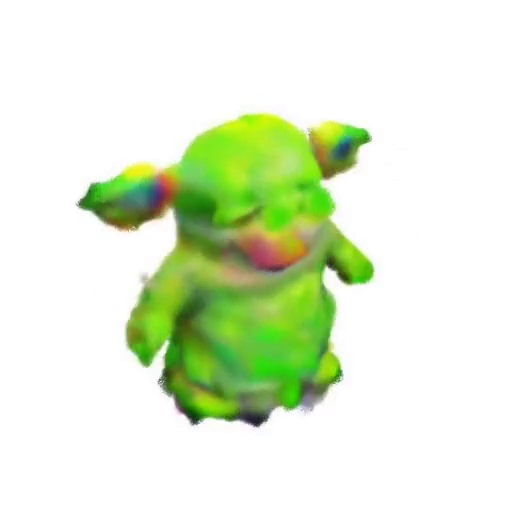} \\

    & \includegraphics[width=0.18\linewidth]{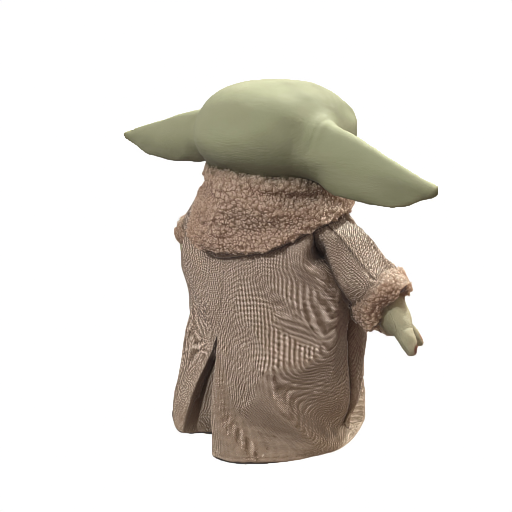} &
    \includegraphics[width=0.18\linewidth]{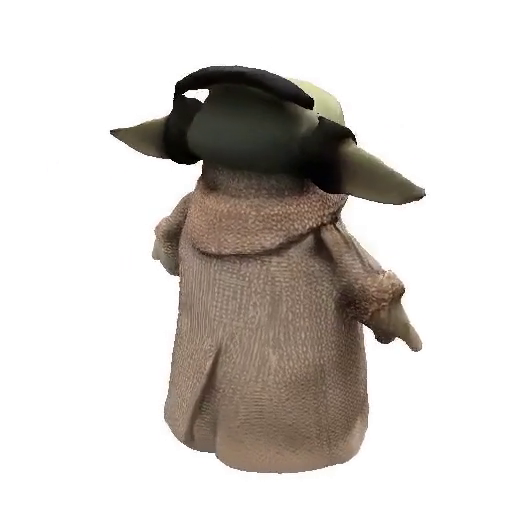} &
    \includegraphics[width=0.18\linewidth]{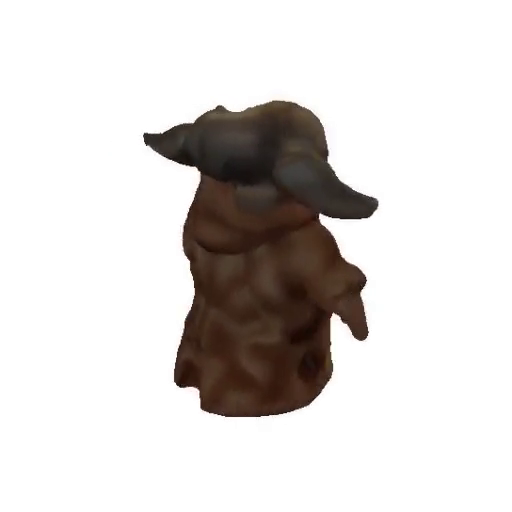} &
    \includegraphics[width=0.18\linewidth, trim={10cm 0cm 10cm 10cm}, clip]{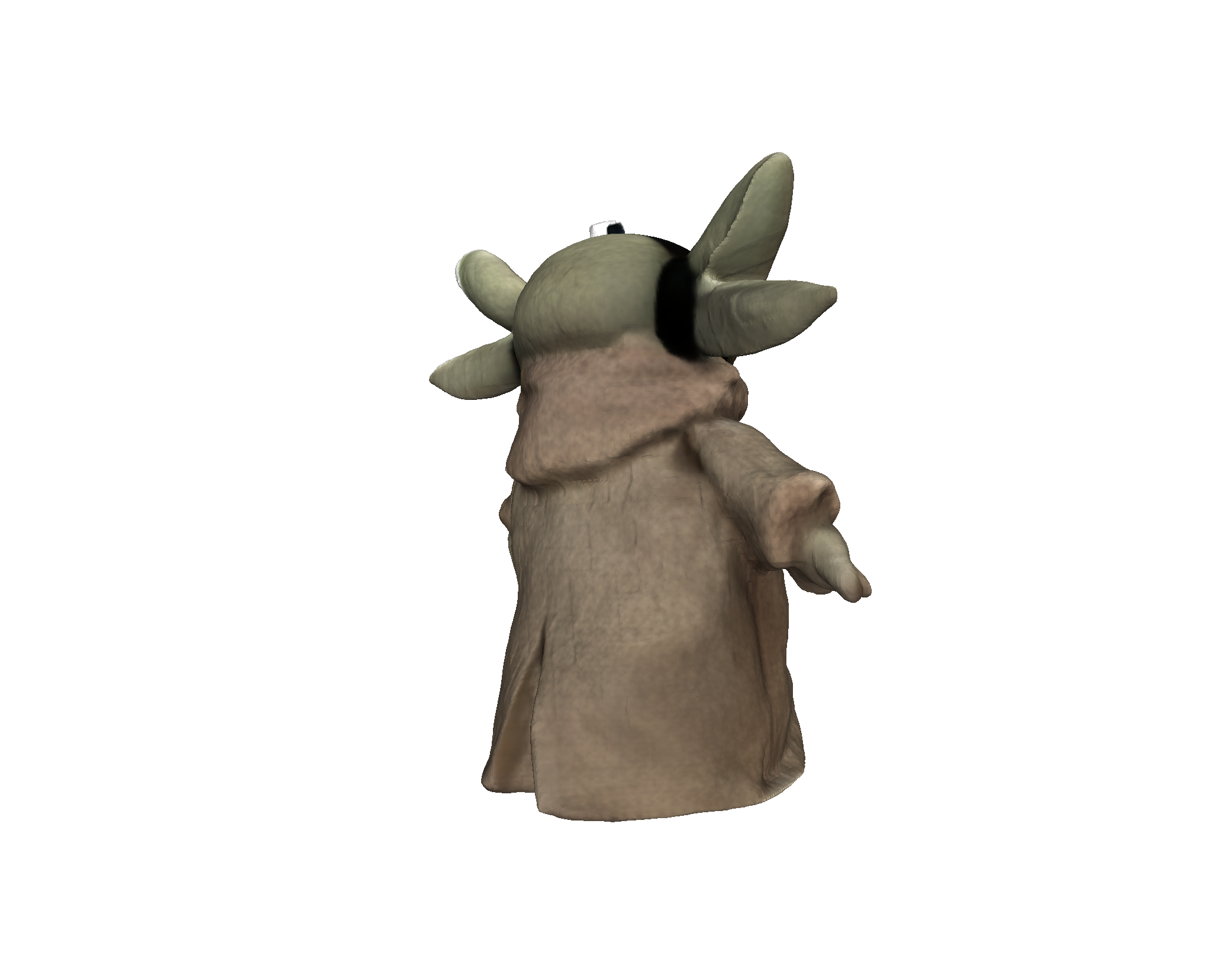} &
    \includegraphics[width=0.18\linewidth]{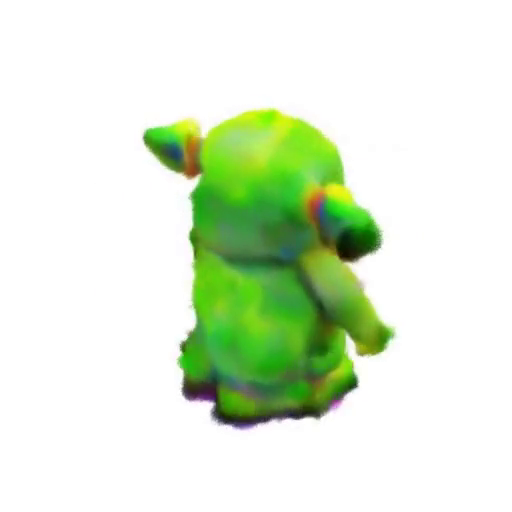} \\

    \multirow{2}{*}{\rotatebox{90}{\small ``with pagoda roof''}} &
    \raisebox{1mm}{\includegraphics[width=0.18\linewidth]{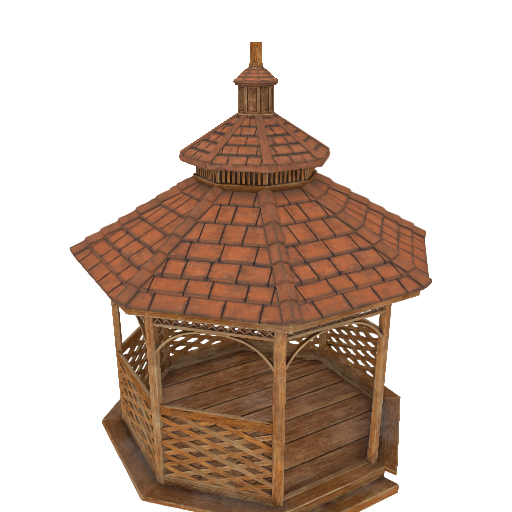}} &
    \includegraphics[width=0.18\linewidth]{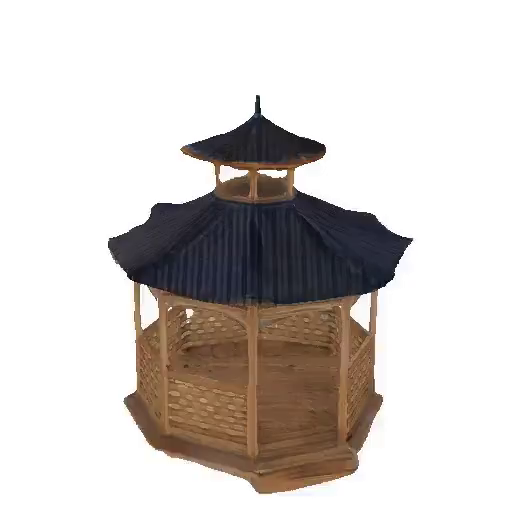} &
    \includegraphics[width=0.18\linewidth,trim={0cm 2cm 0cm 0cm}, clip]{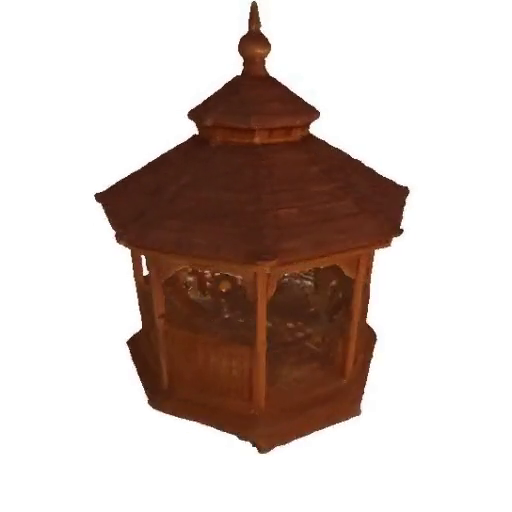} &
    \includegraphics[width=0.18\linewidth, trim={18cm 10cm 18cm 18cm}, clip]{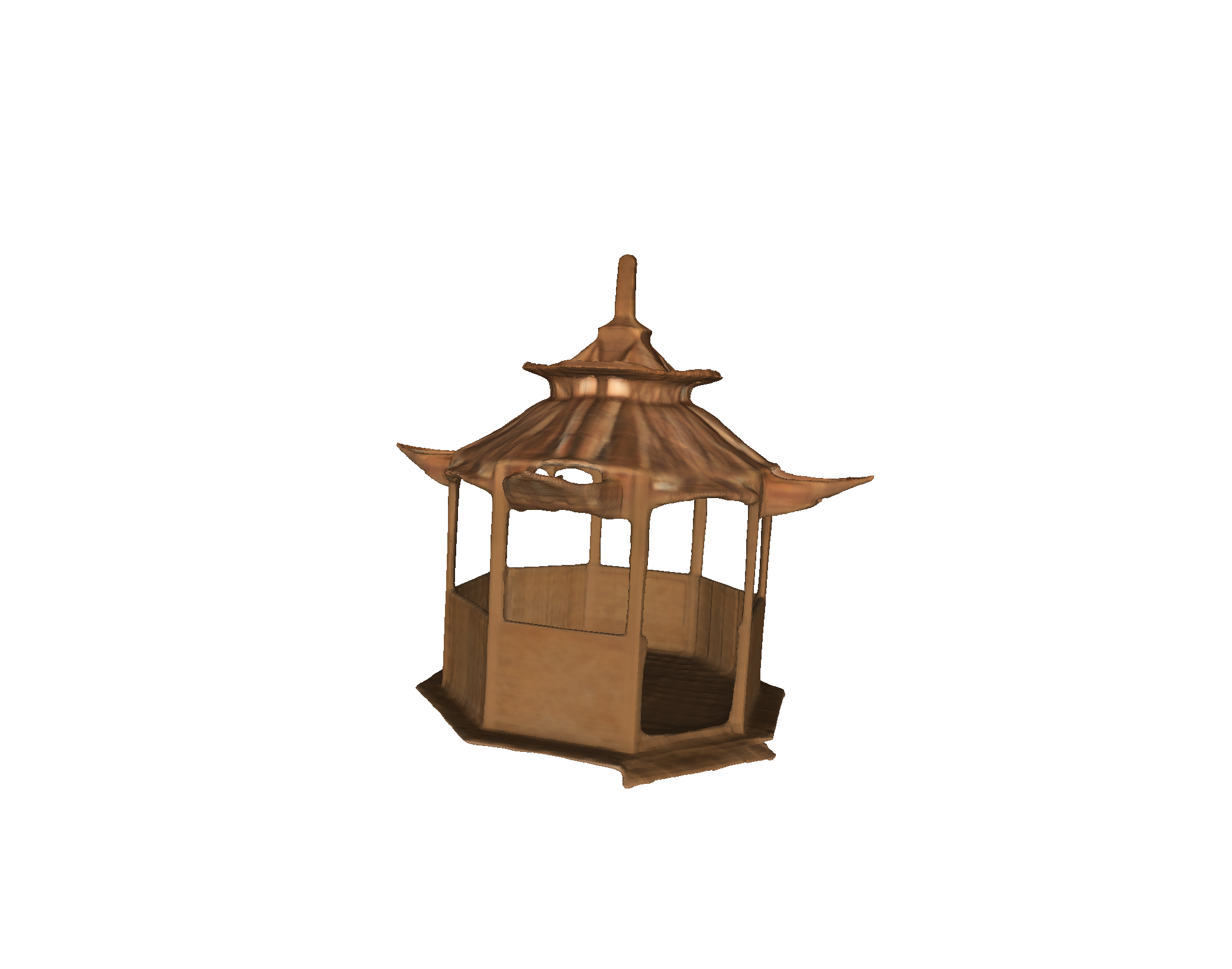} &
    \includegraphics[width=0.18\linewidth]{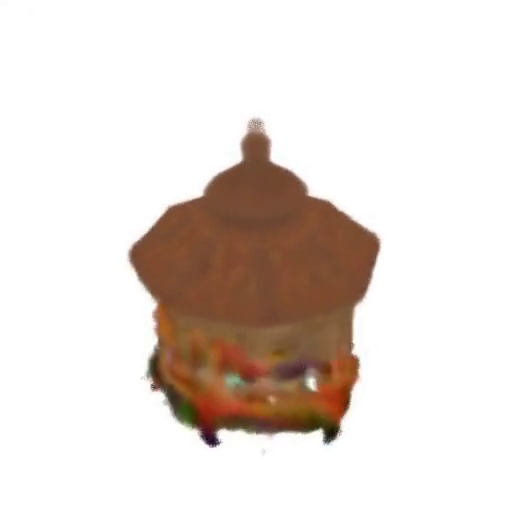} \\

    & \raisebox{1mm}{\includegraphics[width=0.18\linewidth]{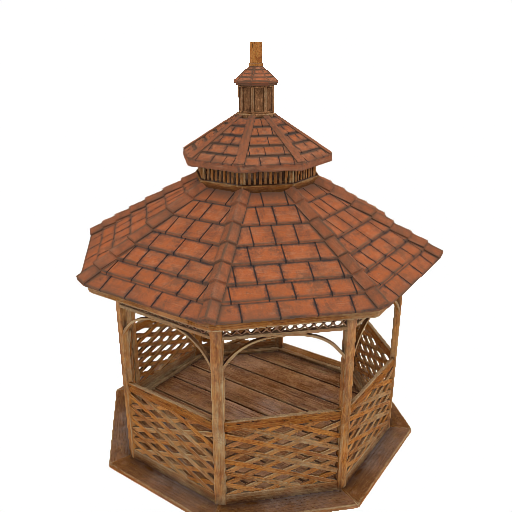}} &
    \includegraphics[width=0.18\linewidth]{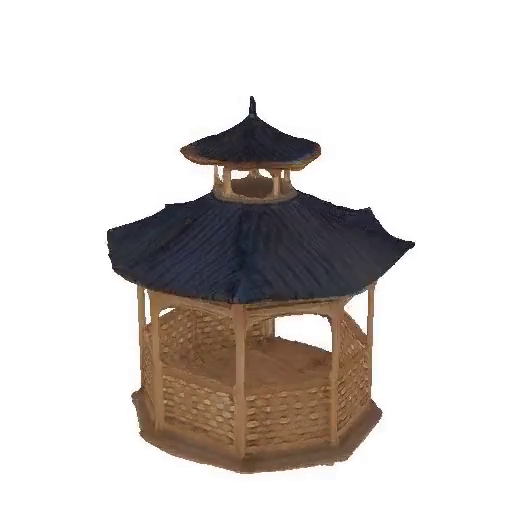} &
    \includegraphics[width=0.18\linewidth,trim={0cm 2cm 0cm 0cm}, clip]{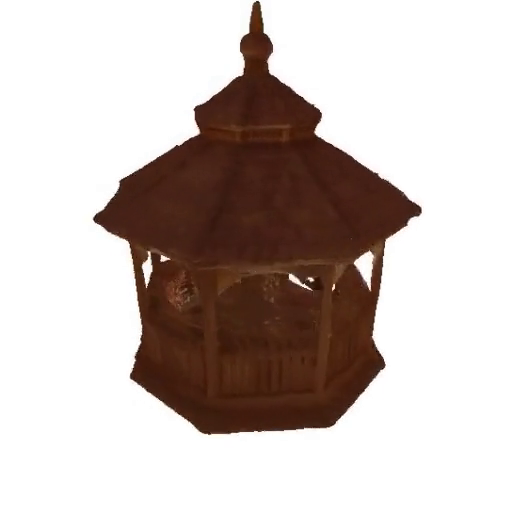} &
    \includegraphics[width=0.18\linewidth, trim={18cm 10cm 18cm 18cm}, clip]{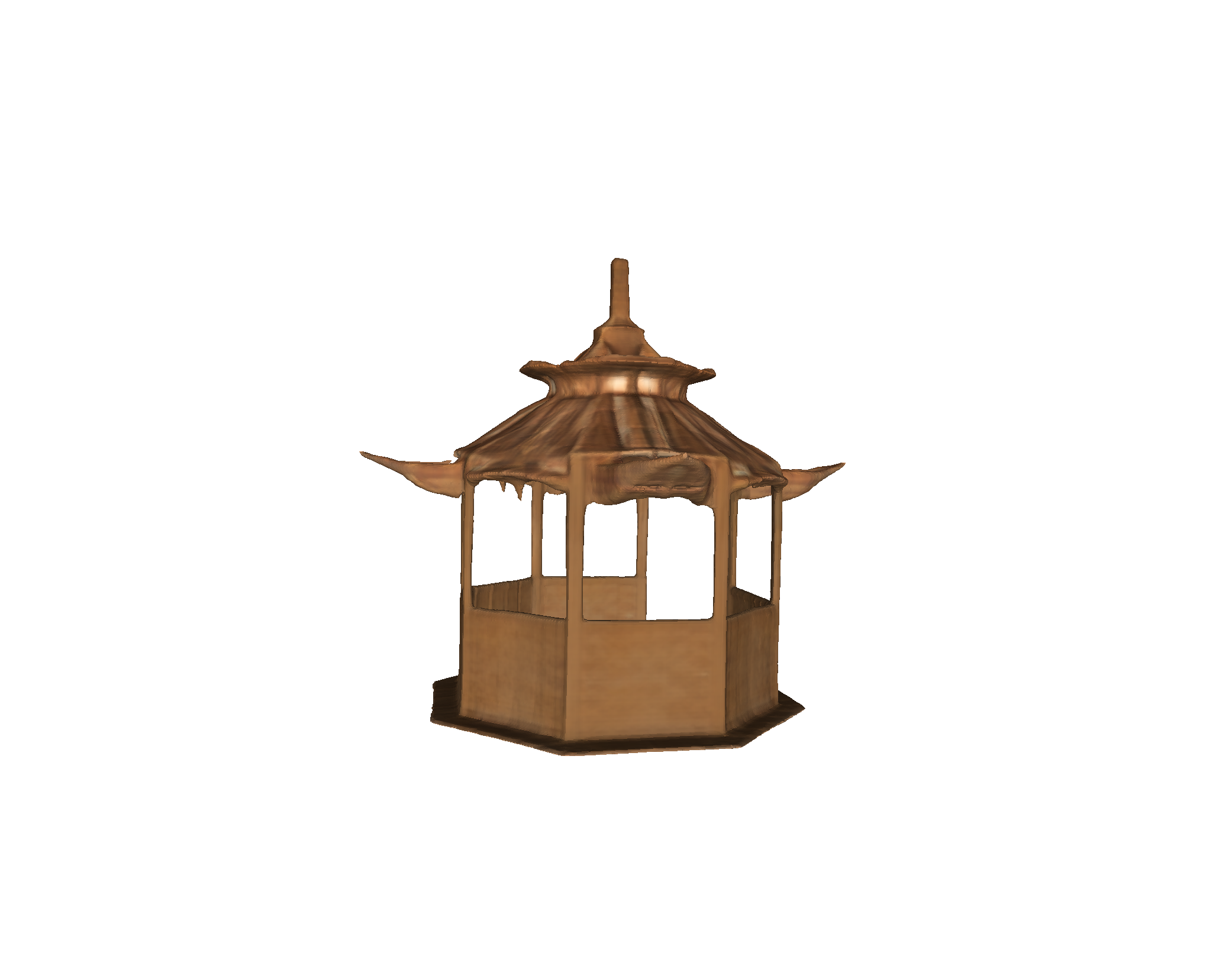} &
    \includegraphics[width=0.18\linewidth]{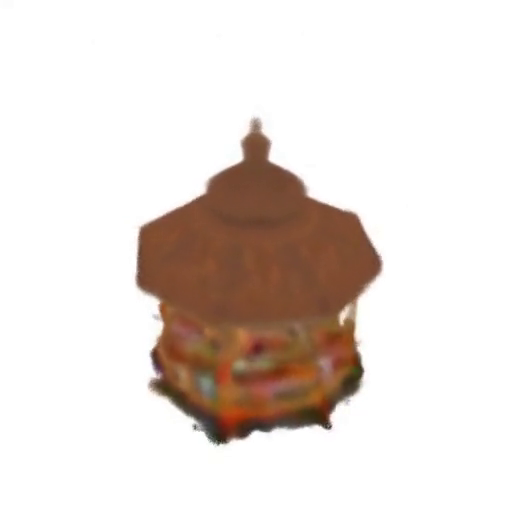} \\

    \multirow{2}{*}{\rotatebox{90}{\small ``cartoonish''}} &
    \raisebox{1mm}{\includegraphics[width=0.18\linewidth, trim={1cm 0cm 3cm 3cm}, clip]{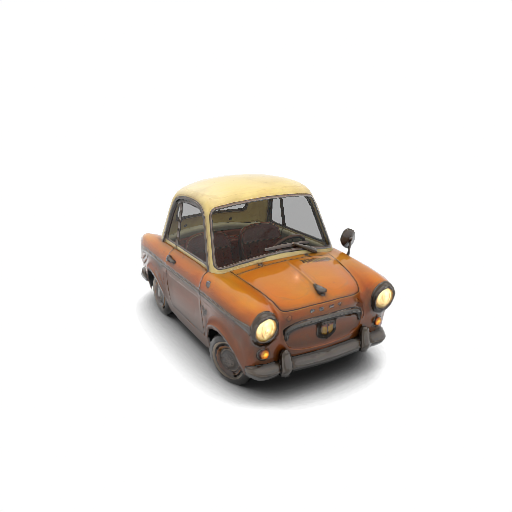}} &
    \includegraphics[width=0.18\linewidth, trim={3.1cm 1.1cm 3.1cm 3.1cm}, clip]{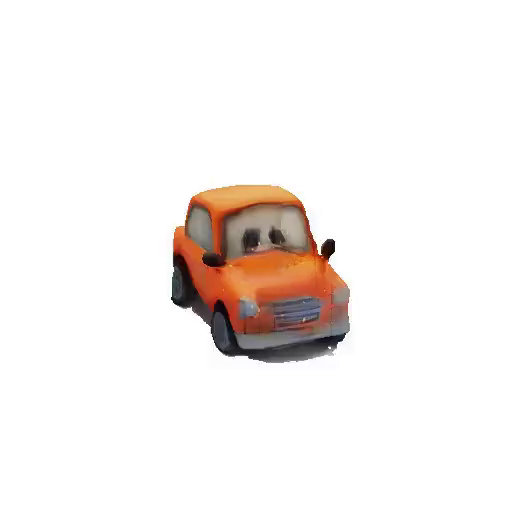} &
    \includegraphics[width=0.18\linewidth, trim={3.1cm 1.1cm 3.1cm 3.1cm}, clip]{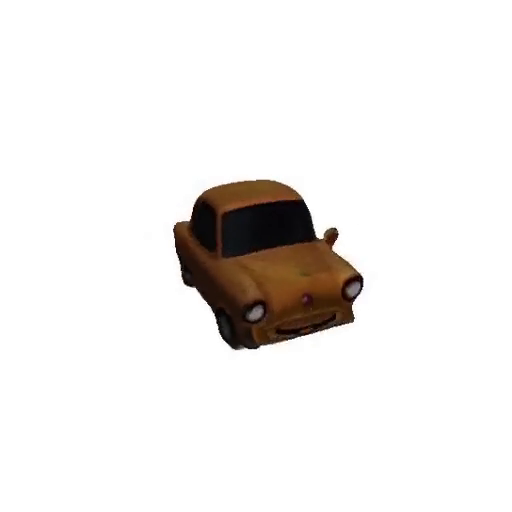} &
    \multicolumn{1}{c}{\raisebox{0.8cm}{N/A}}&
    \includegraphics[width=0.18\linewidth, trim={2cm 2cm 2cm 2cm}, clip]{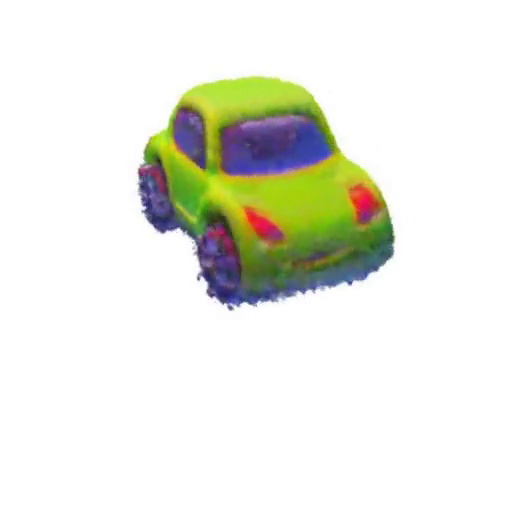} \\

    & \raisebox{1mm}{\includegraphics[width=0.18\linewidth, trim={1cm 0cm 3cm 3cm}, clip]{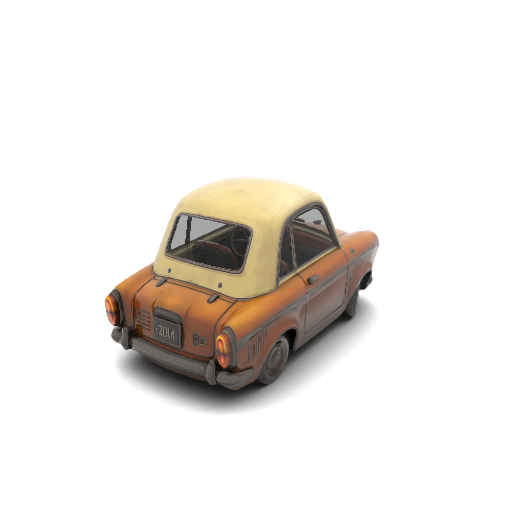}} &
    \includegraphics[width=0.18\linewidth, trim={3.1cm 1.1cm 3.1cm 5cm}, clip]{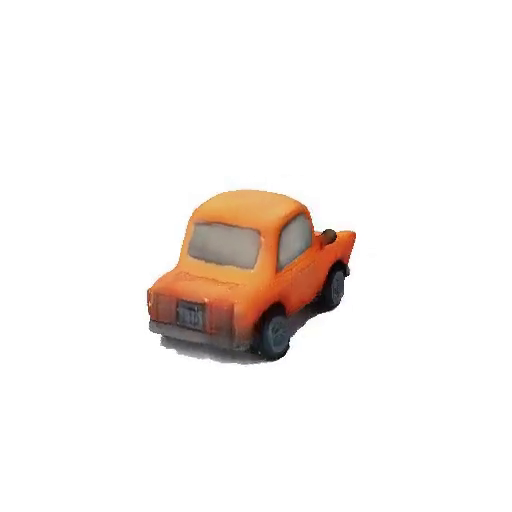} &
    \includegraphics[width=0.18\linewidth, trim={3.1cm 1.1cm 3.1cm 5cm}, clip]{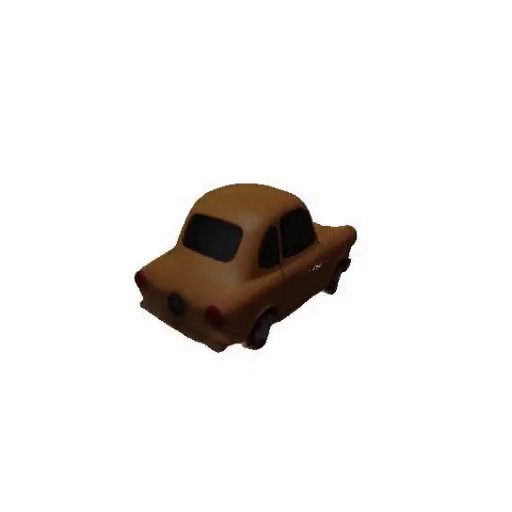} &
    \multicolumn{1}{c}{\raisebox{0.8cm}{N/A}}&
    \includegraphics[width=0.18\linewidth, trim={2cm 2cm 2cm 1cm}, clip]{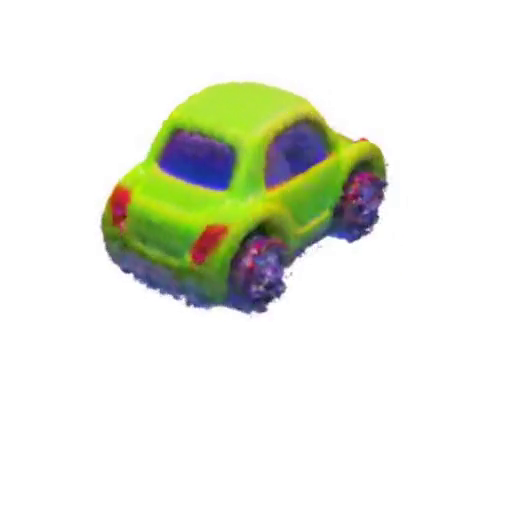} \\

\end{tabular}

    \caption{\textbf{Qualitative Comparison with Baseline 3D Editing Methods.}
The columns correspond to the requested edits  
(\emph{``with headphones''}, \emph{``with pagoda roof''}, \emph{``cartoonish''});  
each cell shows two canonical views of the edited object.
Rows list the original input views and the results produced by
Vox-E, MVEdit, Instant3Dit, and our method.
Instant3Dit is a mask-based local editor and cannot perform a global
style change such as the \emph{cartoonish} car; its entry is therefore
marked ``N/A'' in the last column.
 }
    \label{fig:qualitative_comparison}

\end{figure}

%% file: tables/table_comp.tex
\begin{table}[h!]
    \centering
    \renewcommand{\arraystretch}{1.2}
    \setlength{\tabcolsep}{6pt}
        \caption{Quantitative comparison with mask-free methods.}
    \label{tab:performance_comparison}
    \begin{tabular}{@{}lccc@{}}
        \toprule
        Metric & MVEdit & Vox-E & EditP23 (ours) \\
        \midrule
        CLIP$_\text{Sim} \uparrow$ & 0.84 & 0.78 & \textbf{0.85} \\
        CLIP$_\text{Dir} \uparrow$ & 0.001 & 0.005 & \textbf{0.015} \\
        \bottomrule
    \end{tabular}

\end{table}

%% file: figures/user_study.tex
\begin{figure}[!h]
    \centering
    \includegraphics[width=0.8\columnwidth]{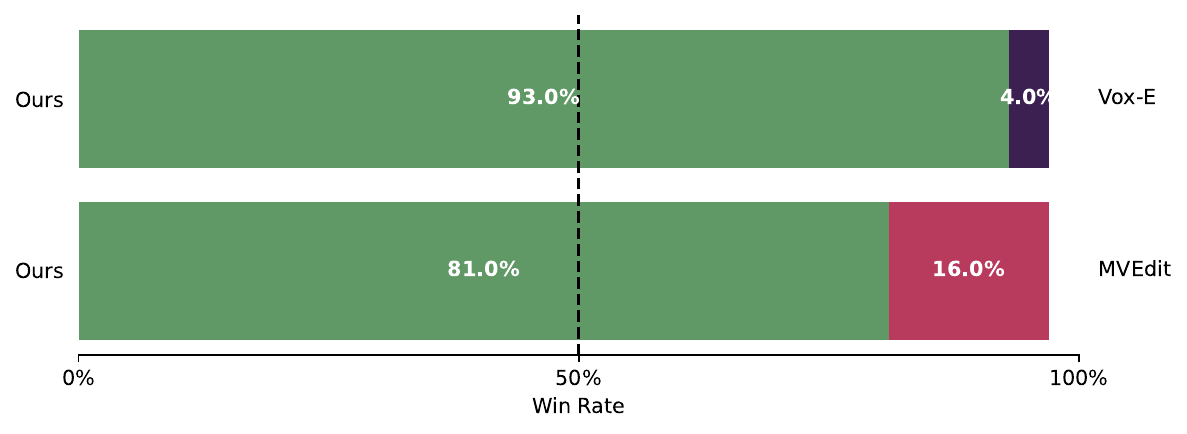}  %
    \vspace{-0.5cm}
    \caption{\textbf{Human Evaluation Study Results.}  \ourmethod{} was compared with two baseline approaches in a 2-alternative. Raters strongly favored \ourmethod{} for better editing. }
    \label{fig:human_study}
    \vspace{-0.3cm}
\end{figure}

%% file: figures/ablation.tex
\renewcommand{\arraystretch}{1.0}
\setlength{\tabcolsep}{2pt}

\newcommand{\imgw}{0.1\textwidth}
\newcommand{\labelw}{0.02\textwidth}
\begin{figure}[htb]
\begin{tabular}{@{}p{\labelw}@{}p{\imgw}@{}p{\imgw}@{}p{\imgw}@{}p{\imgw}@{}}
& \multicolumn{2}{c}{\textbf{-> Cross Arms}} 
& \multicolumn{2}{c}{\textbf{-> Wear Tuxedo}} \\
\midrule
\raisebox{15pt}{\rotatebox{90}{Source} }
& \includegraphics[width=\imgw, trim={0.2cm 0.6cm 0.4cm 0cm}, clip]{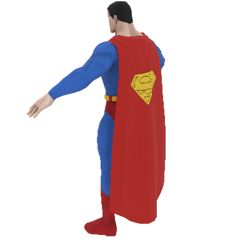} 
& \includegraphics[width=\imgw, trim={0.3cm 0.4cm 0.5cm 0.4cm}, clip]{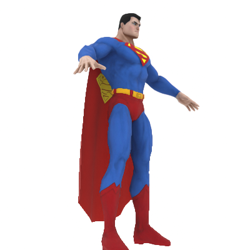} 
& \includegraphics[width=\imgw,  trim={0.7cm 0.7cm 1cm 1cm}, clip]{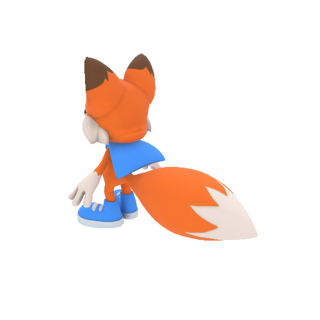}
& \includegraphics[width=\imgw,   trim={0.7cm 0.7cm 1cm 1cm}, clip]{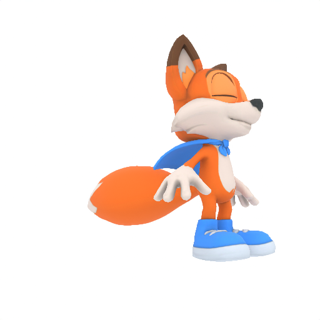} \\
\raisebox{15pt}{\rotatebox{90}{Edit} }
& \multicolumn{2}{c}{\centering\includegraphics[height=0.1\textwidth, trim={1.5cm 2.5cm 1.5cm 0.5cm}, clip]{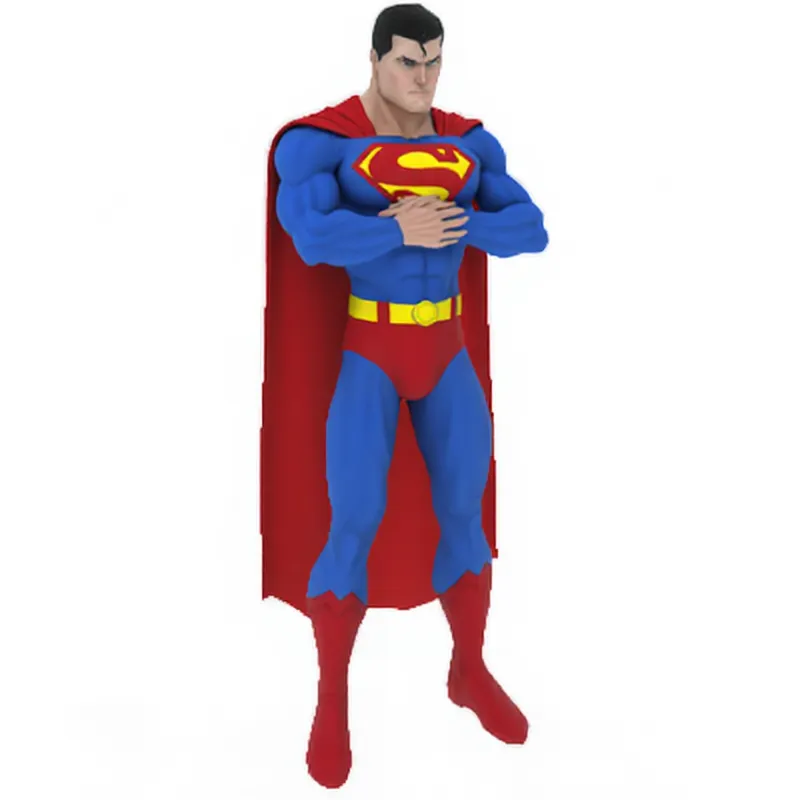}} 
& \multicolumn{2}{c}{\centering\includegraphics[height=0.1\textwidth, trim={5cm 3cm 5cm 7cm}, clip]{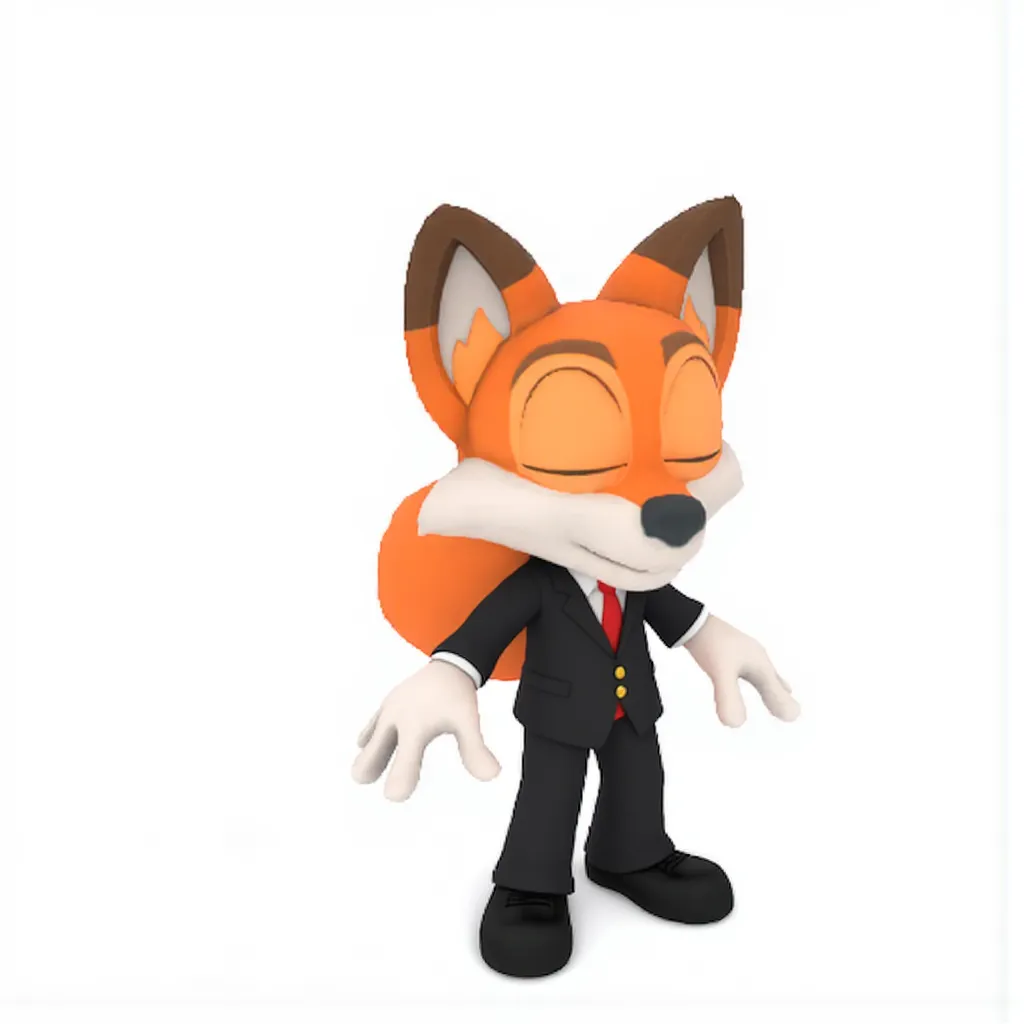}} \\

\\[2pt]  %

\raisebox{20pt}{\rotatebox{90}{Ours} }
& \includegraphics[width=\imgw,  trim={0.2cm 0.6cm 0.4cm 0cm}, clip]{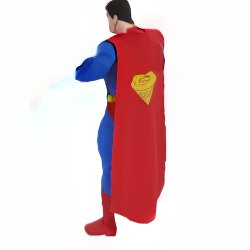} 
& \includegraphics[width=\imgw, trim={0.3cm 0.4cm 0.4cm 0.3cm}, clip]{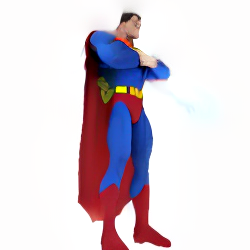} 
& \includegraphics[width=\imgw, trim={0.7cm 0.7cm 0.5cm 0.5cm}, clip]{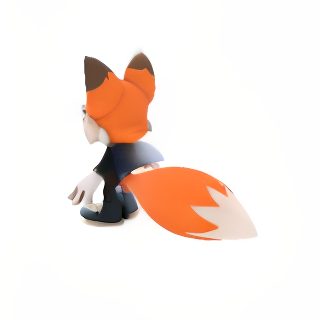} 
& \includegraphics[width=\imgw, trim={0.7cm 0.7cm 0.5cm 0.5cm}, clip]{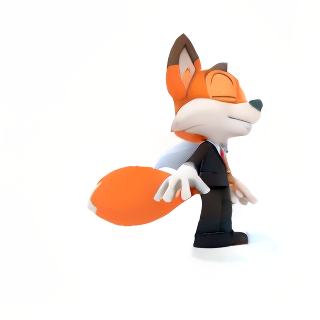} \\
\raisebox{15pt}{\rotatebox{90}{FlowEdit} }
& \includegraphics[width=\imgw, trim={0.2cm 0.6cm 0.4cm 0cm}, clip]{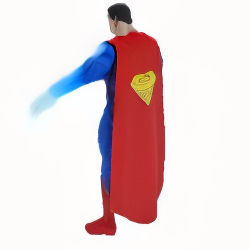} 
& \includegraphics[width=\imgw, trim={0.3cm 0.4cm 0.5cm 0.4cm}, clip]{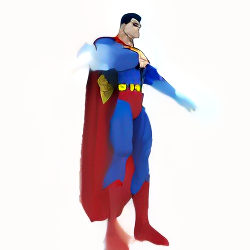} 
& \includegraphics[width=\imgw, trim={0.7cm 0.7cm 0.5cm 0.5cm}, clip]{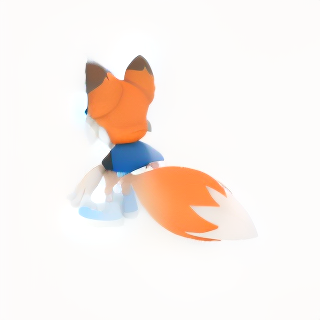} 
& \includegraphics[width=\imgw, trim={0.7cm 0.7cm 0.5cm 0.5cm}, clip]{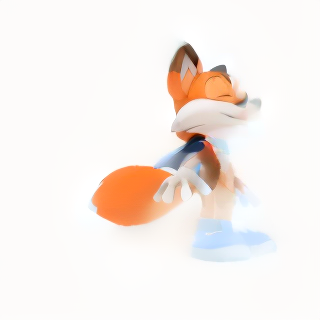} \\

\raisebox{15pt}{\rotatebox{90}{SDEdit}  }
& \includegraphics[width=\imgw, trim={0.2cm 0.6cm 0.4cm 0cm}, clip]{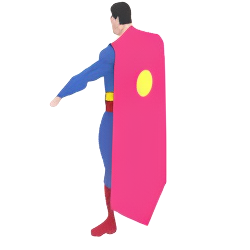} 
& \includegraphics[width=\imgw, trim={0.3cm 0.4cm 0.5cm 0.4cm}, clip]{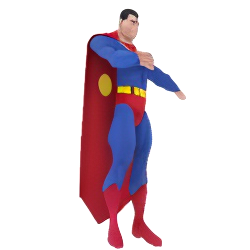} 
& \includegraphics[width=\imgw, trim={0.7cm 0.7cm 0.5cm 0.5cm}, clip]{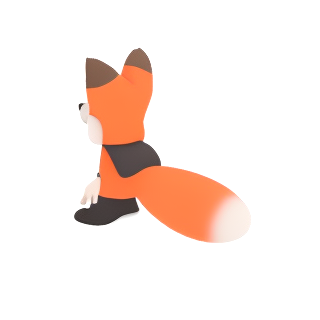} 
& \includegraphics[width=\imgw, trim={0.7cm 0.7cm 0.5cm 0.5cm}, clip]{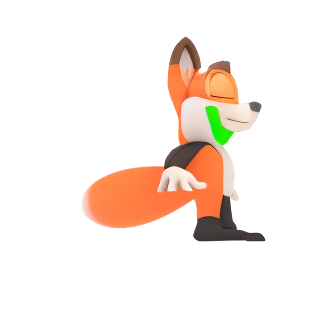} \\
\end{tabular}

    \caption{\textbf{Ablation Study of the Edit-Aware Denoising Mechanism.} This figure compares our full method against two ablated variants: SDEdit and FlowEdit.  
For each edit request (``Cross Arms'' and ``Wear Tuxedo'') we show the target edited view provided to all methods (second row), followed by the source object, rendered from two alternative viewpoints. Rows 4-5 compares the editing results when applying SDEdit, FlowEdit, and our approach on the mv-grid.}
\label{fig:ablation}

\end{figure}

%% file: sections/6_conclusion.tex
\section{Conclusions and Limitations}\label{sec:conclusion}
We present a 3D object editing technique where users modify a single 2D view and the edit propagates across views for consistent 3D modification. The key challenge is ensuring 3D consistency from inherently 2D input. We address this by leveraging the strong geometric coherence prior of a pre-trained diffusion model. The proposed edit-aware mechanism, conditioned on image prompts, provides high-fidelity, mask-free support for both local and global edits.

While our approach performs well across a wide range of scenarios, some limitations remain. Large object removals may leave residual artifacts, though these can often be addressed with post-processing. Additionally, challenging edits requiring high $\text{CFG}_{\text{tar}}$ scale, can produce intermediate visual artifacts like over-saturation. As shown in \cref{fig:limitation}, these issues are substantially mitigated by the final 3D reconstruction module.

The underlying principle of propagating edits from a lower-dimensional space could extend beyond 3D assets. We believe similar image-prompted strategies could benefit other domains like video, animation, or scene editing, where local modifications must be generalized coherently across space or time.
\input{figures/limitation}

%% file: figures/limitation.tex
\begin{figure}[hbtp]
\centering
\setlength{\tabcolsep}{2pt}
\renewcommand{\arraystretch}{0.7}
\begin{tabular}{@{}ccccc@{}}
& \textbf{Cond. View} & \textbf{View 1} & \textbf{View 2} & \textbf{View 3} \\

\raisebox{20pt}{\rotatebox[origin=c]{90}{Original}} &
\includegraphics[width=0.17\linewidth]{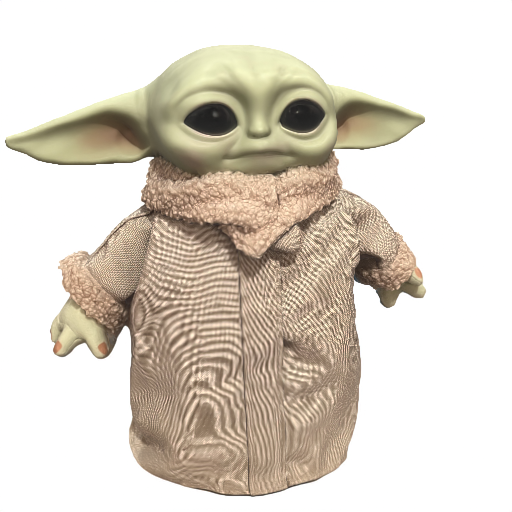} &
\includegraphics[width=0.22\linewidth]{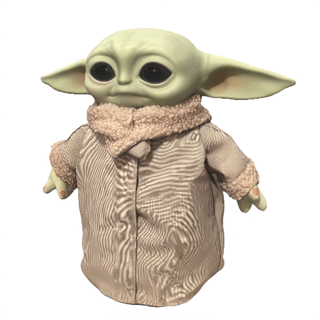} &
\includegraphics[width=0.22\linewidth]{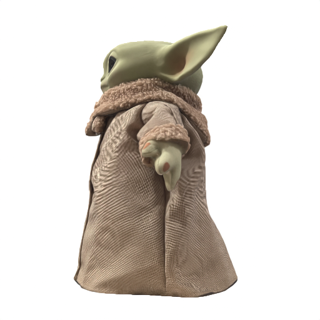} &
\includegraphics[width=0.22\linewidth]{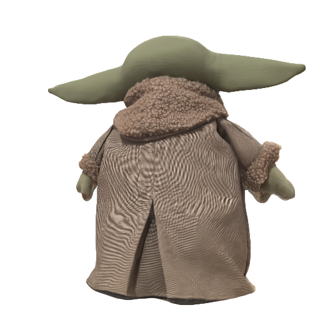} \\

\raisebox{20pt}{\rotatebox[origin=c]{90}{Edited}} &
\raisebox{12pt}{\multirow{2}{*}{\includegraphics[width=0.17\linewidth]{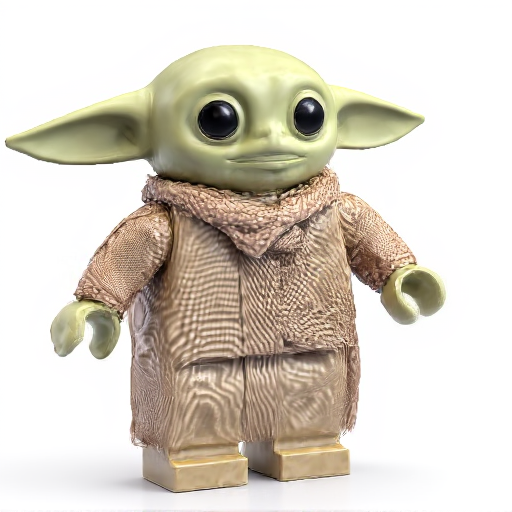}}} &
\includegraphics[width=0.22\linewidth]{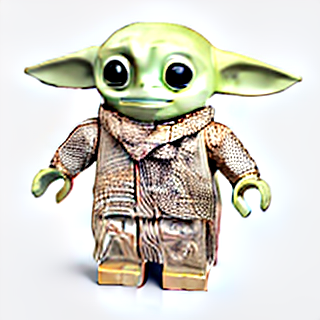} &
\includegraphics[width=0.22\linewidth]{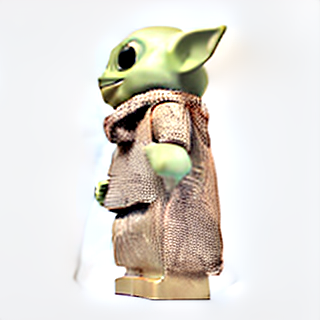} &
\includegraphics[width=0.22\linewidth]{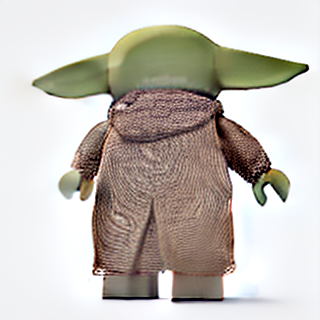} \\

\raisebox{20pt}{\rotatebox[origin=c]{90}{Recon.}} &
& \includegraphics[width=0.22\linewidth]{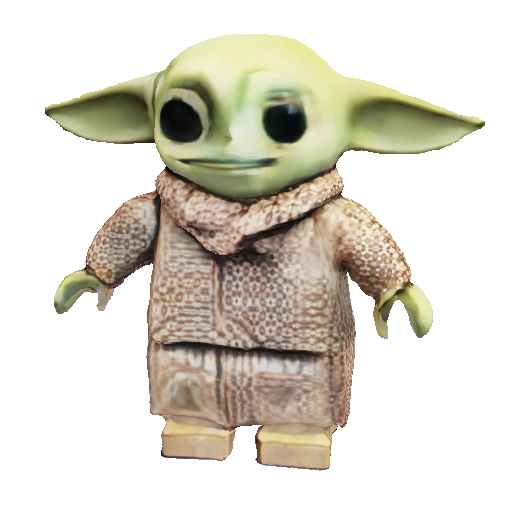} &
\includegraphics[width=0.22\linewidth]{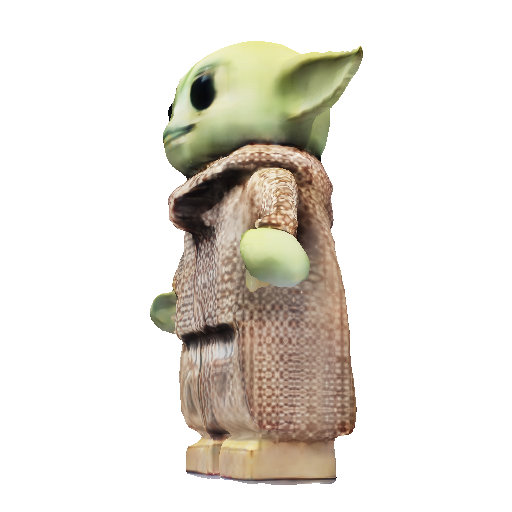} &
\includegraphics[width=0.22\linewidth]{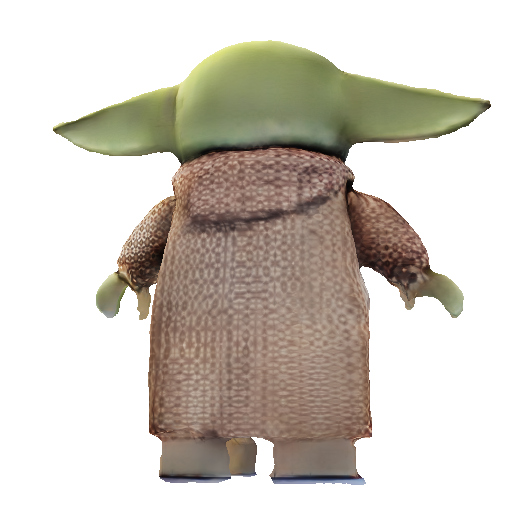} \\
\end{tabular}
\caption{\textbf{Limitations of \ourmethod{}.}  Challenging edits \eg transforming Grogu into a LEGO figure requires a high target guidance scale ($\text{CFG}_\text{tar}$). This can cause artifacts in the intermediate multi-view propagation, such as over-saturation and inconsistent backgrounds (middle row, ``Edited''). However, these artifacts are substantially mitigated during the final 3D reconstruction, which produces a more coherent result (bottom row, ``Recon.'').}
\label{fig:limitation}

\end{figure}

%% file: sections/7_additional_res.tex
\input{figures/additional_res}

\input{figures/rendered}

%% file: figures/additional_res.tex
\begin{figure*}[htbp]
\centering

\setlength{\tabcolsep}{2pt}
\renewcommand{\arraystretch}{0.6} 

\begin{minipage}[t]{0.49\textwidth}
    \centering
    \begin{tabular}{@{}cccc@{}}
        & \textbf{Cond. View} & \textbf{View 1} & \textbf{View 2} \\
        \raisebox{25pt}{\rotatebox[origin=c]{90}{Original}} &
        \includegraphics[width=0.27\linewidth]{images/additional_res/more_views/grogu/lego_fig/src.png} &
        \includegraphics[width=0.28\linewidth]{images/additional_res/more_views/grogu/lego_fig/src_mv_0_0.png} &
        \includegraphics[width=0.28\linewidth]{images/additional_res/more_views/grogu/lego_fig/src_mv_1_1.png} \\
        
        \raisebox{25pt}{\rotatebox[origin=c]{90}{Edited}} &
        \includegraphics[width=0.27\linewidth]{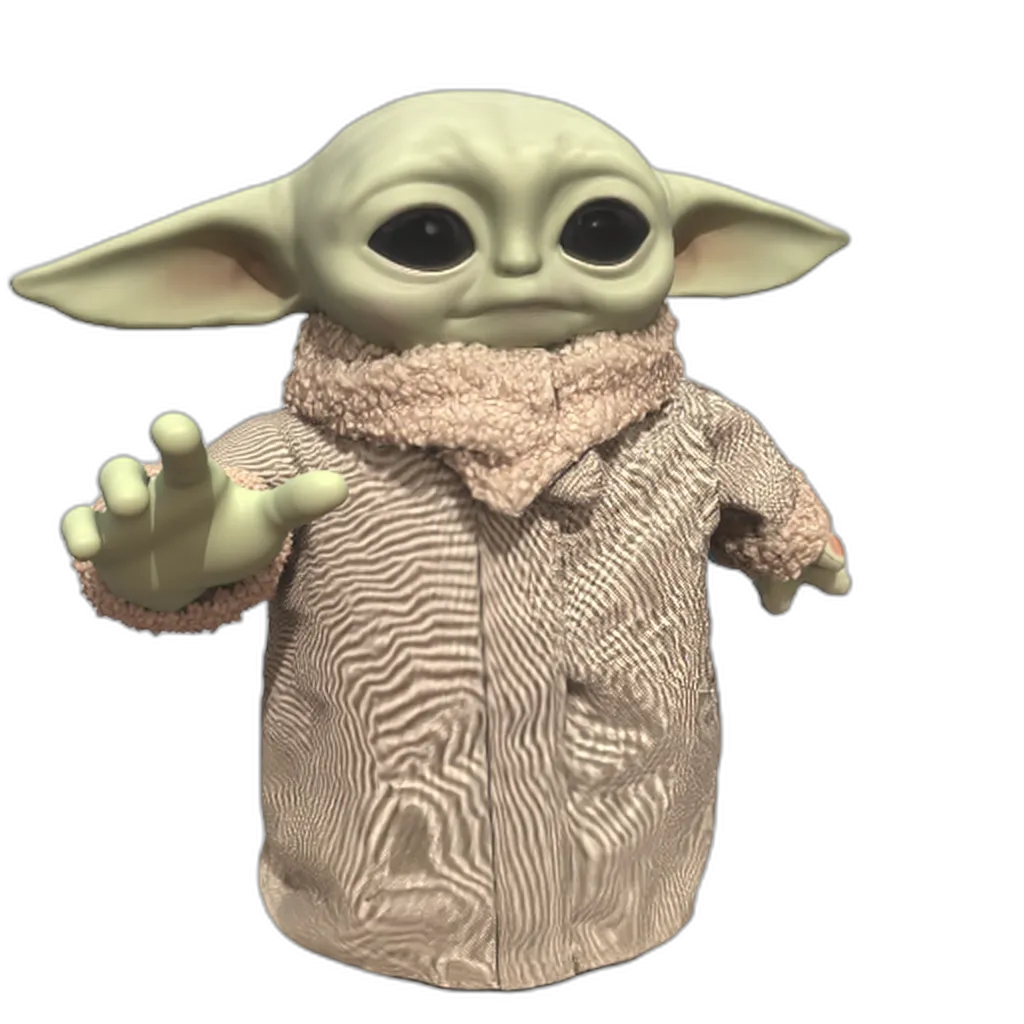} &
        \includegraphics[width=0.28\linewidth]{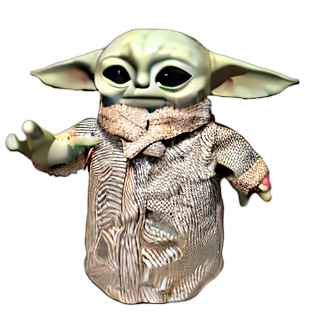} &
        \includegraphics[width=0.28\linewidth]{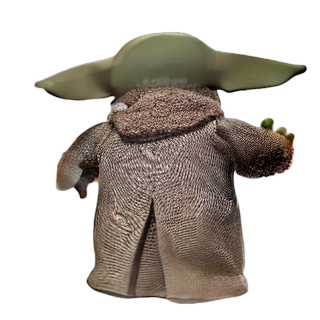} \\ %
        
        \raisebox{25pt}{\rotatebox[origin=c]{90}{Original}} &
        \includegraphics[width=0.27\linewidth]{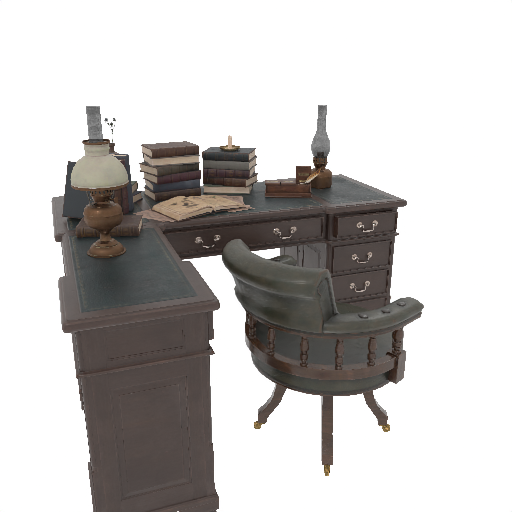} &
        \includegraphics[width=0.28\linewidth]{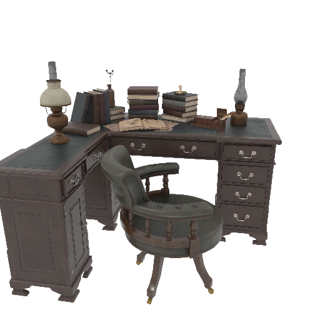} &
        \includegraphics[width=0.28\linewidth]{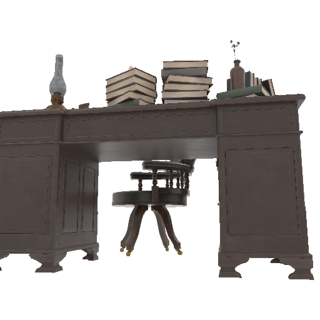} \\

        \raisebox{25pt}{\rotatebox[origin=c]{90}{Edited}} &
        \includegraphics[width=0.27\linewidth]{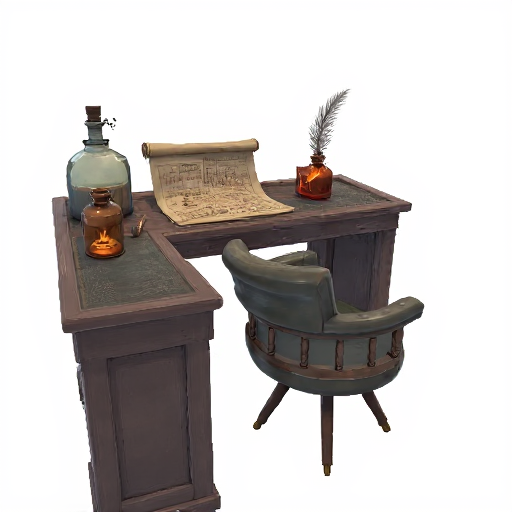} &
        \includegraphics[width=0.28\linewidth]{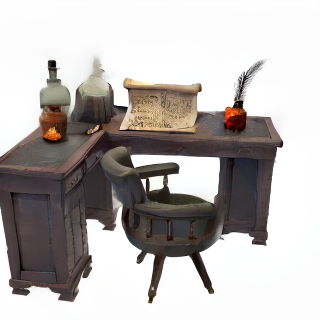} &
        \includegraphics[width=0.28\linewidth]{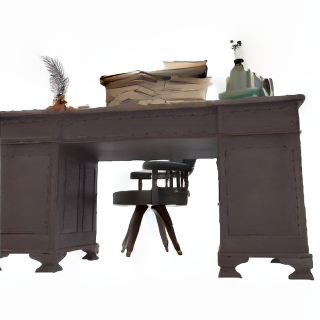} \\ %
        
        \raisebox{25pt}{\rotatebox[origin=c]{90}{Original}} &
        \includegraphics[width=0.27\linewidth, trim={2cm 2cm 2cm 2cm}, clip]{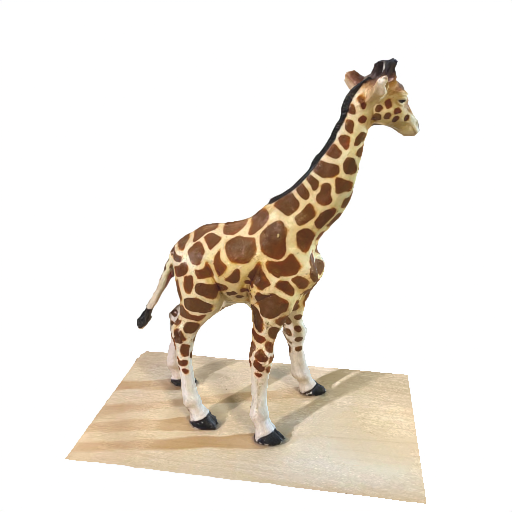} &
        \includegraphics[width=0.28\linewidth, trim={1.5cm 2cm 1.5cm 1cm}, clip]{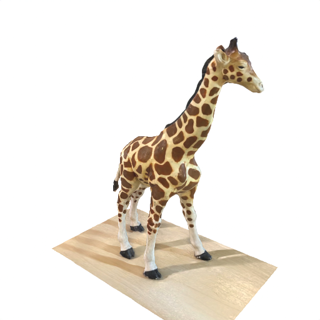} &
            \begin{tikzpicture}[spy using outlines={circle, myspycolor, magnification=2, size=0.6cm, connect spies}]
    \node {\includegraphics[width=0.28\linewidth, trim={2cm 2cm 2cm 2cm}, clip]{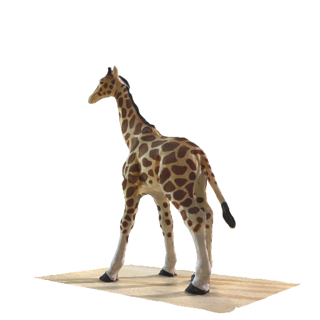}};
    \spy on (-0.1,0) in node [left] at (1.45,1); %
    \end{tikzpicture}  \\

        \raisebox{25pt}{\rotatebox[origin=c]{90}{Edited}} &
        \includegraphics[width=0.27\linewidth, trim={2cm 2cm 2cm 2cm}, clip]{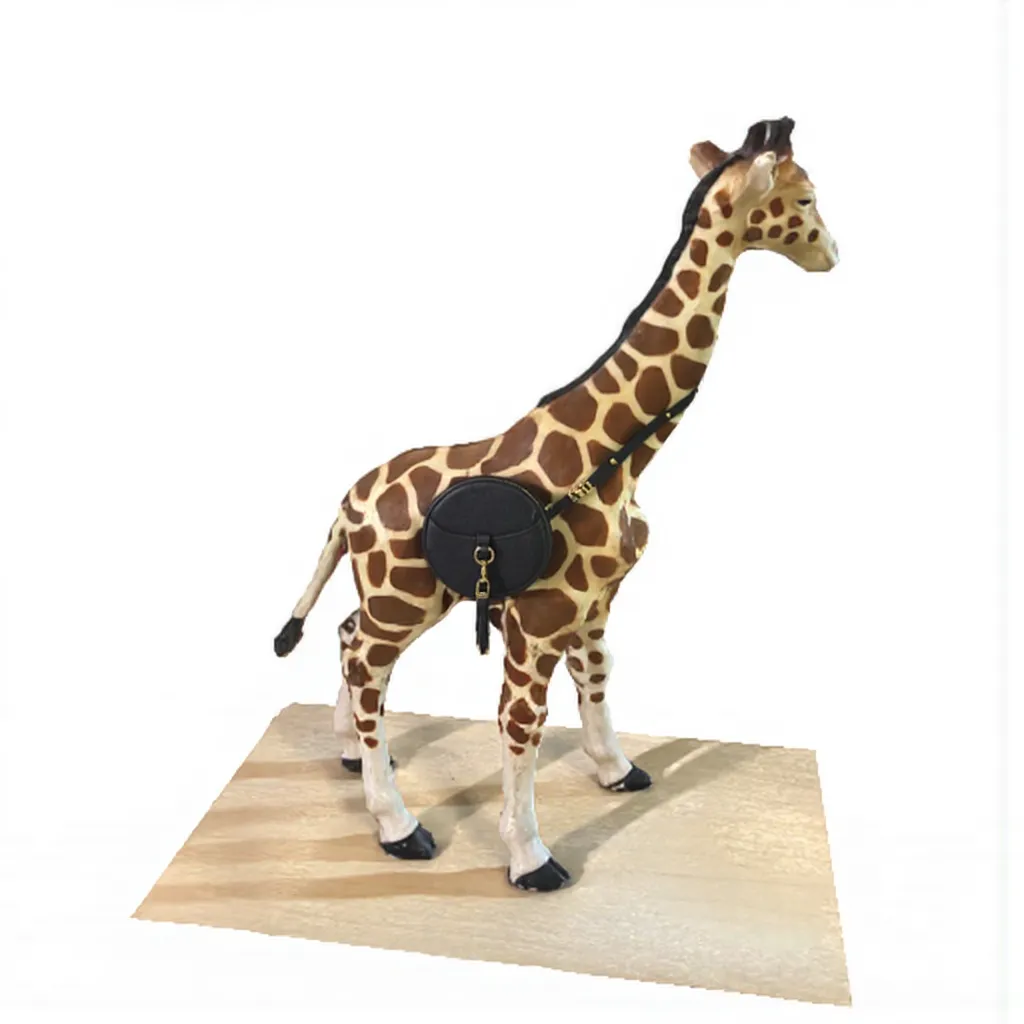} &
        \includegraphics[width=0.28\linewidth, trim={1.5cm 2cm 1.5cm 1cm}, clip]{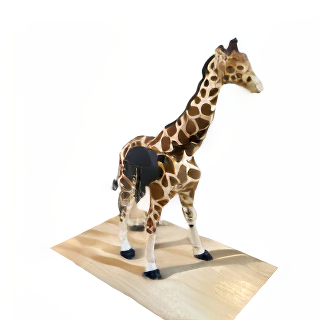} &
            \begin{tikzpicture}[spy using outlines={circle, myspycolor, magnification=2, size=0.6cm, connect spies}]
    \node {\includegraphics[width=0.28\linewidth, trim={2cm 2cm 2cm 2cm}, clip]{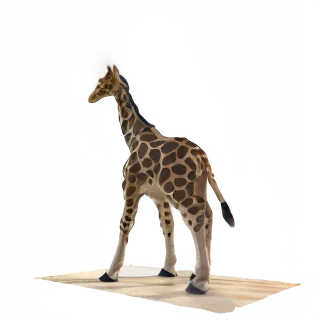}};
    \spy on (-0.1,0) in node [left] at (1.45,1); %
    \end{tikzpicture} \\ %
        
        \raisebox{25pt}{\rotatebox[origin=c]{90}{Original}} &
        \includegraphics[width=0.27\linewidth]{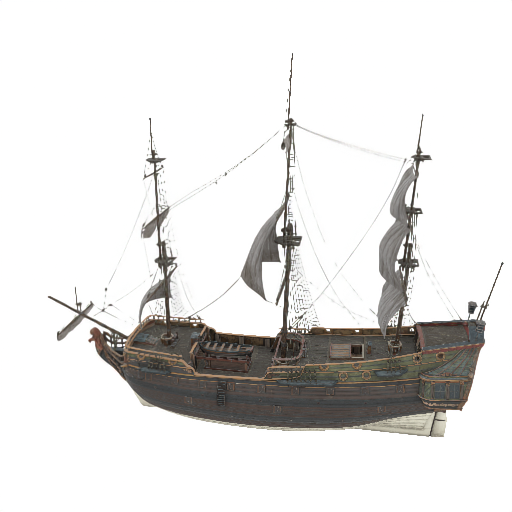} &
        \includegraphics[width=0.28\linewidth]{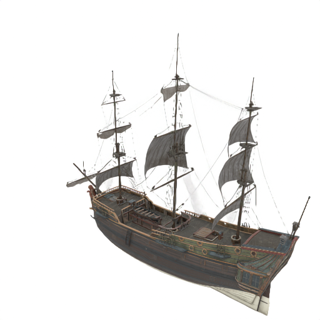} &
        \includegraphics[width=0.28\linewidth]{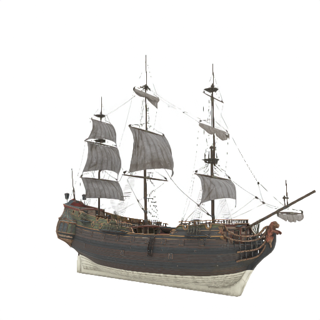} \\

        \raisebox{25pt}{\rotatebox[origin=c]{90}{Edited}} &
        \includegraphics[width=0.27\linewidth]{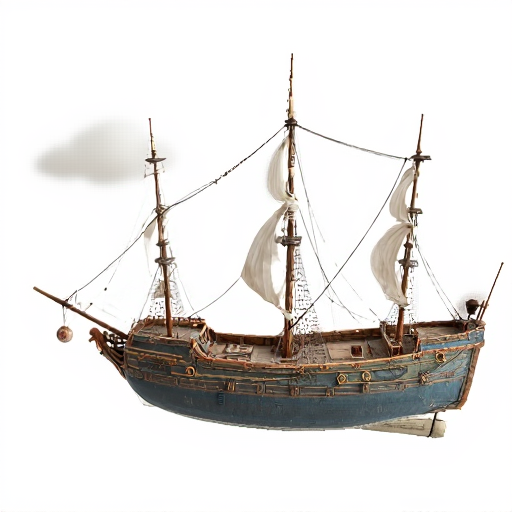} &
        \includegraphics[width=0.28\linewidth]{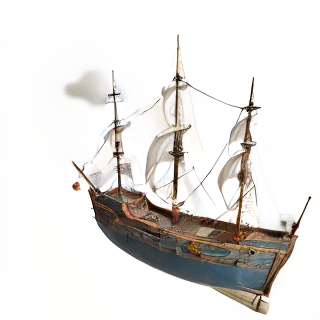} &
        \includegraphics[width=0.28\linewidth]{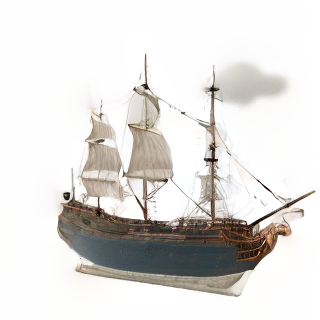} \\
    \end{tabular}
\end{minipage}%
\hfill %
\begin{minipage}[t]{0.49\textwidth}
    \centering
    \begin{tabular}{@{}cccc@{}}
        & \textbf{Cond. View} & \textbf{View 1} & \textbf{View 2} \\
        \raisebox{25pt}{\rotatebox[origin=c]{90}{Original}} &
        \includegraphics[width=0.27\linewidth, trim={1cm 1cm 1cm 1cm}, clip]{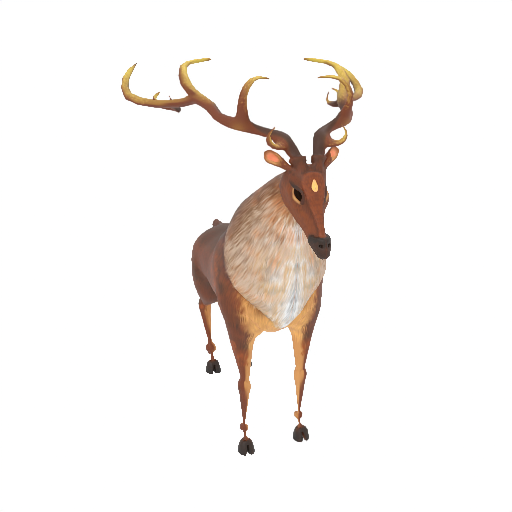} &
        \includegraphics[width=0.28\linewidth, trim={0.5cm 1cm 0.5cm 0cm}, clip]{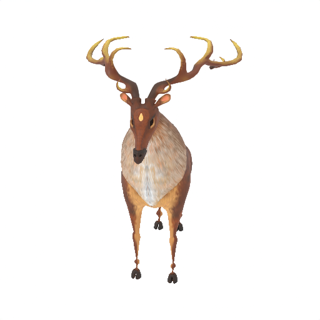} &
        \includegraphics[width=0.28\linewidth, trim={0.5cm 1cm 0.5cm 0cm}, clip]{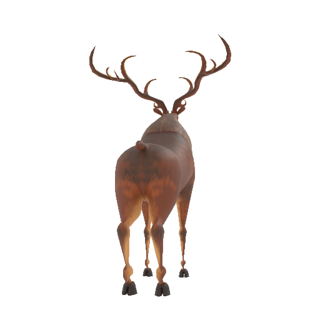} \\

        \raisebox{25pt}{\rotatebox[origin=c]{90}{Edited}} &
        \includegraphics[width=0.27\linewidth, trim={1cm 1cm 1cm 1cm}, clip]{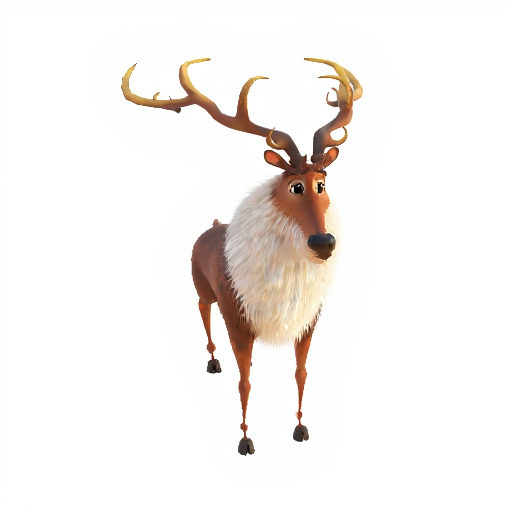} &
        \includegraphics[width=0.28\linewidth, trim={0.5cm 1cm 0.5cm 0cm}, clip]{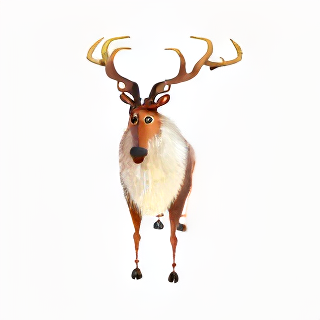} &
        \includegraphics[width=0.28\linewidth, trim={0.5cm 1cm 0.5cm 0cm}, clip]{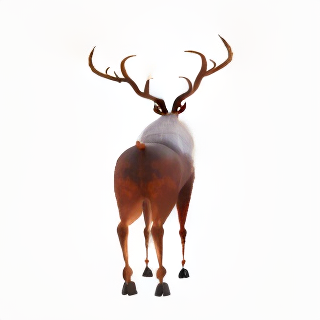} \\
        
        \raisebox{25pt}{\rotatebox[origin=c]{90}{Edited}} &
        \includegraphics[width=0.27\linewidth, trim={1cm 1cm 1cm 1cm}, clip]{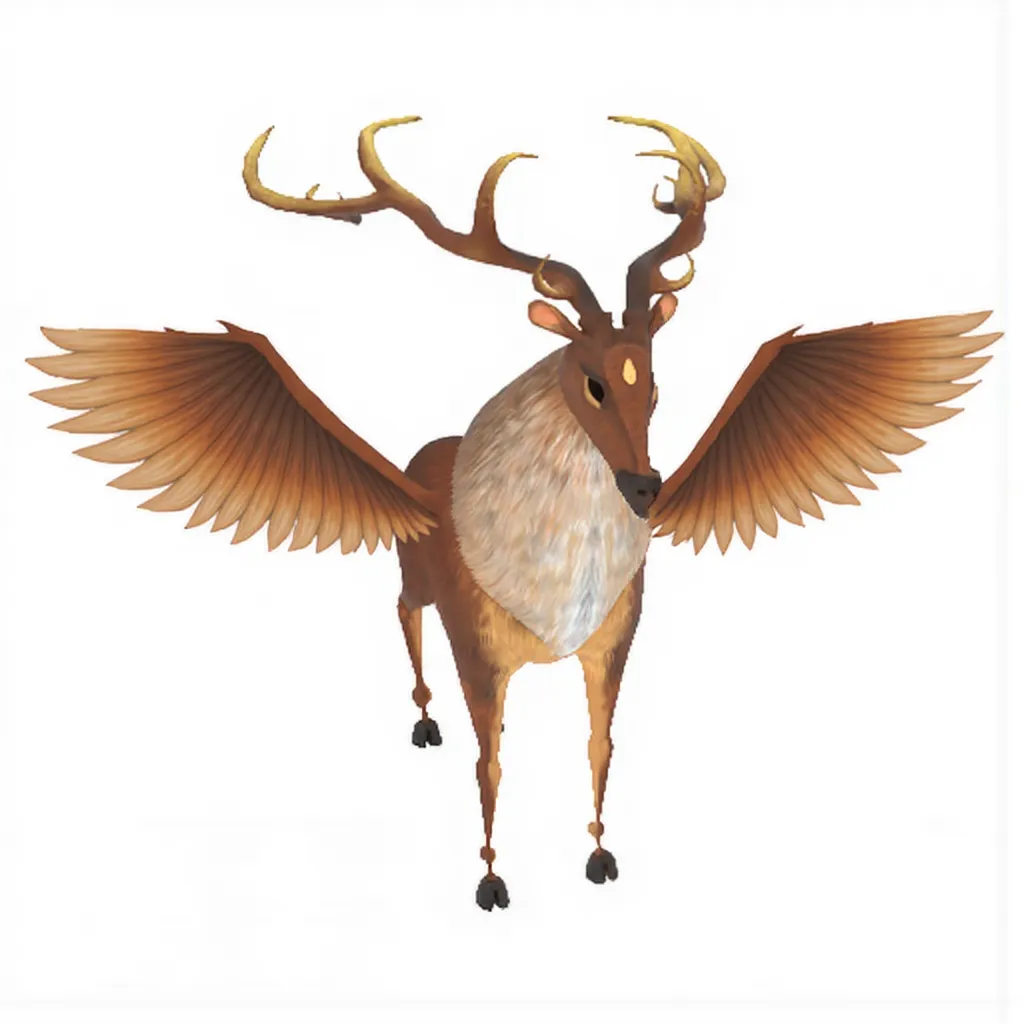} &
        \includegraphics[width=0.28\linewidth, trim={0.5cm 1cm 0.5cm 0cm}, clip]{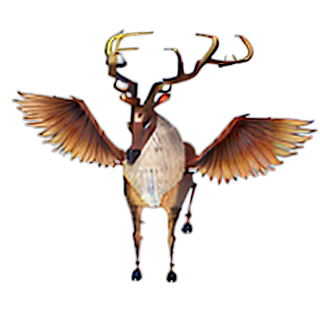} &
        \includegraphics[width=0.28\linewidth, trim={0.5cm 1cm 0.5cm 0cm}, clip]{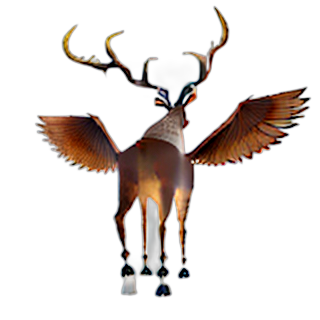} \\

        \raisebox{25pt}{\rotatebox[origin=c]{90}{Original}} &
        \includegraphics[width=0.27\linewidth, trim={2cm 2cm 2cm 2cm}, clip]{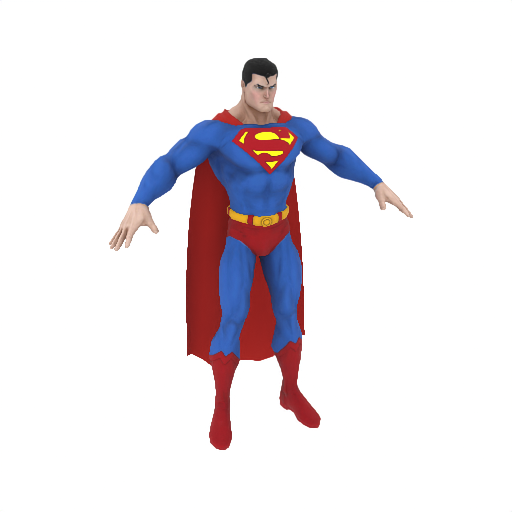} &
        \includegraphics[width=0.28\linewidth, trim={0.8cm 0.8cm 0.8cm 0.8cm}, clip]{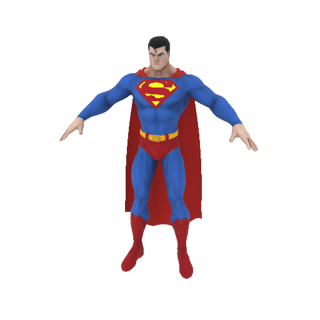} &
        \includegraphics[width=0.28\linewidth, trim={0.8cm 0.8cm 0.8cm 0.8cm}, clip]{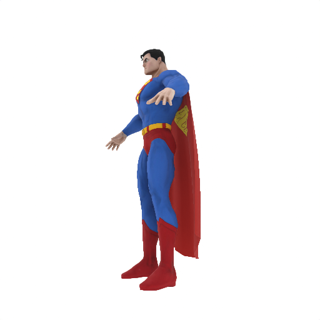} \\

        \raisebox{25pt}{\rotatebox[origin=c]{90}{Edited}} &
        \includegraphics[width=0.27\linewidth, trim={2cm 2cm 2cm 2cm}, clip]{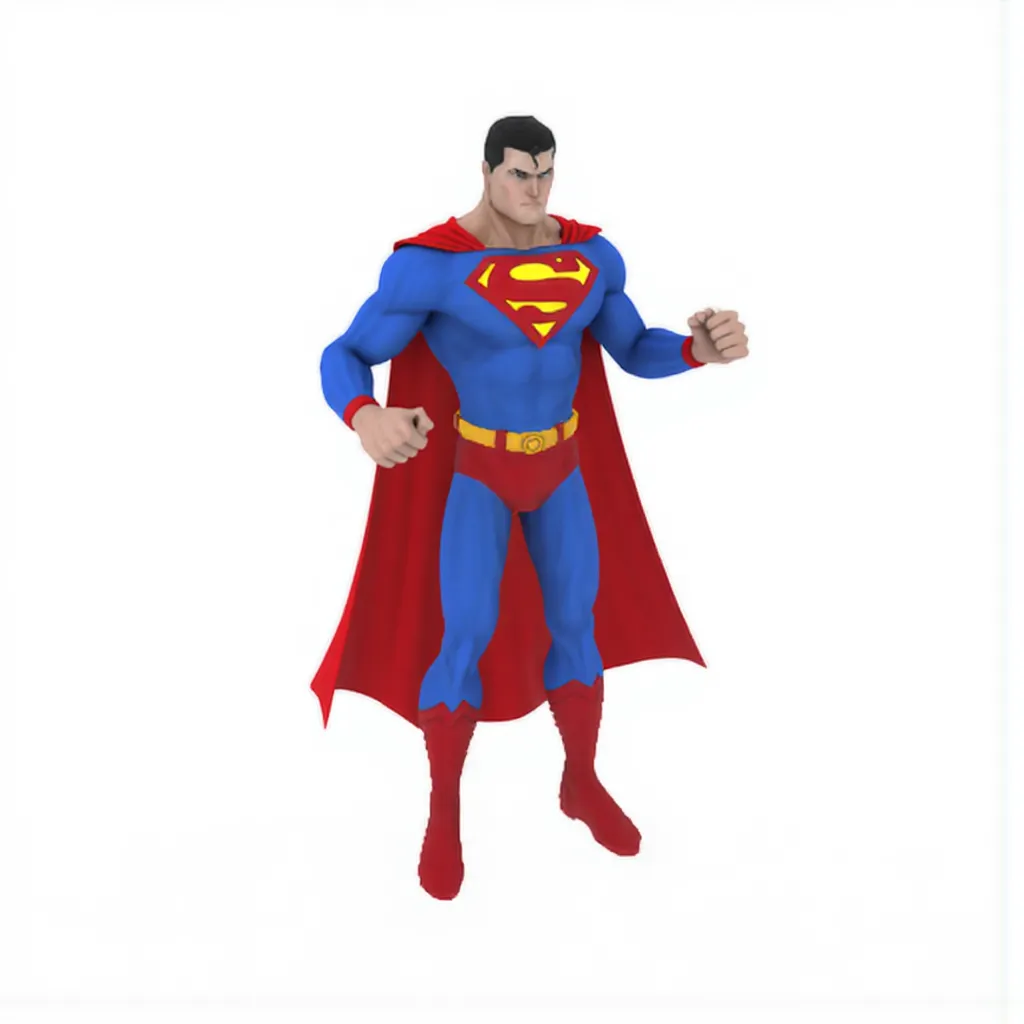} &
        \includegraphics[width=0.28\linewidth, trim={0.4cm 0.8cm 0.4cm 0cm}, clip]{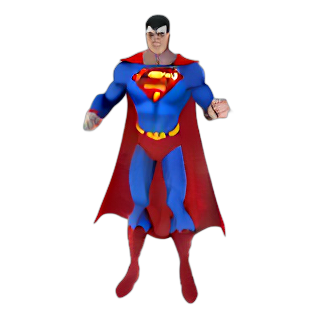} &
        \includegraphics[width=0.28\linewidth, trim={0.7cm 0.8cm 0.7cm 0.6cm}, clip]{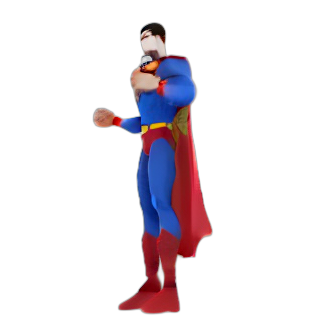} \\

        \raisebox{25pt}{\rotatebox[origin=c]{90}{Original}} &
        \includegraphics[width=0.27\linewidth, trim={2cm 2cm 2cm 2cm}, clip]{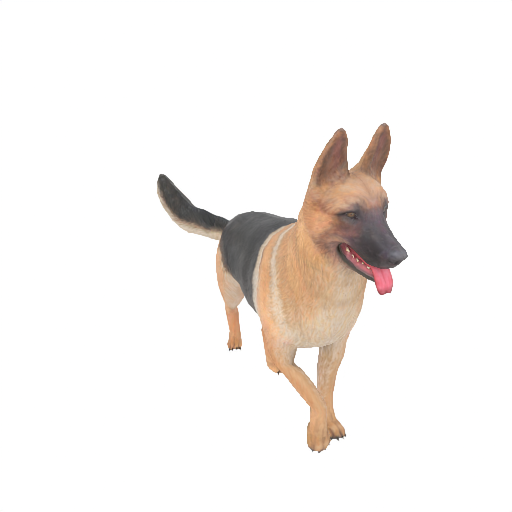} &
        \includegraphics[width=0.28\linewidth, trim={1.5cm 1cm 1.5cm 2cm}, clip]{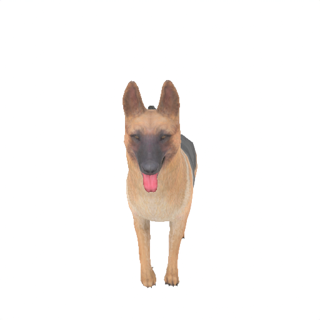} &
        \includegraphics[width=0.28\linewidth, trim={0.7cm 0.5cm 1.4cm 1.6cm}, clip]{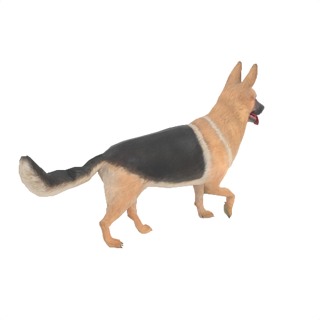} \\

        \raisebox{25pt}{\rotatebox[origin=c]{90}{Edited}} &
        \includegraphics[width=0.27\linewidth, trim={2cm 2cm 2cm 2cm}, clip]{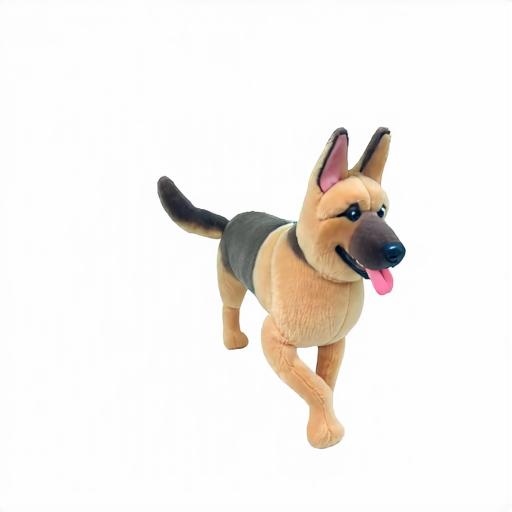} &
        \includegraphics[width=0.28\linewidth, trim={1.3cm 0.8cm 1.5cm 2cm}, clip]{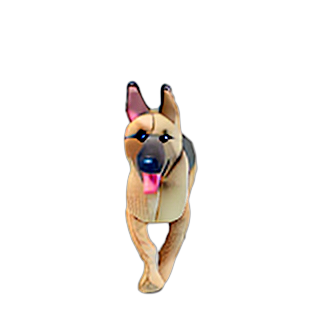} &
        \includegraphics[width=0.28\linewidth,trim={0.7cm 0.5cm 1.4cm 1.6cm}, clip]{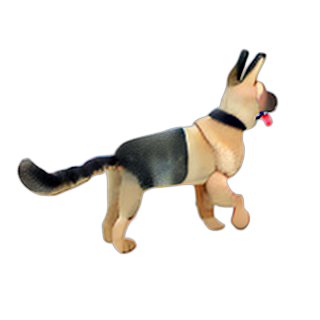} \\

        \raisebox{25pt}{\rotatebox[origin=c]{90}{Edited}} &
        \includegraphics[width=0.27\linewidth, trim={2cm 2cm 2cm 2cm}, clip]{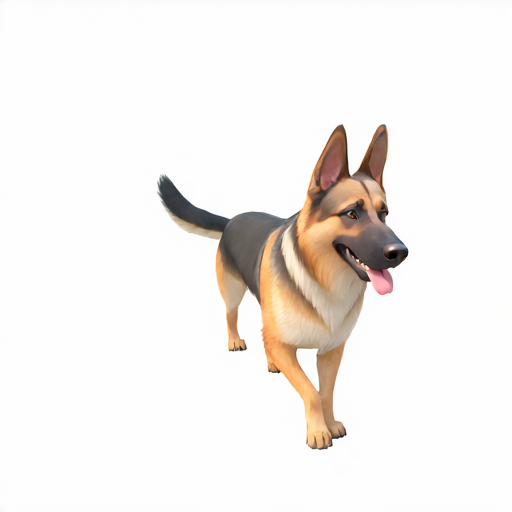} &
        \includegraphics[width=0.28\linewidth, trim={1.5cm 1cm 1.5cm 2cm}, clip]{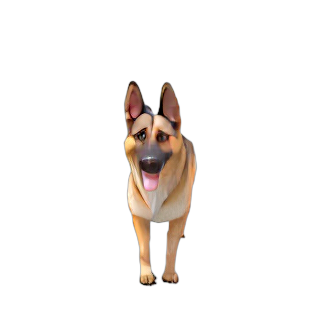} &
        \includegraphics[width=0.28\linewidth, trim={0.7cm 0.5cm 1.4cm 1.6cm}, clip]{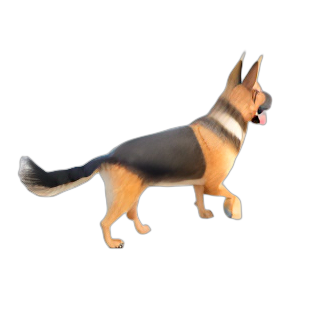} \\
    \end{tabular}
\end{minipage}

\caption{\textbf{Examples of Multi-View Grid Editing.} Each block shows an original object (top) and its edited result (bottom). The leftmost column contains the conditioning views (source and target), while the other columns display the propagated edit from novel viewpoints.}
\label{fig:additional_results1}
\end{figure*}

\begin{figure*}[htbp]
\centering

\setlength{\tabcolsep}{2pt}
\renewcommand{\arraystretch}{0.6} 

\begin{minipage}[t]{0.49\textwidth}
    \centering
    \begin{tabular}{@{}cccc@{}}
        & \textbf{Cond. View} & \textbf{View 1} & \textbf{View 2} \\
        \raisebox{25pt}{\rotatebox[origin=c]{90}{Original}} &
\includegraphics[width=0.27\linewidth]{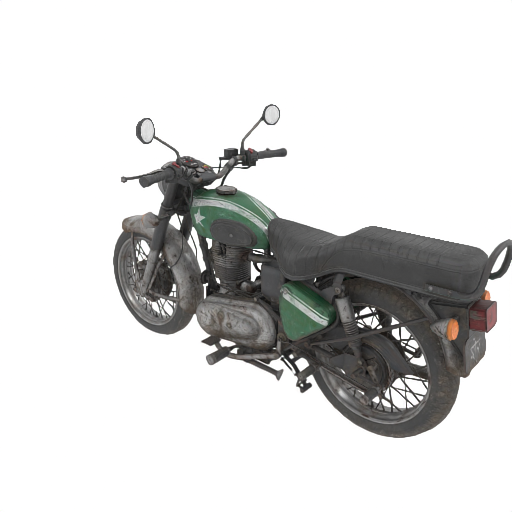} & 
\includegraphics[width=0.28\linewidth]{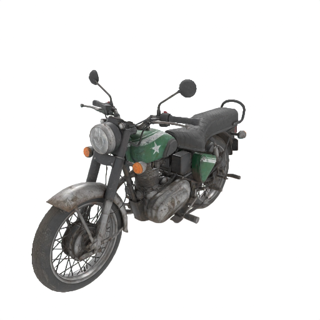} & 
\includegraphics[width=0.28\linewidth]{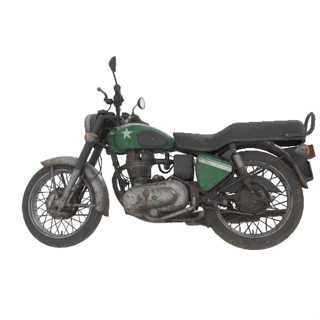} \\

\raisebox{25pt}{\rotatebox[origin=c]{90}{Edited}} &
\includegraphics[width=0.27\linewidth]{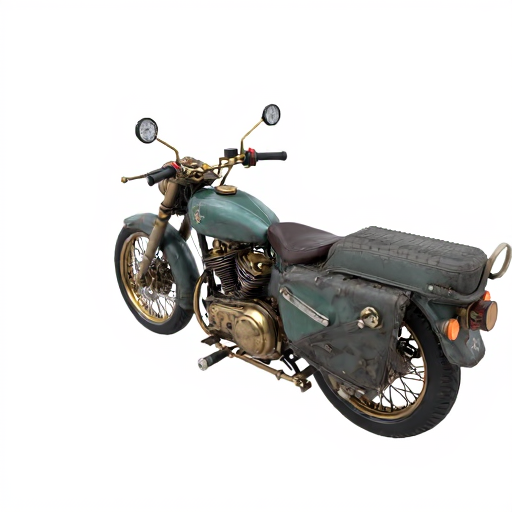} & 
\includegraphics[width=0.28\linewidth]{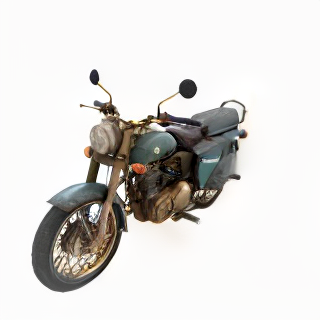} & 
\includegraphics[width=0.28\linewidth]{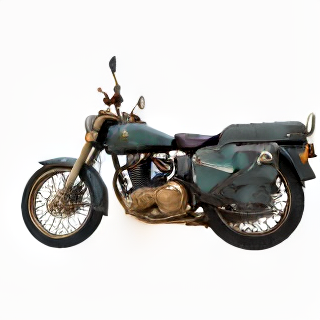} \\
\raisebox{25pt}{\rotatebox[origin=c]{90}{Original}} &
\includegraphics[width=0.27\linewidth, trim={2cm 2cm 2cm 2cm}, clip]{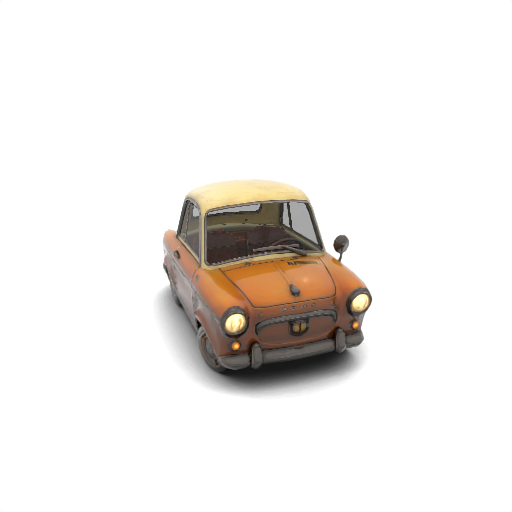} &
\includegraphics[width=0.28\linewidth, trim={2cm 2cm 2cm 2cm}, clip]{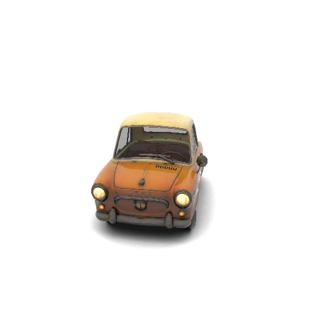} &
  \begin{tikzpicture}[spy using outlines={circle, myspycolor, magnification=1.2, size=0.5cm, connect spies}]
    \node {\includegraphics[width=0.28\linewidth, trim={2cm 2cm 2cm 2cm}, clip]{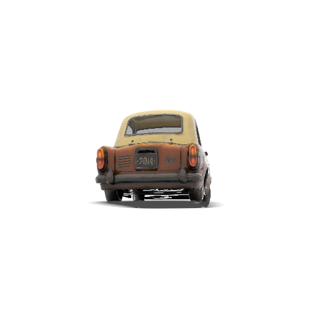}};
    \spy on (-0.1,0) in node [left] at (1.45,1); %
    \end{tikzpicture} \\

\raisebox{25pt}{\rotatebox[origin=c]{90}{Edited}} &
\includegraphics[width=0.27\linewidth, trim={2cm 2cm 2cm 2cm}, clip]{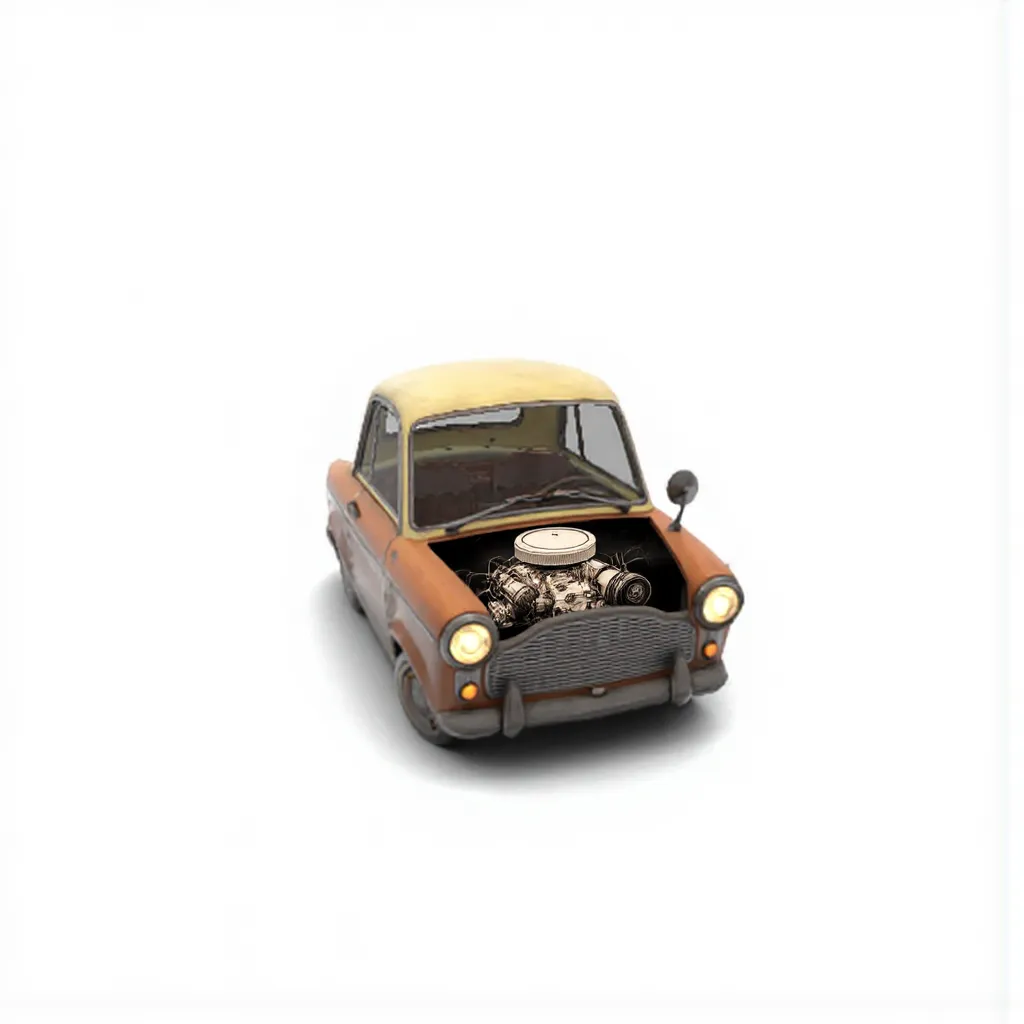} &
\includegraphics[width=0.28\linewidth, trim={2cm 2cm 2cm 2cm}, clip]{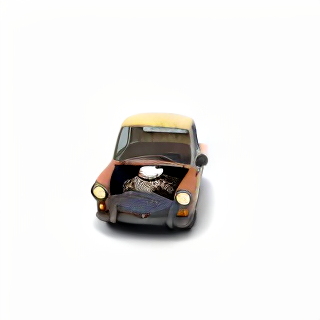} &
  \begin{tikzpicture}[spy using outlines={circle, myspycolor, magnification=1.2, size=0.5cm, connect spies}]
    \node {\includegraphics[width=0.28\linewidth, trim={2cm 2cm 2cm 2cm}, clip]{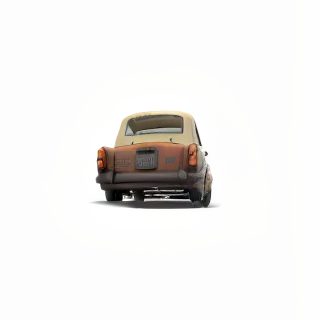}};
    \spy on (-0.1,0) in node [left] at (1.45,1); %
    \end{tikzpicture} \\
\raisebox{25pt}{\rotatebox[origin=c]{90}{Original}} &
\includegraphics[width=0.27\linewidth]{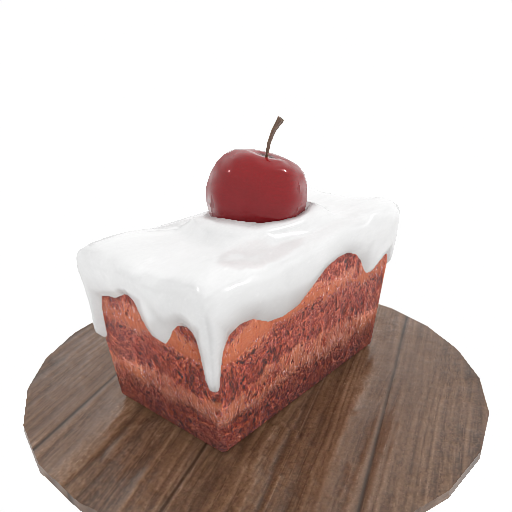} &
\includegraphics[width=0.28\linewidth]{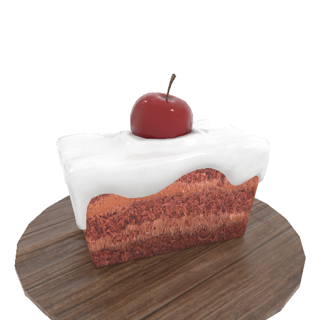} &
\includegraphics[width=0.28\linewidth]{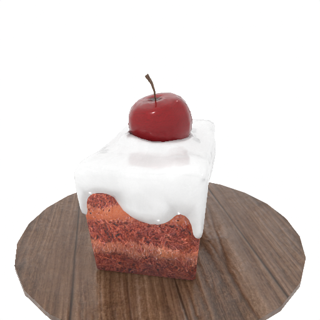} \\

\raisebox{25pt}{\rotatebox[origin=c]{90}{Edited}} &
\includegraphics[width=0.27\linewidth]{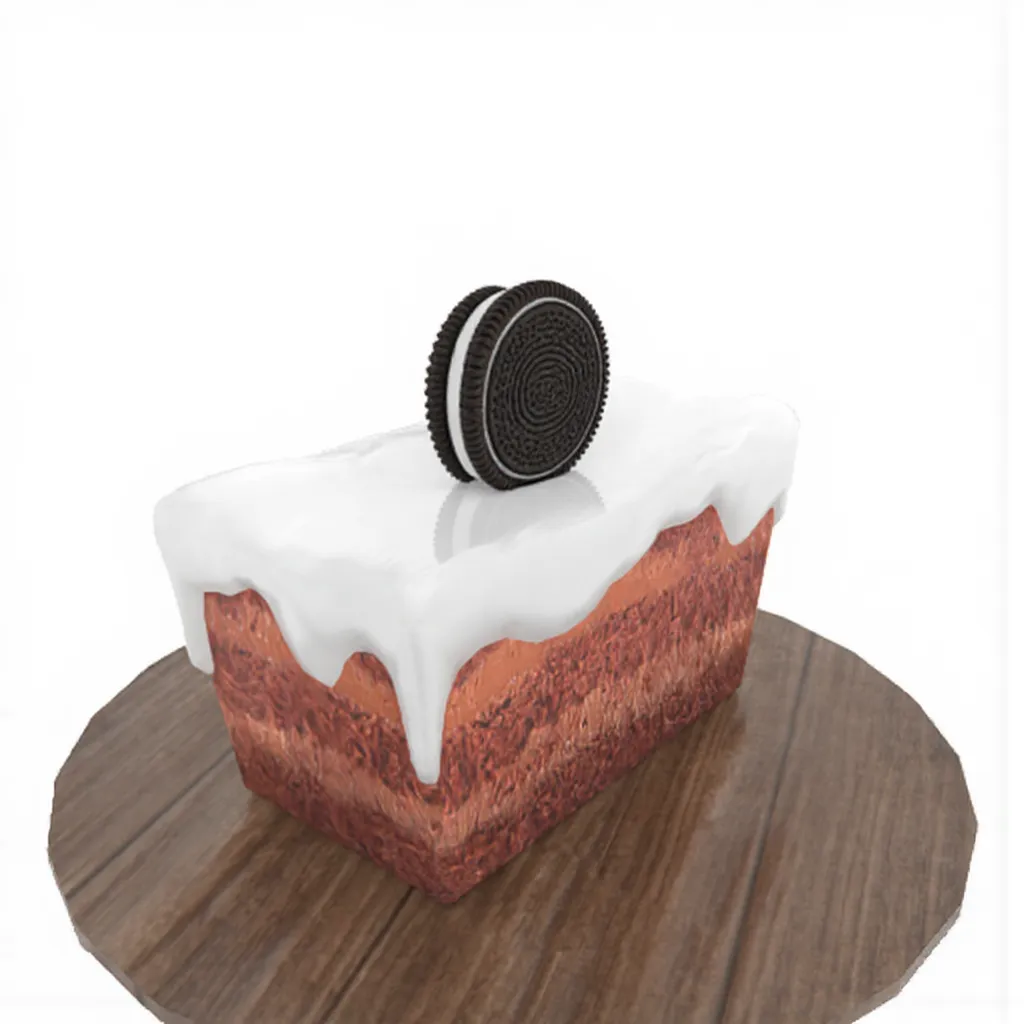} &
\includegraphics[width=0.28\linewidth]{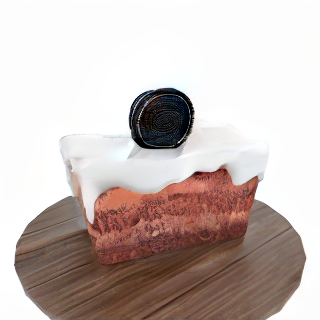} &
\includegraphics[width=0.28\linewidth]{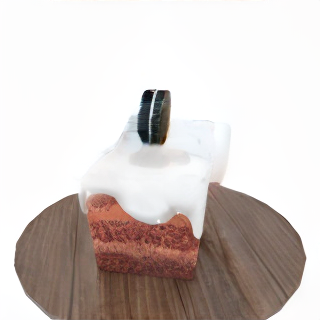} \\
\raisebox{25pt}{\rotatebox[origin=c]{90}{Original}} &
\includegraphics[width=0.27\linewidth, trim={0.8cm 0.6cm 0.8cm 1cm}, clip]{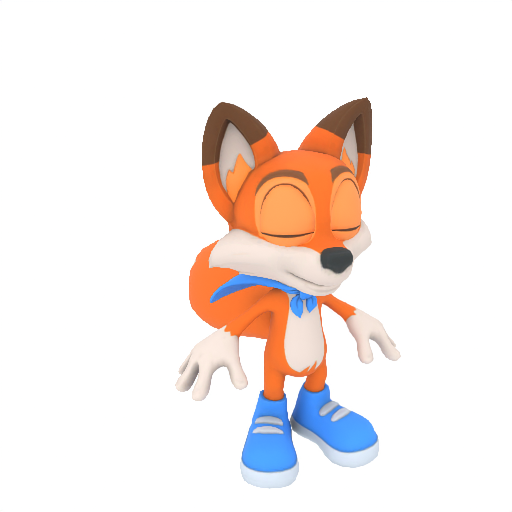} & 
\includegraphics[width=0.28\linewidth, trim={0.8cm 0.6cm 0.8cm 1cm}, clip]{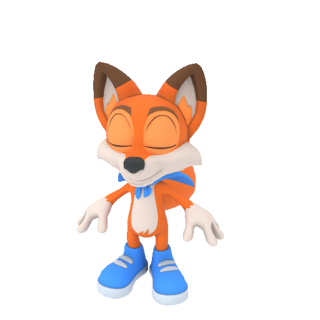} & 
\includegraphics[width=0.28\linewidth, trim={2cm 2cm 2cm 2cm}, clip]{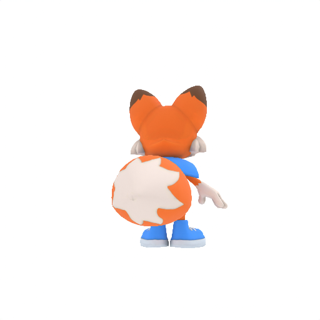} \\
\raisebox{25pt}{\rotatebox[origin=c]{90}{Edited}} &
\includegraphics[width=0.27\linewidth, trim={0.8cm 0.6cm 0.8cm 1cm}, clip]{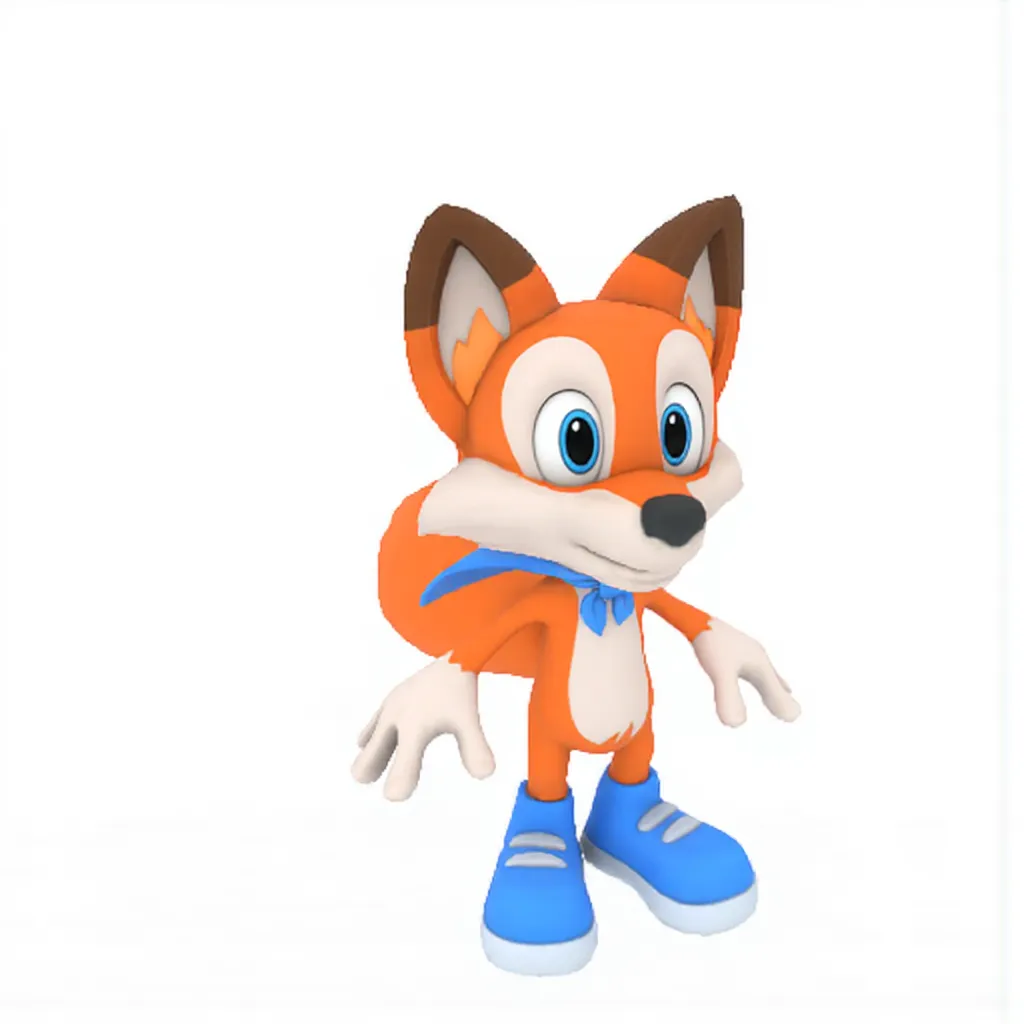} & 
\includegraphics[width=0.28\linewidth, trim={0.8cm 0.6cm 0.8cm 1cm}, clip]{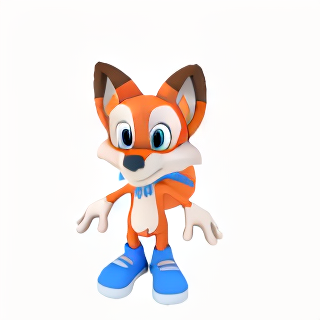} & 
\includegraphics[width=0.28\linewidth, trim={2cm 2cm 2cm 2cm}, clip]{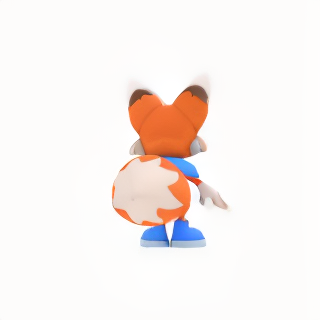} \\

    \end{tabular}
\end{minipage}%
\hfill %
\begin{minipage}[t]{0.49\textwidth}
    \centering
    \begin{tabular}{@{}cccc@{}}
        & \textbf{Cond. View} & \textbf{View 1} & \textbf{View 2} \\
        \raisebox{25pt}{\rotatebox[origin=c]{90}{Original}} &
\includegraphics[width=0.27\linewidth, trim={2cm 2cm 2cm 2cm}, clip]{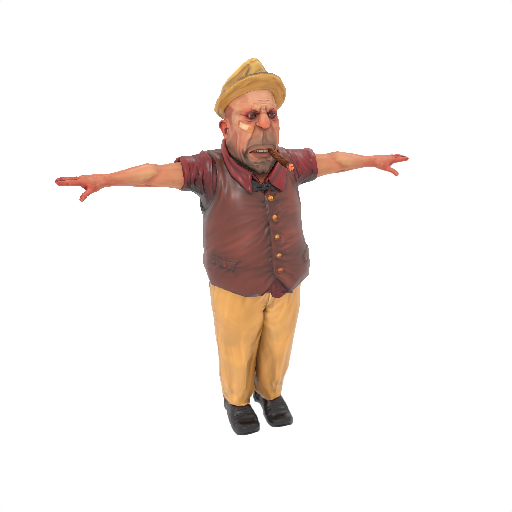} &
\includegraphics[width=0.28\linewidth, trim={1.5cm 1.5cm 1.5cm 1.5cm}, clip]{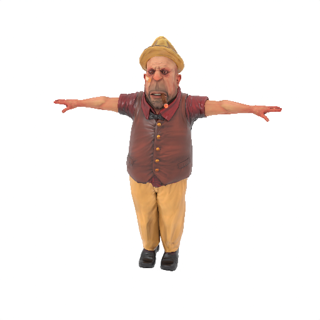} &
\includegraphics[width=0.28\linewidth, trim={1.5cm 1.5cm 1.5cm 1.5cm}, clip]{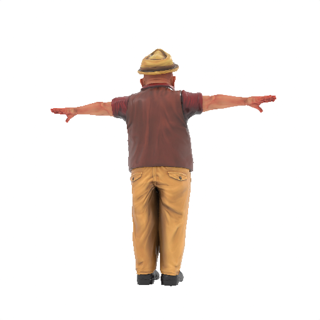} \\

\raisebox{25pt}{\rotatebox[origin=c]{90}{Edited}} &
\includegraphics[width=0.27\linewidth, trim={2cm 2cm 2cm 2cm}, clip]{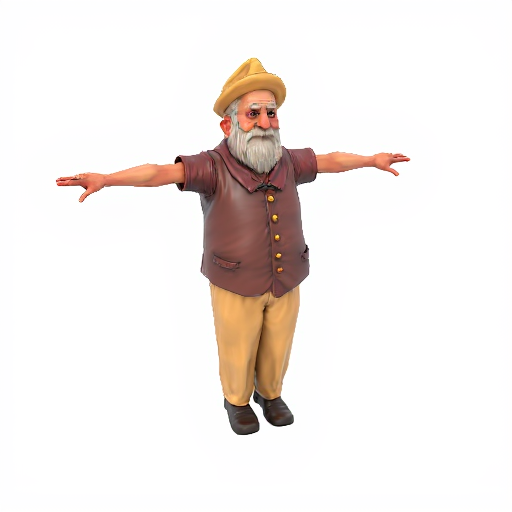} &
\includegraphics[width=0.28\linewidth, trim={1.5cm 1.5cm 1.5cm 1.5cm}, clip]{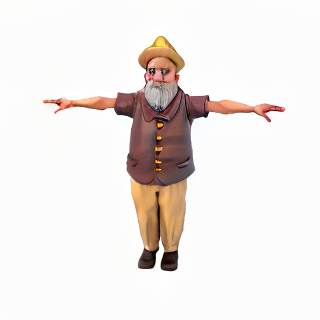} &
\includegraphics[width=0.28\linewidth, trim={1.5cm 1.5cm 1.5cm 1.5cm}, clip]{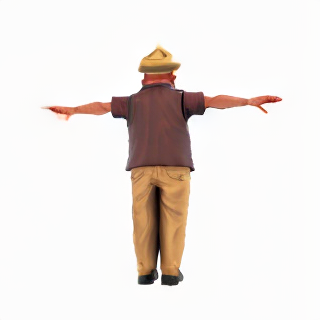} \\

\raisebox{25pt}{\rotatebox[origin=c]{90}{Original}} &
\includegraphics[width=0.27\linewidth, trim={2cm 2cm 2cm 2cm}, clip]{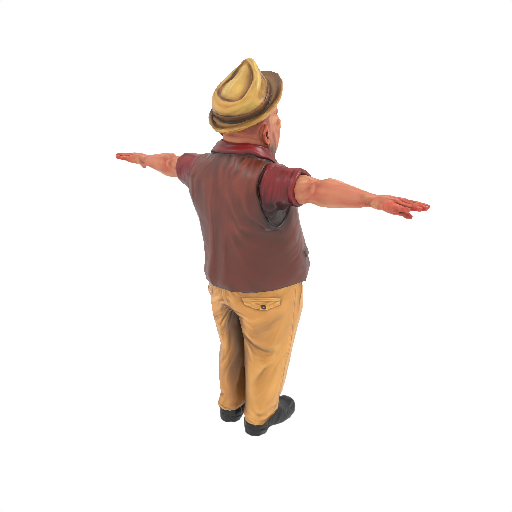} &
\includegraphics[width=0.28\linewidth, trim={1.5cm 1.5cm 1.5cm 1.5cm}, clip]{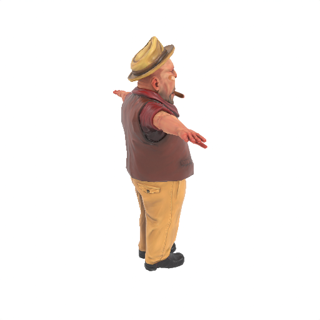} &
            \begin{tikzpicture}[spy using outlines={circle, myspycolor, magnification=2, size=0.5cm, connect spies}]
    \node {\includegraphics[width=0.28\linewidth, trim={1.5cm 1.5cm 1.5cm 1.5cm}, clip]{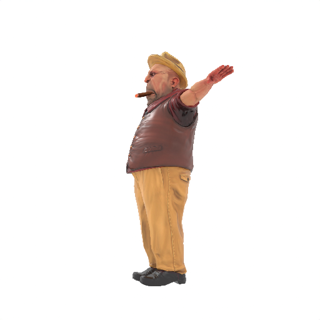}};
    \spy on (-0.15,0.7) in node [left] at (1.45,1); %
    \end{tikzpicture} \\

\raisebox{25pt}{\rotatebox[origin=c]{90}{Edited}} &
\includegraphics[width=0.27\linewidth, trim={2cm 2cm 2cm 2cm}, clip]{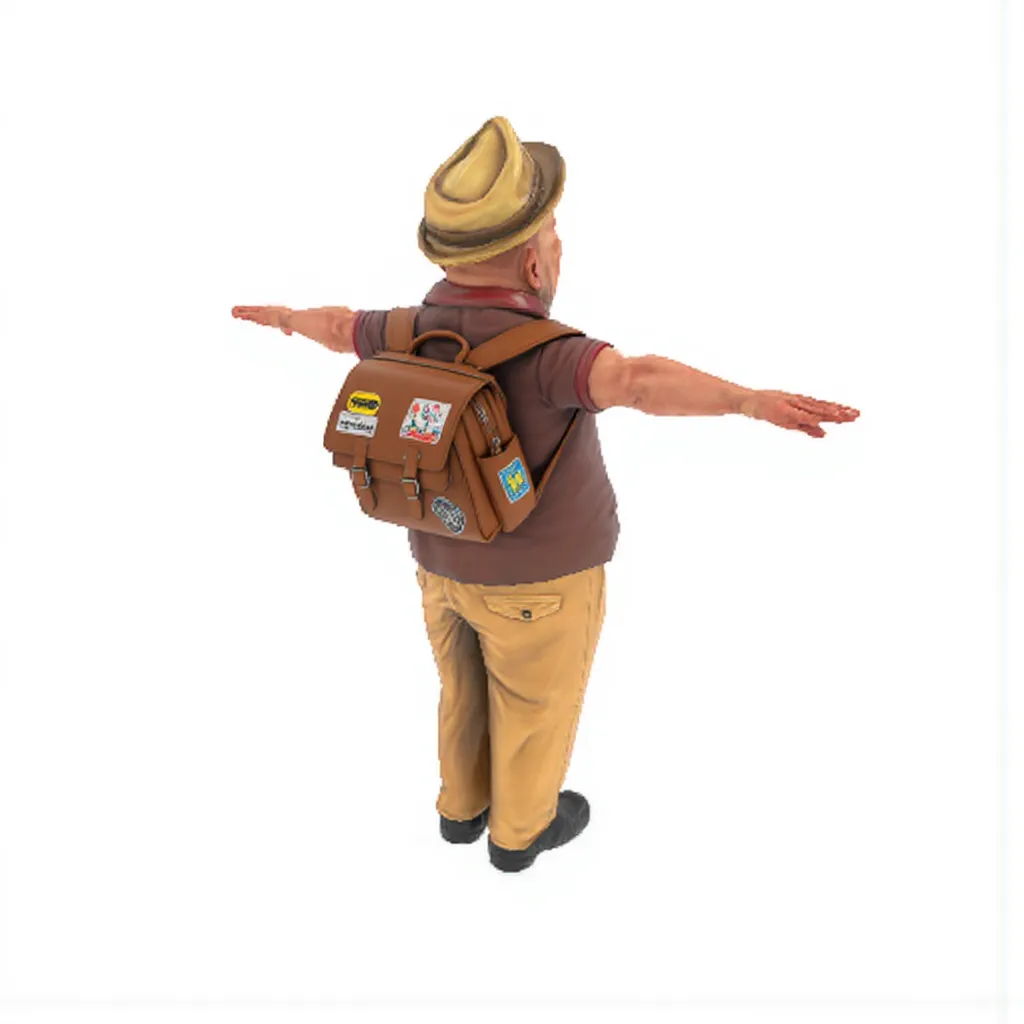} &
\includegraphics[width=0.28\linewidth, trim={1.5cm 1.5cm 1.5cm 1.5cm}, clip]{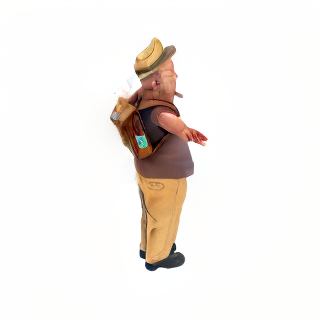} &
            \begin{tikzpicture}[spy using outlines={circle, myspycolor, magnification=2, size=0.5cm, connect spies}]
    \node {\includegraphics[width=0.28\linewidth, trim={1.5cm 1.5cm 1.5cm 1.5cm}, clip]{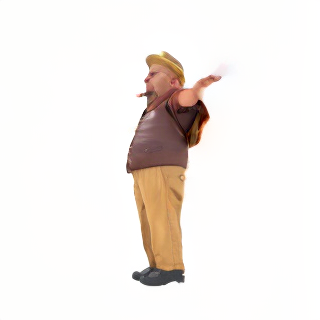}};
    \spy on (-0.15,0.7) in node [left] at (1.45,1); %
    \end{tikzpicture} \\
 
\raisebox{25pt}{\rotatebox[origin=c]{90}{Original}} &
\includegraphics[width=0.27\linewidth, trim={1cm 1cm 1cm 1cm}, clip]{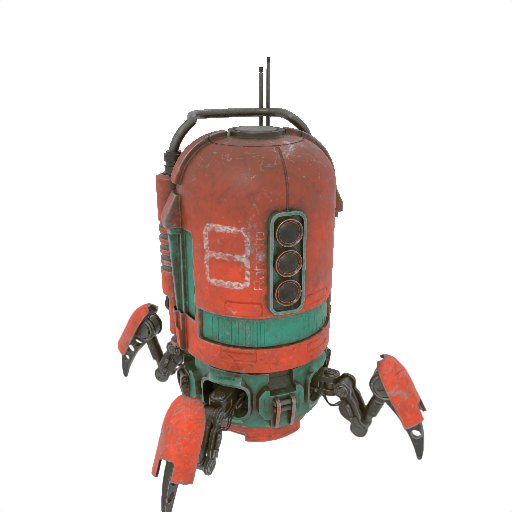} &
\includegraphics[width=0.28\linewidth, trim={0.5cm 0.5cm 0.5cm 0.5cm}, clip]{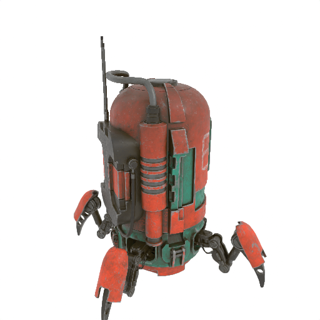} &
\includegraphics[width=0.28\linewidth, trim={1cm 1cm 1cm 1cm}, clip]{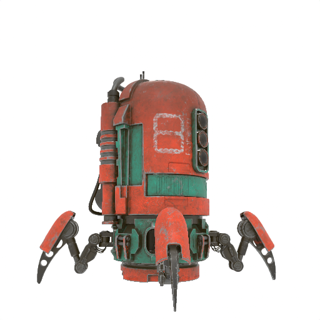} \\

\raisebox{25pt}{\rotatebox[origin=c]{90}{Edited}} &
\includegraphics[width=0.27\linewidth, trim={1cm 1cm 1cm 1cm}, clip]{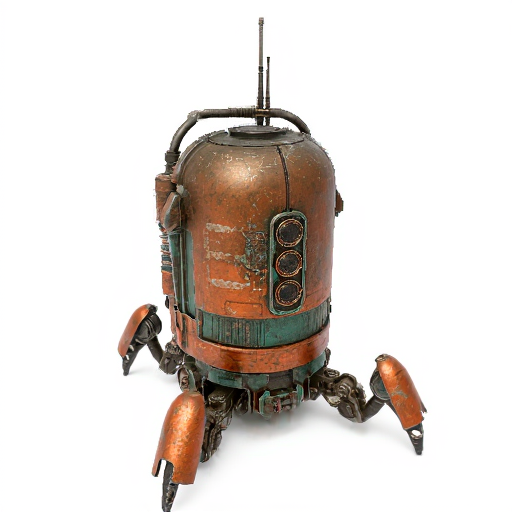} &
\includegraphics[width=0.28\linewidth, trim={1cm 1cm 1cm 1cm}, clip]{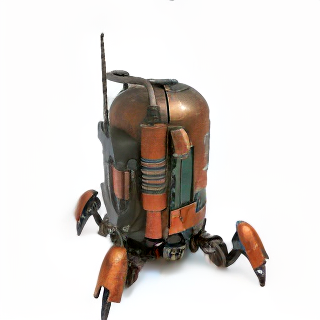} &
\includegraphics[width=0.28\linewidth, trim={1cm 1cm 1cm 1cm}, clip]{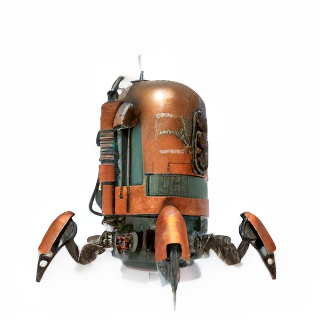} \\

\raisebox{25pt}{\rotatebox[origin=c]{90}{Original}} &
\includegraphics[width=0.27\linewidth, trim={1cm 1cm 1cm 1cm}, clip]{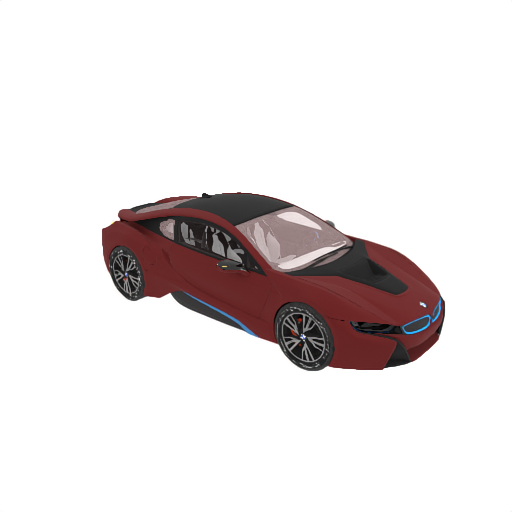} & 
\includegraphics[width=0.28\linewidth, trim={1cm 1cm 1cm 1cm}, clip]{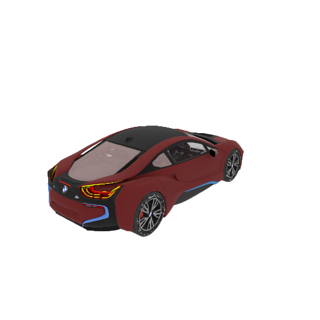} & 
\includegraphics[width=0.28\linewidth, trim={1cm 1cm 1cm 1cm}, clip]{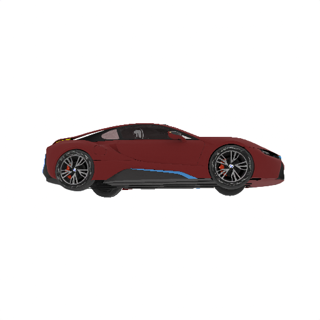} \\

\raisebox{25pt}{\rotatebox[origin=c]{90}{Edited}} &
\includegraphics[width=0.27\linewidth, trim={1cm 1cm 1cm 1cm}, clip]{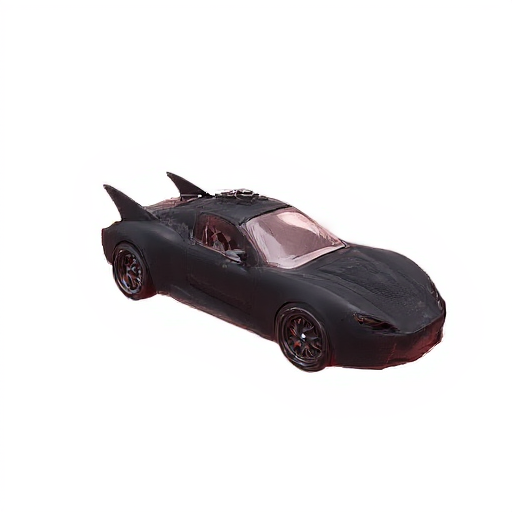} & 
\includegraphics[width=0.28\linewidth, trim={1cm 1cm 1cm 1cm}, clip]{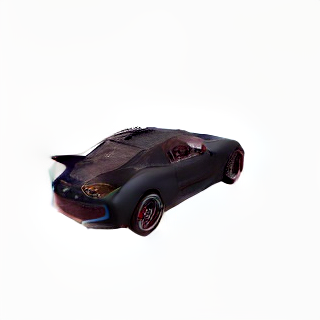} & 
\includegraphics[width=0.28\linewidth, trim={1cm 1cm 1cm 1cm}, clip]{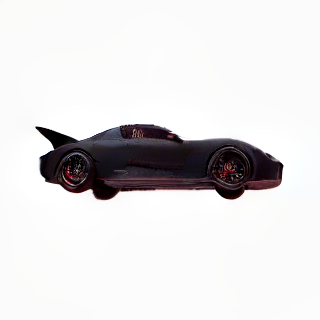} \\
    \end{tabular}
\end{minipage}

\caption{\textbf{Examples of Multi-View Grid Editing.} Each block shows an original object (top) and its edited result (bottom). The leftmost column contains the conditioning views (source and target), while the other columns display the propagated edit from novel viewpoints.}
\label{fig:additional_results2}
\end{figure*}

%% file: figures/rendered.tex
\begin{figure*}[hbtp]
\centering

\begin{tabular}{c c c c c c c c c}
&
\multicolumn{2}{c}{\includegraphics[width=0.1\textwidth, trim={1cm 1cm 1cm 1cm}, clip, valign=c]{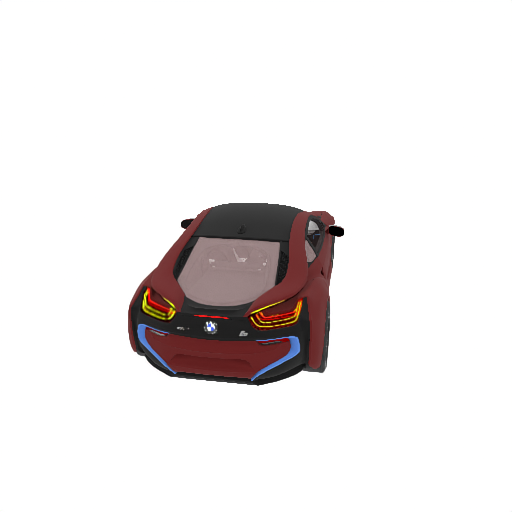} $\rightarrow$ \includegraphics[width=0.1\textwidth, trim={1cm 1cm 1cm 1cm}, clip, valign=c]{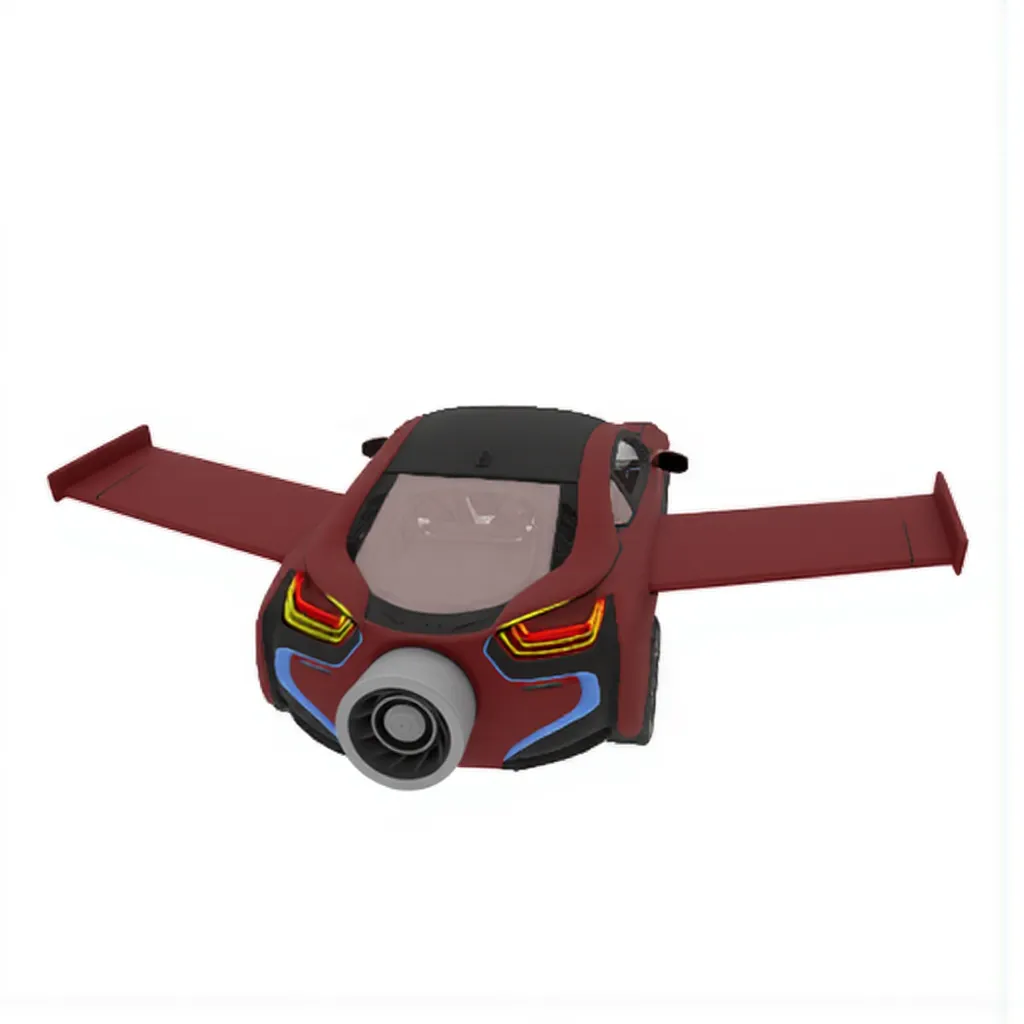}} &
\multicolumn{2}{c}{\includegraphics[width=0.1\textwidth, trim={1cm 1cm 0.5cm 0.5cm}, clip, valign=c]{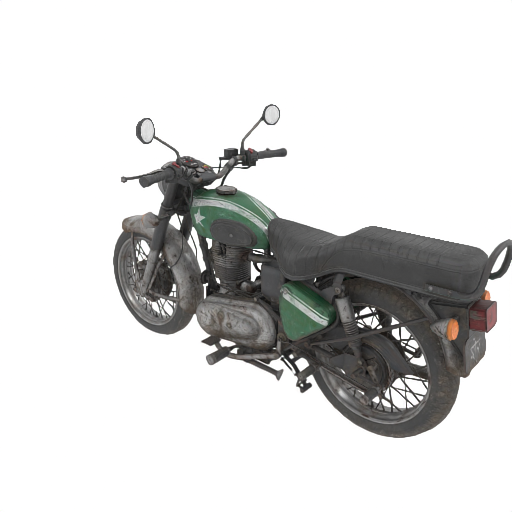} $\rightarrow$ \includegraphics[width=0.1\textwidth, trim={1cm 1cm 0.5cm 0.5cm}, clip, valign=c]{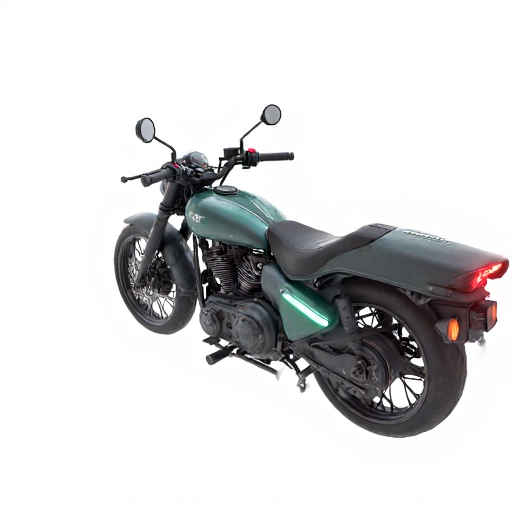}} &
\multicolumn{2}{c}{\includegraphics[width=0.1\textwidth, trim={1cm 1cm 1cm 1cm}, clip, valign=c]{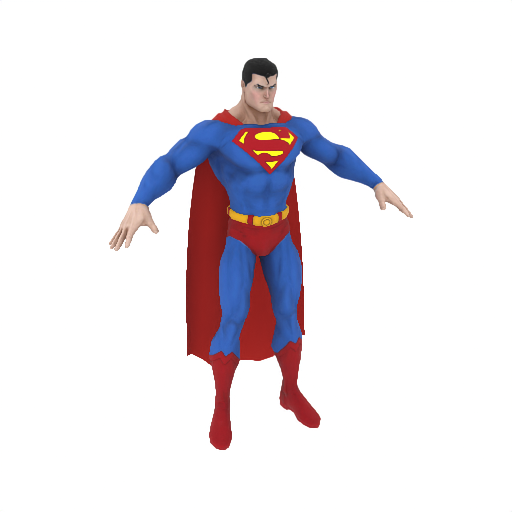} $\rightarrow$ \includegraphics[width=0.1\textwidth, trim={1cm 1cm 1cm 1cm}, clip, valign=c]{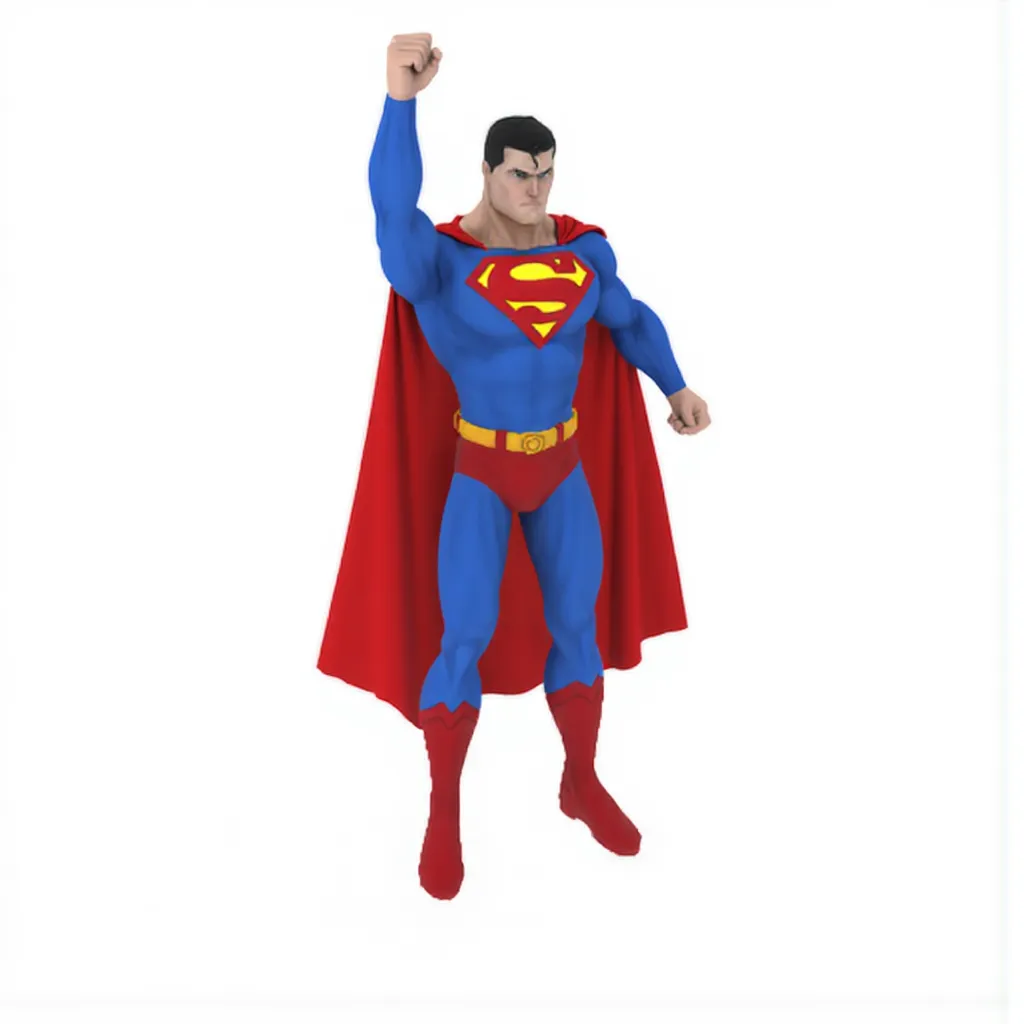}} &
\multicolumn{2}{c}{\includegraphics[width=0.1\textwidth, trim={0.6cm 1cm 0.2cm 0.2cm}, clip, valign=c]{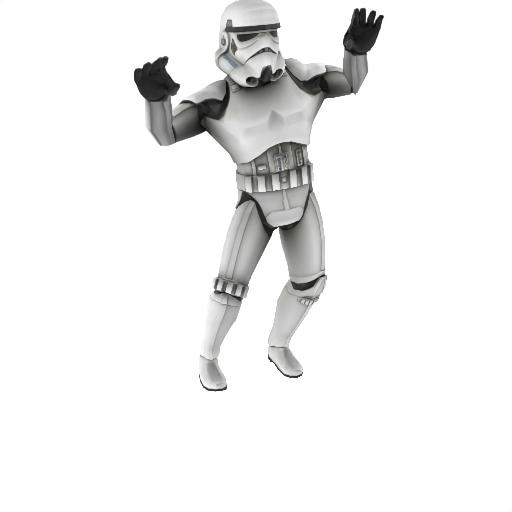} $\rightarrow$ \includegraphics[width=0.1\textwidth, trim={0.6cm 1cm 0.6cm 0.2cm}, clip, valign=c]{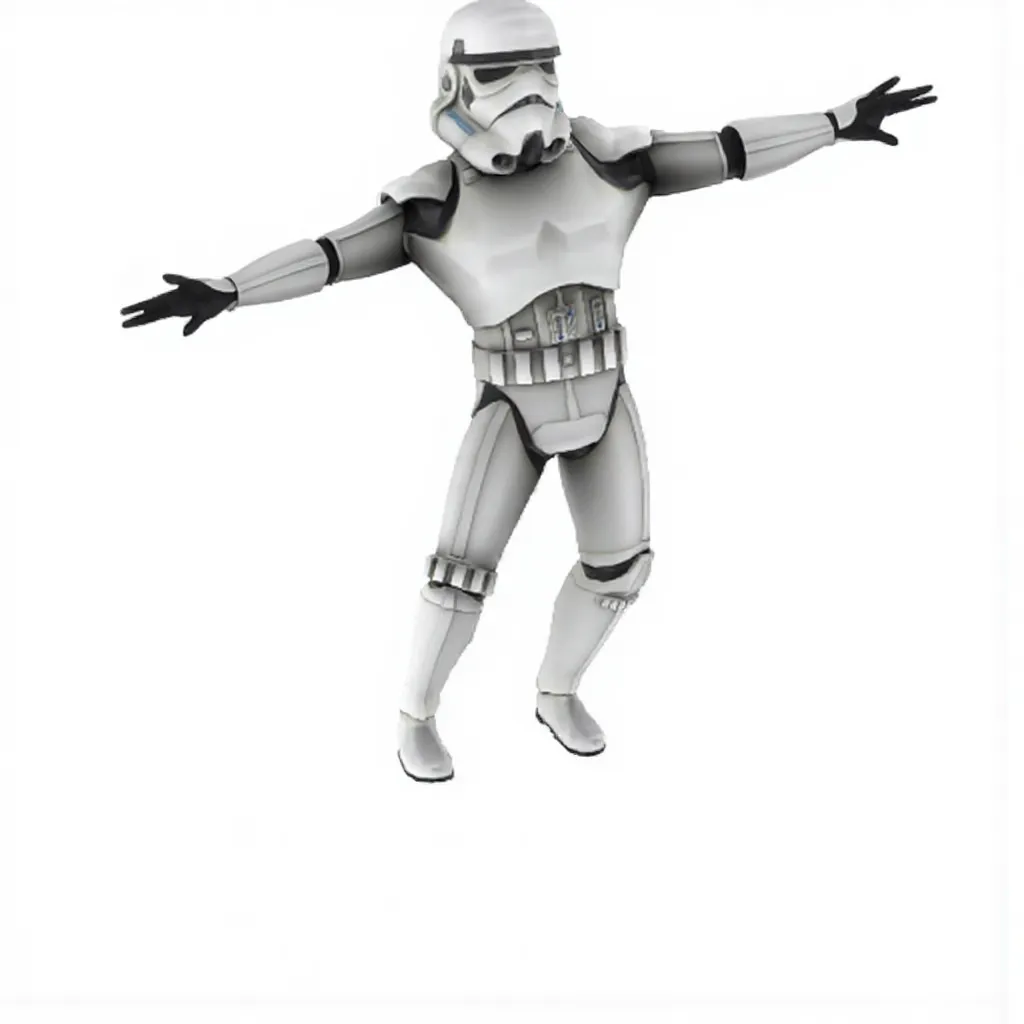}} \\
\addlinespace[1.5em]
& \textbf{Textured} & \textbf{Untextured} & \textbf{Textured} & \textbf{Untextured} & \textbf{Textured} & \textbf{Untextured} & \textbf{Textured} & \textbf{Untextured} \\
\addlinespace[0.5em]
& \includegraphics[width=0.1\textwidth, trim={1cm 1cm 1cm 1cm}, clip]{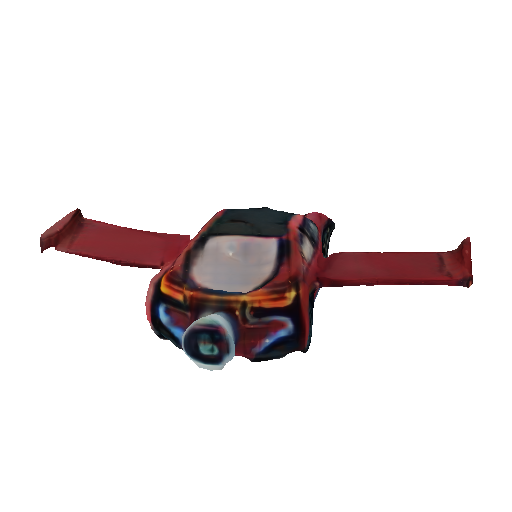} & \includegraphics[width=0.1\textwidth, trim={1cm 1cm 1cm 1cm}, clip]{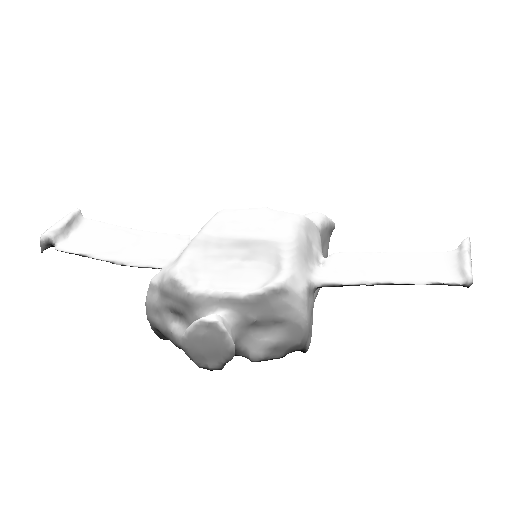} & \includegraphics[width=0.1\textwidth, trim={1cm 1cm 1cm 1cm}, clip]{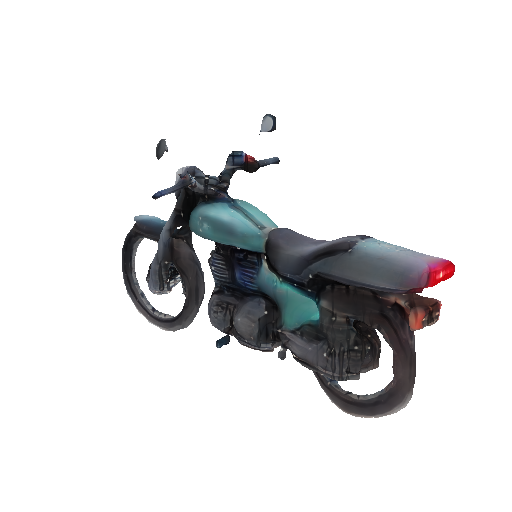} & \includegraphics[width=0.1\textwidth, trim={1cm 1cm 1cm 1cm}, clip]{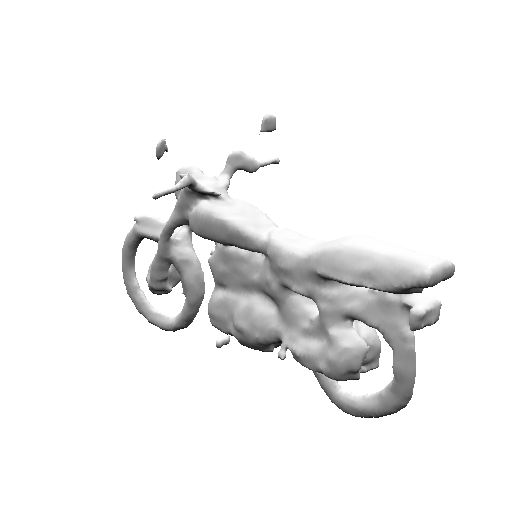} & \includegraphics[width=0.1\textwidth, trim={1cm 1cm 1cm 1cm}, clip]{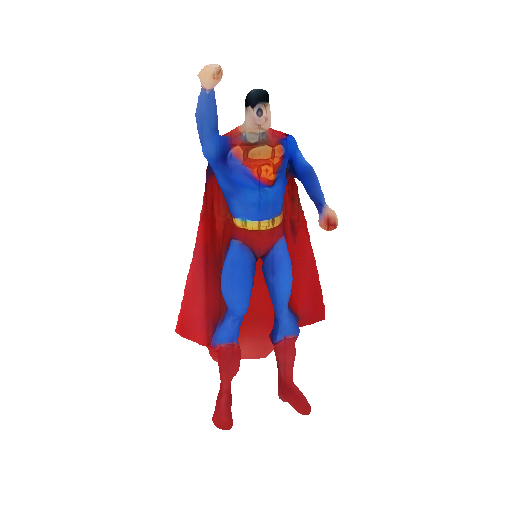} & \includegraphics[width=0.1\textwidth, trim={1cm 1cm 1cm 1cm}, clip]{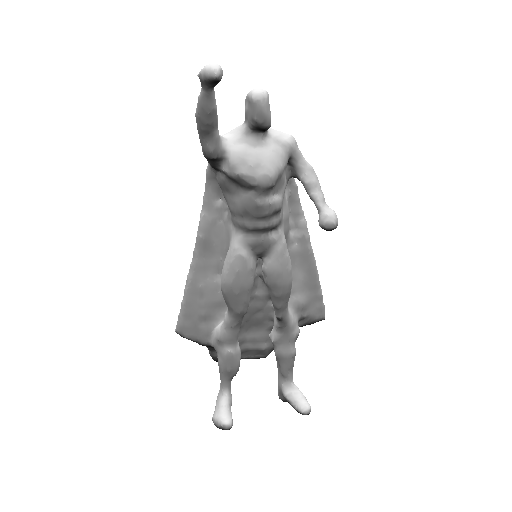} & \includegraphics[width=0.1\textwidth, trim={1cm 1cm 1cm 1cm}, clip]{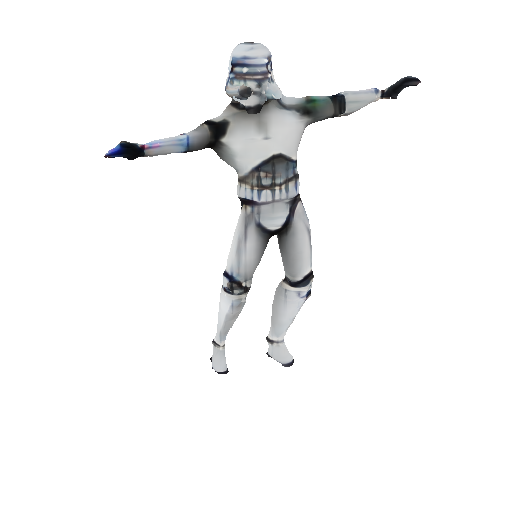} & \includegraphics[width=0.1\textwidth, trim={1cm 1cm 1cm 1cm}, clip]{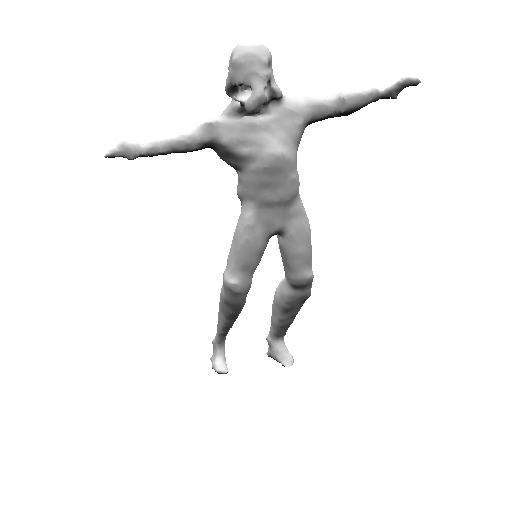} \\
& \includegraphics[width=0.1\textwidth, trim={1cm 1cm 1cm 1cm}, clip]{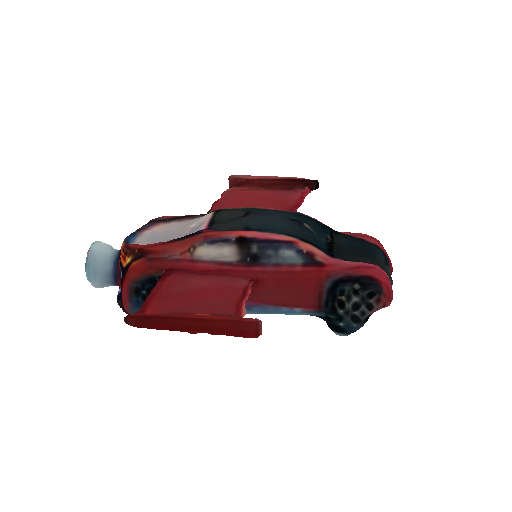} & \includegraphics[width=0.1\textwidth, trim={1cm 1cm 1cm 1cm}, clip]{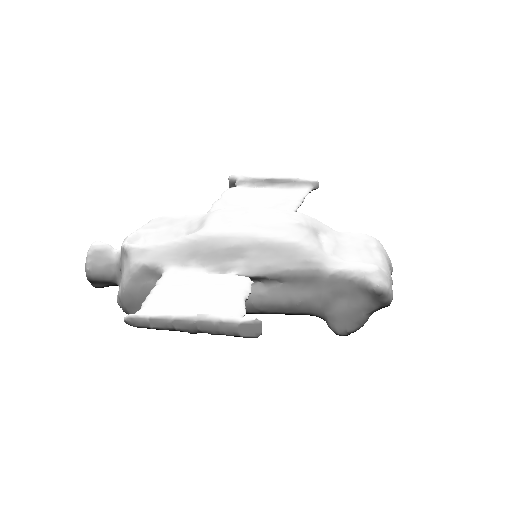} & \includegraphics[width=0.1\textwidth, trim={1cm 1cm 1cm 1cm}, clip]{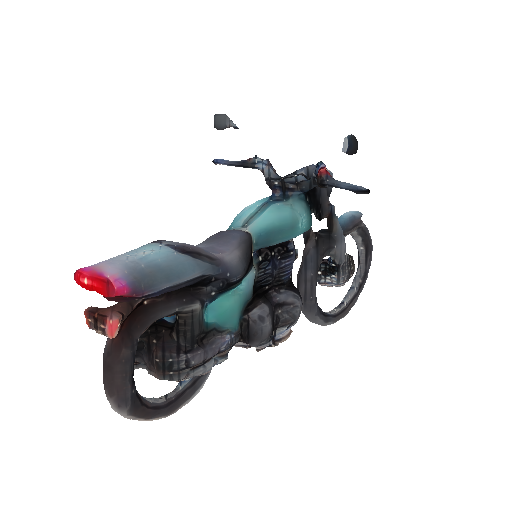} & \includegraphics[width=0.1\textwidth, trim={1cm 1cm 1cm 1cm}, clip]{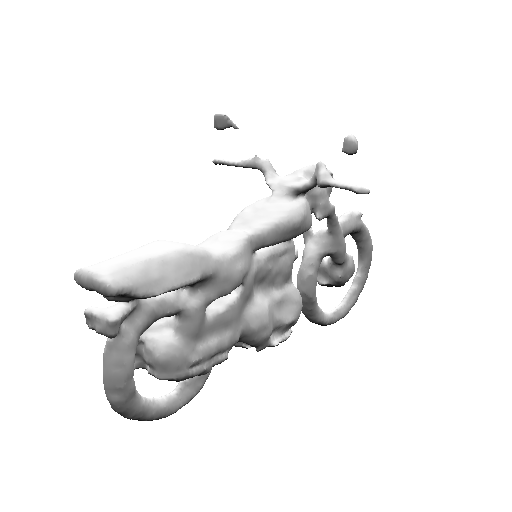} & \includegraphics[width=0.1\textwidth, trim={1cm 1cm 1cm 1cm}, clip]{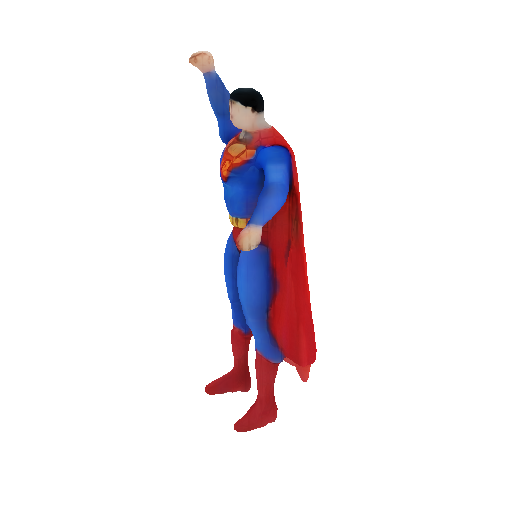} & \includegraphics[width=0.1\textwidth, trim={1cm 1cm 1cm 1cm}, clip]{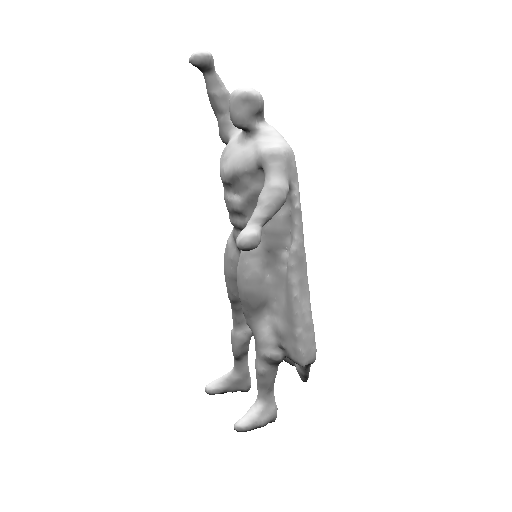} & \includegraphics[width=0.1\textwidth, trim={1cm 1cm 1cm 1cm}, clip]{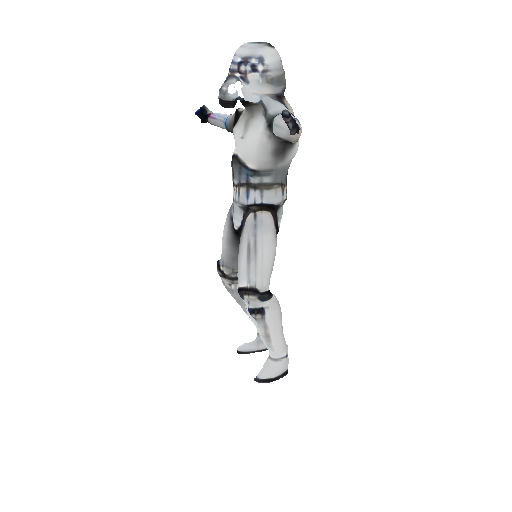} & \includegraphics[width=0.1\textwidth, trim={1cm 1cm 1cm 1cm}, clip]{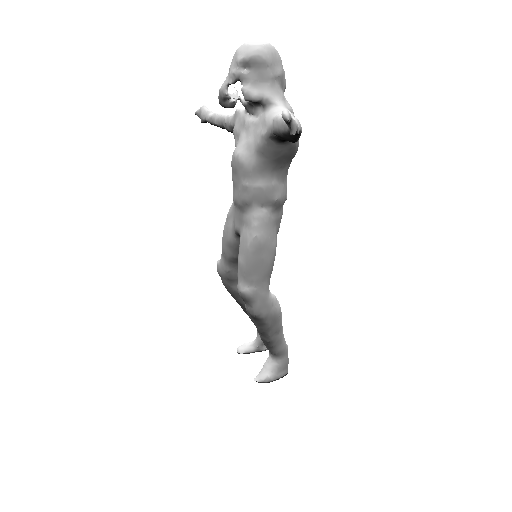} \\
& \includegraphics[width=0.1\textwidth, trim={1cm 1cm 1cm 1cm}, clip]{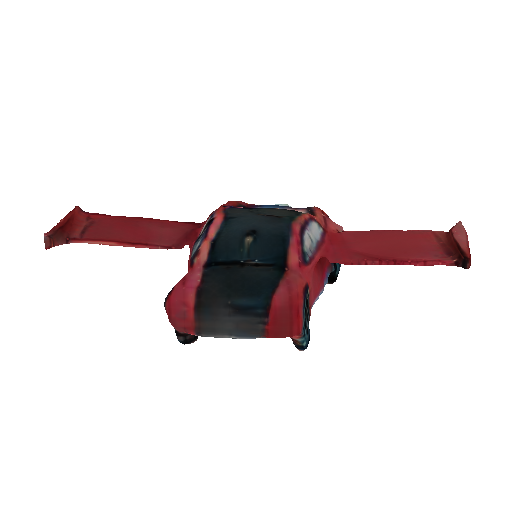} & \includegraphics[width=0.1\textwidth, trim={1cm 1cm 1cm 1cm}, clip]{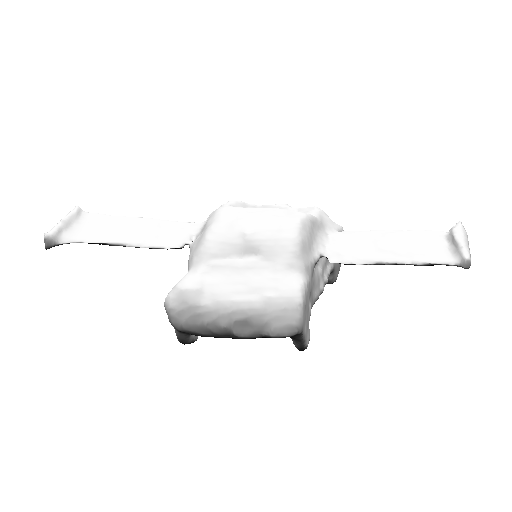} & \includegraphics[width=0.1\textwidth, trim={1cm 1cm 1cm 1cm}, clip]{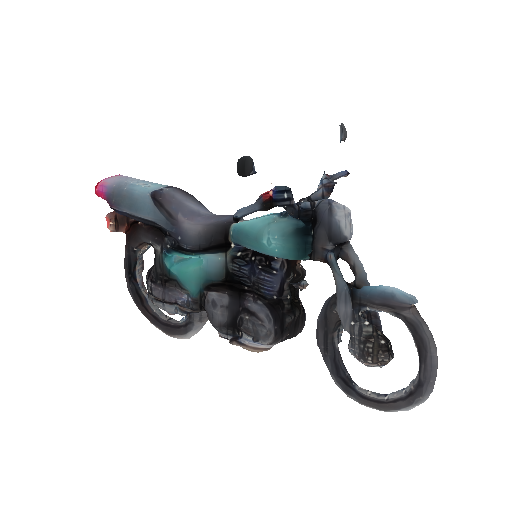} & \includegraphics[width=0.1\textwidth, trim={1cm 1cm 1cm 1cm}, clip]{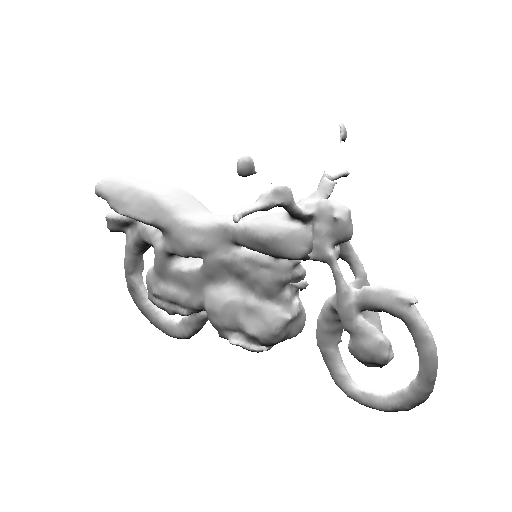} & \includegraphics[width=0.1\textwidth, trim={1cm 1cm 1cm 1cm}, clip]{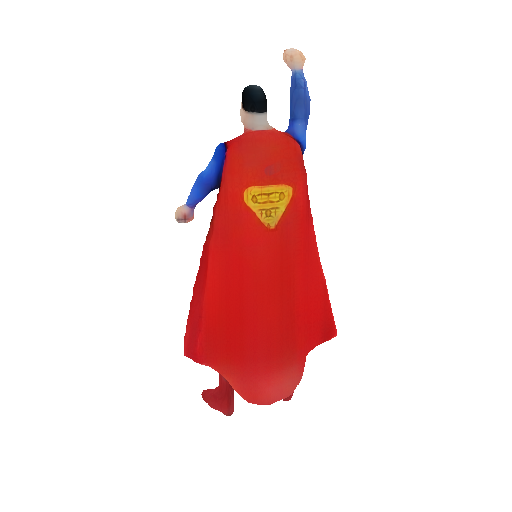} & \includegraphics[width=0.1\textwidth, trim={1cm 1cm 1cm 1cm}, clip]{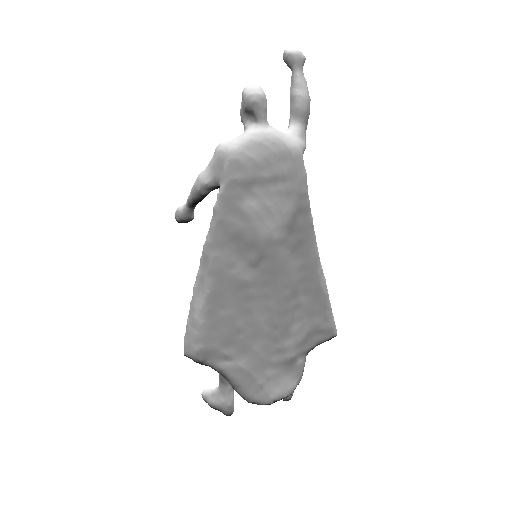} & \includegraphics[width=0.1\textwidth, trim={1cm 1cm 1cm 1cm}, clip]{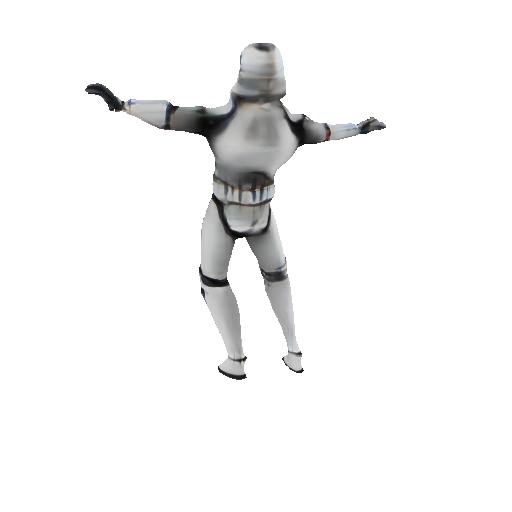} & \includegraphics[width=0.1\textwidth, trim={1cm 1cm 1cm 1cm}, clip]{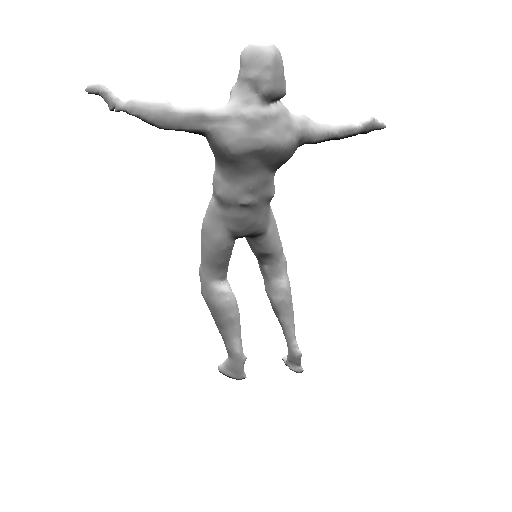} \\
& \includegraphics[width=0.1\textwidth, trim={1cm 1cm 1cm 1cm}, clip]{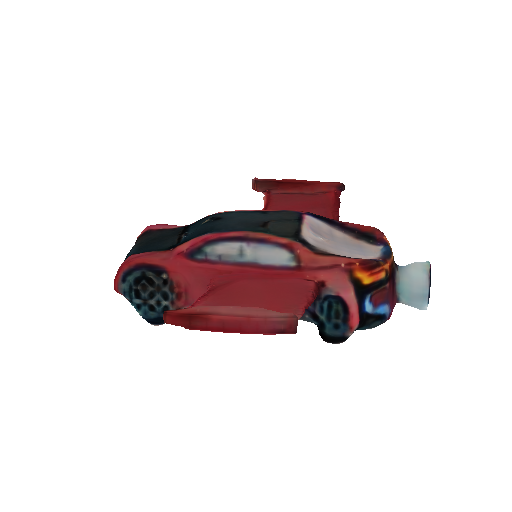} & \includegraphics[width=0.1\textwidth, trim={1cm 1cm 1cm 1cm}, clip]{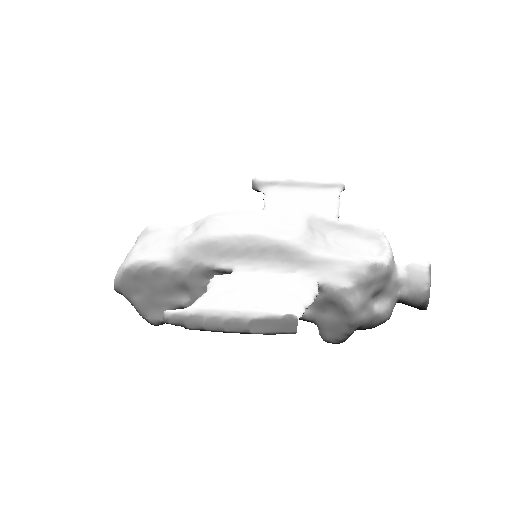} & \includegraphics[width=0.1\textwidth, trim={1cm 1cm 1cm 1cm}, clip]{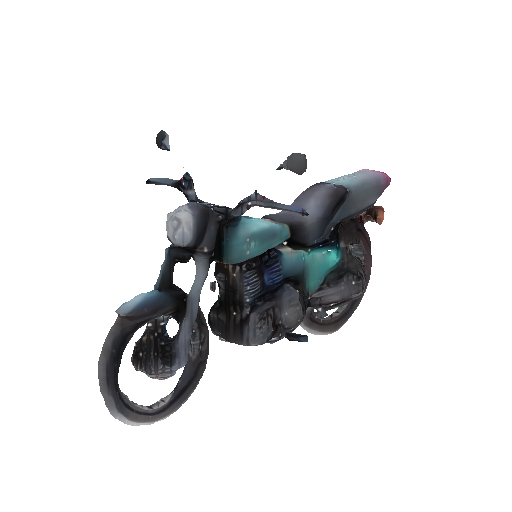} & \includegraphics[width=0.1\textwidth, trim={1cm 1cm 1cm 1cm}, clip]{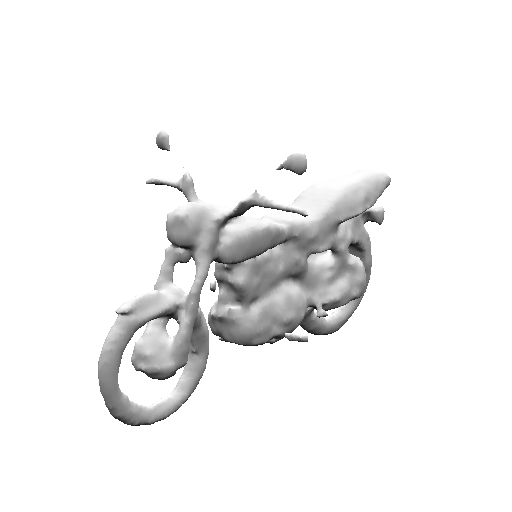} & \includegraphics[width=0.1\textwidth, trim={1cm 1cm 1cm 1cm}, clip]{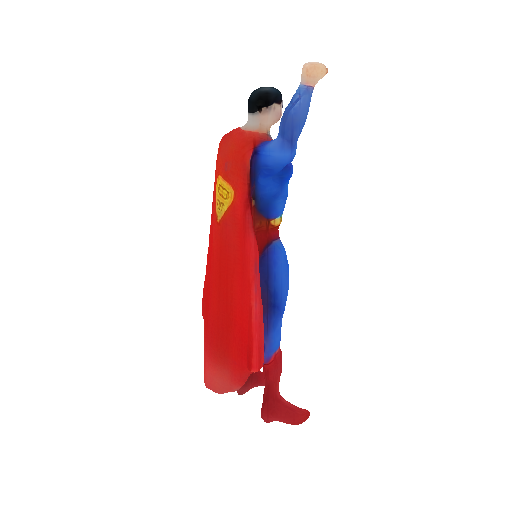} & \includegraphics[width=0.1\textwidth, trim={1cm 1cm 1cm 1cm}, clip]{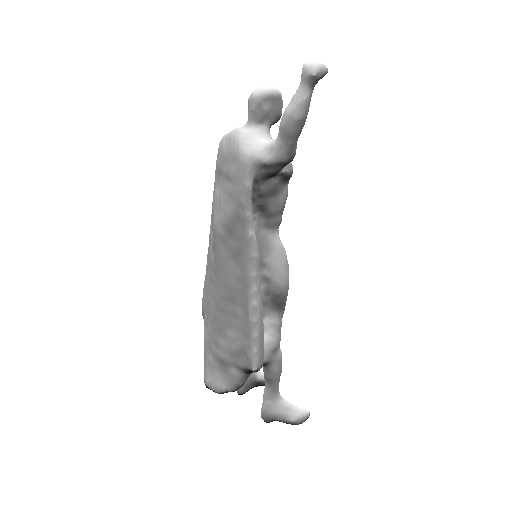} & \includegraphics[width=0.1\textwidth, trim={1cm 1cm 1cm 1cm}, clip]{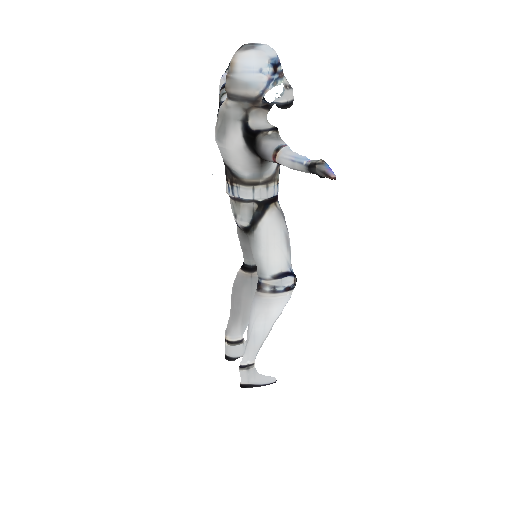} & \includegraphics[width=0.1\textwidth, trim={1cm 1cm 1cm 1cm}, clip]{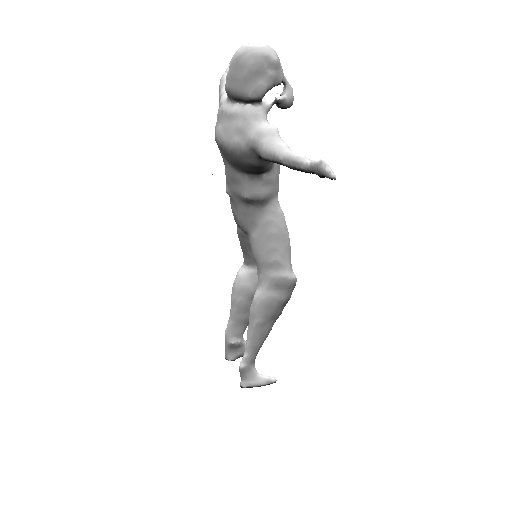}
\end{tabular}
\caption{\textbf{Textured and Untextured Edits After 3D Reconstruction.} The top row shows the 2D edit that guides the process (source $\rightarrow$ edited). The rows below present novel views of the final reconstructed 3D mesh, displaying both the textured and untextured geometry. The untextured results confirm that the edits are modifications to the shape itself and not just surface effects.}
\label{fig:reconstructed_editing_results}

\vspace{2em}

\begin{minipage}{0.48\textwidth}
    \centering
    \begin{tabular}{c c c c}
         & \textbf{Cond. View} & \textbf{View 1} & \textbf{View 2} \\
        \addlinespace[0.3em]
        \raisebox{25pt}{\rotatebox{90}{Original}} &
        \includegraphics[width=0.25\linewidth, trim={0.5cm 1.5cm 1.5cm 0.5cm}, clip]{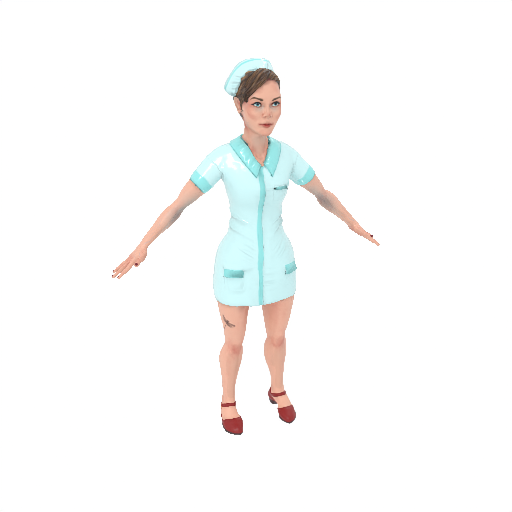} &
        \includegraphics[width=0.25\linewidth, trim={0.5cm 1.5cm 1.5cm 0.5cm}, clip]{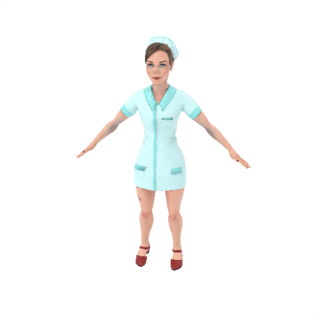} &
        \includegraphics[width=0.25\linewidth, trim={0.5cm 1.5cm 1.5cm 0.5cm}, clip]{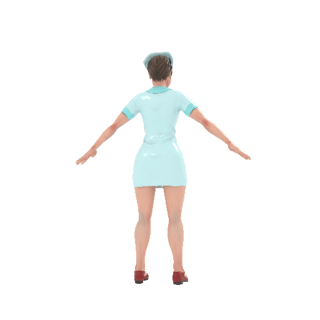} \\
        \addlinespace[0.3em]
        \raisebox{25pt}{\rotatebox{90}{Edited}} &
        \includegraphics[width=0.25\linewidth, trim={0.5cm 1.5cm 1.5cm 0.5cm}, clip]{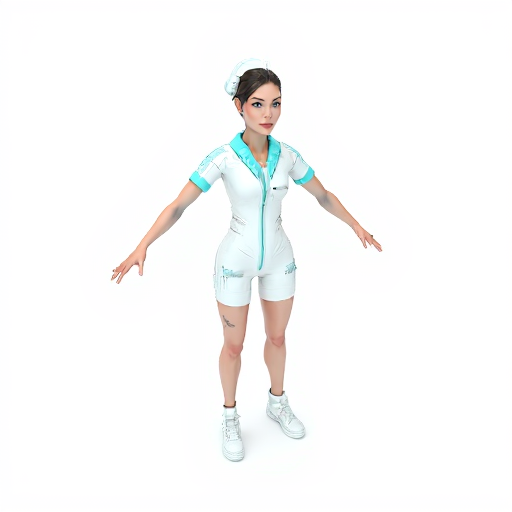} &
        \includegraphics[width=0.25\linewidth, trim={0.5cm 1.5cm 1.5cm 0.5cm}, clip]{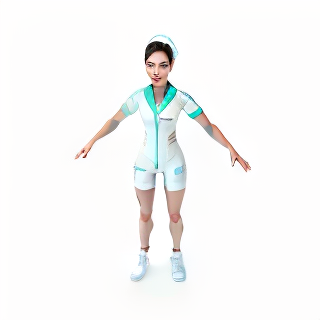} &
        \includegraphics[width=0.25\linewidth, trim={0.5cm 1.5cm 1.5cm 0.5cm}, clip]{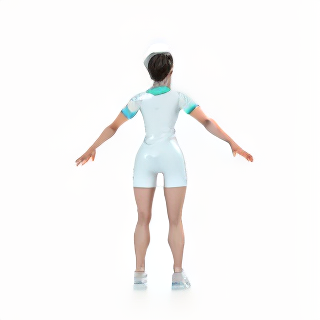}
    \end{tabular}
\end{minipage}
\hfill %
\begin{minipage}{0.48\textwidth}
    \centering
    \begin{tabular}{c c c c}
         & \textbf{Cond. View} & \textbf{View 1} & \textbf{View 2} \\
        \addlinespace[0.3em]
         \raisebox{25pt}{\rotatebox{90}{{Original}}} &
        \includegraphics[width=0.27\linewidth]{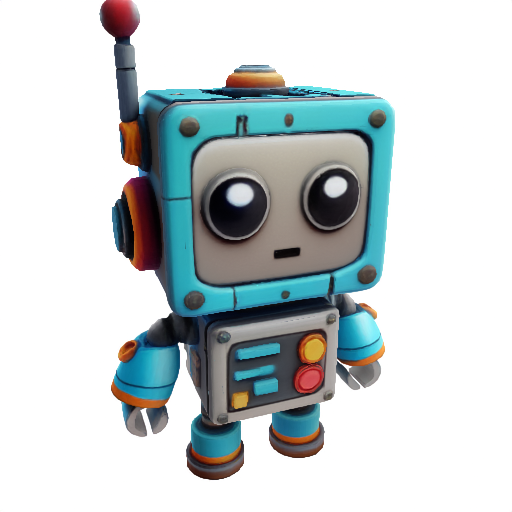} & 
        \includegraphics[width=0.28\linewidth]{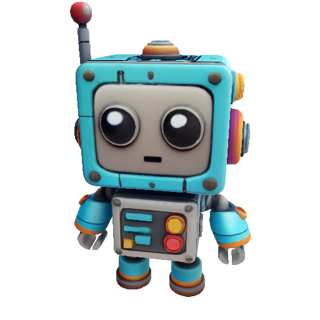} & 
        \includegraphics[width=0.28\linewidth]{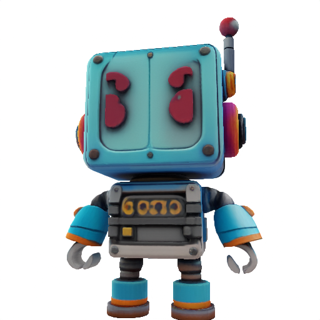} \\
        \addlinespace[0.3em]
        \raisebox{25pt}{\rotatebox{90}{Edited}} &
        \includegraphics[width=0.27\linewidth]{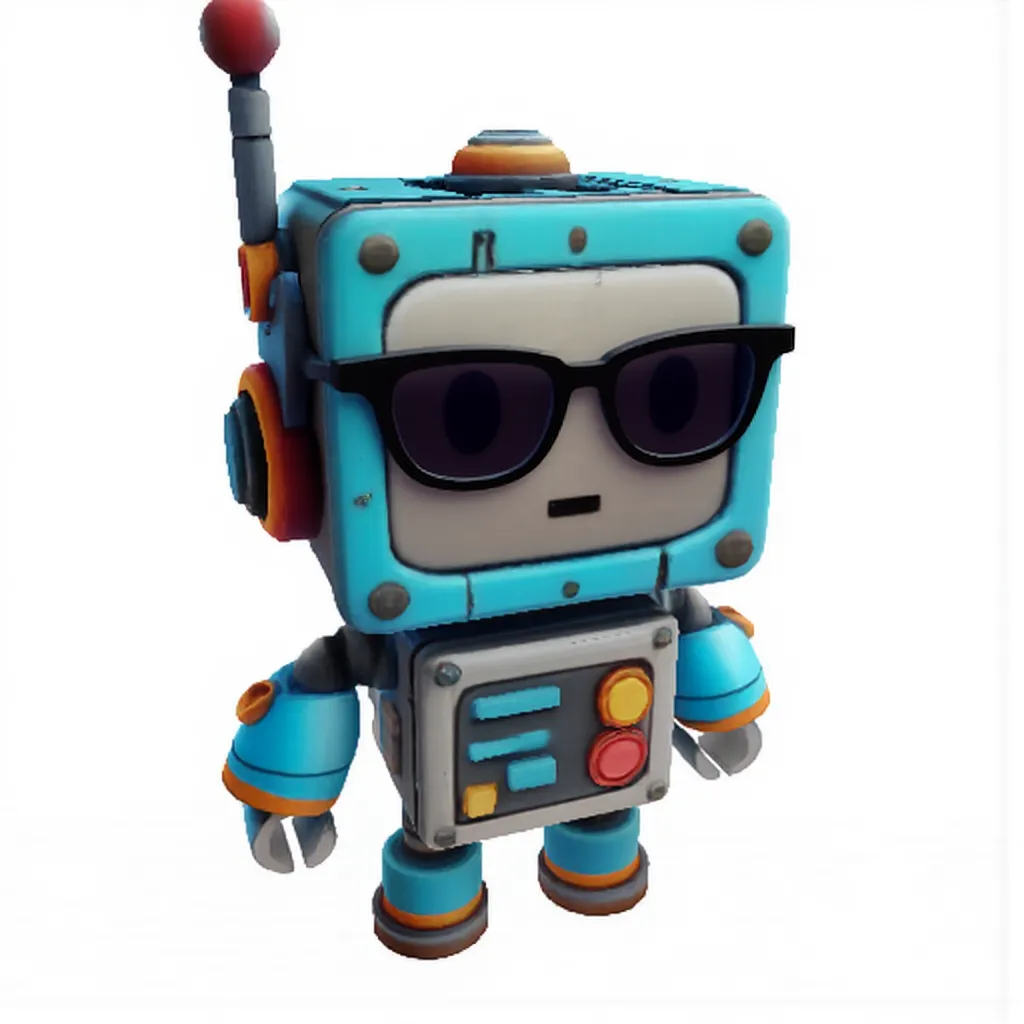} & 
        \includegraphics[width=0.28\linewidth]{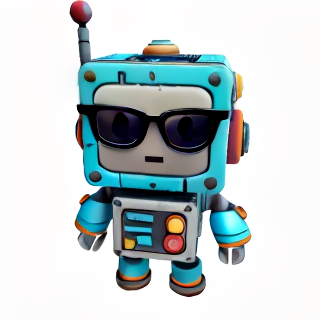} & 
        \includegraphics[width=0.28\linewidth]{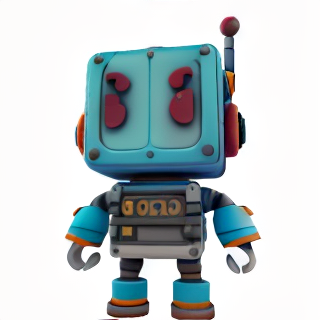} 
    \end{tabular}
\end{minipage}

\captionof{figure}{\textbf{Examples of Multi-View Grid Editing.} Each block shows an original object (top) and its edited result (bottom). The leftmost column contains the conditioning views (source and target), while the other columns display the propagated edit from novel viewpoints.}
\label{fig:side_by_side_results} %

\end{figure*}

%% file: main.bbl

\begin{thebibliography}{43}


\ifx \showCODEN    \undefined \def \showCODEN     #1{\unskip}     \fi
\ifx \showDOI      \undefined \def \showDOI       #1{#1}\fi
\ifx \showISBNx    \undefined \def \showISBNx     #1{\unskip}     \fi
\ifx \showISBNxiii \undefined \def \showISBNxiii  #1{\unskip}     \fi
\ifx \showISSN     \undefined \def \showISSN      #1{\unskip}     \fi
\ifx \showLCCN     \undefined \def \showLCCN      #1{\unskip}     \fi
\ifx \shownote     \undefined \def \shownote      #1{#1}          \fi
\ifx \showarticletitle \undefined \def \showarticletitle #1{#1}   \fi
\ifx \showURL      \undefined \def \showURL       {\relax}        \fi
\providecommand\bibfield[2]{#2}
\providecommand\bibinfo[2]{#2}
\providecommand\natexlab[1]{#1}
\providecommand\showeprint[2][]{arXiv:#2}

\bibitem[Avrahami et~al\mbox{.}(2022)]%
        {avrahami2022blended}
\bibfield{author}{\bibinfo{person}{Omri Avrahami}, \bibinfo{person}{Dani Lischinski}, {and} \bibinfo{person}{Ohad Fried}.} \bibinfo{year}{2022}\natexlab{}.
\newblock \showarticletitle{Blended Diffusion for Text-driven Editing of Natural Images}. In \bibinfo{booktitle}{\emph{Proceedings of the IEEE/CVF Conference on Computer Vision and Pattern Recognition}}. \bibinfo{pages}{18208--18218}.
\newblock


\bibitem[Barda et~al\mbox{.}(2025)]%
        {barda2024instant3dit}
\bibfield{author}{\bibinfo{person}{Amir Barda}, \bibinfo{person}{Matheus Gadelha}, \bibinfo{person}{Vladimir~G Kim}, \bibinfo{person}{Noam Aigerman}, \bibinfo{person}{Amit~H Bermano}, {and} \bibinfo{person}{Thibault Groueix}.} \bibinfo{year}{2025}\natexlab{}.
\newblock \showarticletitle{Instant3dit: Multiview inpainting for fast editing of 3d objects}. In \bibinfo{booktitle}{\emph{Proceedings of the Computer Vision and Pattern Recognition Conference}}. \bibinfo{pages}{16273--16282}.
\newblock
\urldef\tempurl%
\url{https://amirbarda.github.io/Instant3dit.github.io/}
\showURL{%
\tempurl}


\bibitem[Brack et~al\mbox{.}(2024)]%
        {ledits++}
\bibfield{author}{\bibinfo{person}{Manuel Brack}, \bibinfo{person}{Felix Friedrich}, \bibinfo{person}{Katharia Kornmeier}, \bibinfo{person}{Linoy Tsaban}, \bibinfo{person}{Patrick Schramowski}, \bibinfo{person}{Kristian Kersting}, {and} \bibinfo{person}{Apolin{\'a}rio Passos}.} \bibinfo{year}{2024}\natexlab{}.
\newblock \showarticletitle{Ledits++: Limitless image editing using text-to-image models}. In \bibinfo{booktitle}{\emph{Proceedings of the IEEE/CVF conference on computer vision and pattern recognition}}. \bibinfo{pages}{8861--8870}.
\newblock


\bibitem[Brooks et~al\mbox{.}(2023)]%
        {brooks2023instructpix2pix}
\bibfield{author}{\bibinfo{person}{Tim Brooks}, \bibinfo{person}{Aleksander Holynski}, {and} \bibinfo{person}{Alexei~A Efros}.} \bibinfo{year}{2023}\natexlab{}.
\newblock \showarticletitle{InstructPix2Pix: Learning to Follow Image Editing Instructions}. In \bibinfo{booktitle}{\emph{Proceedings of the IEEE/CVF Conference on Computer Vision and Pattern Recognition}}. \bibinfo{pages}{18392--18400}.
\newblock


\bibitem[Cao et~al\mbox{.}(2023)]%
        {masactrl}
\bibfield{author}{\bibinfo{person}{Mingdeng Cao}, \bibinfo{person}{Xintao Wang}, \bibinfo{person}{Zhongang Qi}, \bibinfo{person}{Ying Shan}, \bibinfo{person}{Xiaohu Qie}, {and} \bibinfo{person}{Yinqiang Zheng}.} \bibinfo{year}{2023}\natexlab{}.
\newblock \showarticletitle{Masactrl: Tuning-free mutual self-attention control for consistent image synthesis and editing}. In \bibinfo{booktitle}{\emph{Proceedings of the IEEE/CVF international conference on computer vision}}. \bibinfo{pages}{22560--22570}.
\newblock


\bibitem[Chen et~al\mbox{.}(2024b)]%
        {chen2024mvedit}
\bibfield{author}{\bibinfo{person}{Hansheng Chen}, \bibinfo{person}{Yujun Zhang}, \bibinfo{person}{Yuliang Liu}, \bibinfo{person}{Qiangeng Zhang}, \bibinfo{person}{Thomas Funkhouser}, \bibinfo{person}{Anima Anandkumar}, {et~al\mbox{.}}} \bibinfo{year}{2024}\natexlab{b}.
\newblock \showarticletitle{MVEdit: Generic 3D Diffusion Adapter Using Controlled Multi-View Editing}.
\newblock \bibinfo{journal}{\emph{arXiv preprint arXiv:2403.12032}} (\bibinfo{year}{2024}).
\newblock
\urldef\tempurl%
\url{https://hanshengchen.com/mvedit/}
\showURL{%
\tempurl}


\bibitem[Chen et~al\mbox{.}(2025)]%
        {dge}
\bibfield{author}{\bibinfo{person}{Minghao Chen}, \bibinfo{person}{Iro Laina}, {and} \bibinfo{person}{Andrea Vedaldi}.} \bibinfo{year}{2025}\natexlab{}.
\newblock \showarticletitle{DGE: Direct Gaussian 3D Editing by Consistent Multi-view Editing}. In \bibinfo{booktitle}{\emph{Computer Vision -- ECCV 2024}}, \bibfield{editor}{\bibinfo{person}{Ale{\v{s}} Leonardis}, \bibinfo{person}{Elisa Ricci}, \bibinfo{person}{Stefan Roth}, \bibinfo{person}{Olga Russakovsky}, \bibinfo{person}{Torsten Sattler}, {and} \bibinfo{person}{G{\"u}l Varol}} (Eds.). \bibinfo{publisher}{Springer Nature Switzerland}, \bibinfo{address}{Cham}, \bibinfo{pages}{74--92}.
\newblock
\showISBNx{978-3-031-72904-1}


\bibitem[Chen et~al\mbox{.}(2024a)]%
        {shapeditor}
\bibfield{author}{\bibinfo{person}{Minghao Chen}, \bibinfo{person}{Junyu Xie}, \bibinfo{person}{Iro Laina}, {and} \bibinfo{person}{Andrea Vedaldi}.} \bibinfo{year}{2024}\natexlab{a}.
\newblock \showarticletitle{SHAP-EDITOR: Instruction-guided Latent 3D Editing in Seconds}. In \bibinfo{booktitle}{\emph{CVPR}}.
\newblock


\bibitem[Deitke et~al\mbox{.}(2023a)]%
        {objaverseXL}
\bibfield{author}{\bibinfo{person}{Matt Deitke}, \bibinfo{person}{Ruoshi Liu}, \bibinfo{person}{Matthew Wallingford}, \bibinfo{person}{Huong Ngo}, \bibinfo{person}{Oscar Michel}, \bibinfo{person}{Aditya Kusupati}, \bibinfo{person}{Alan Fan}, \bibinfo{person}{Christian Laforte}, \bibinfo{person}{Vikram Voleti}, \bibinfo{person}{Samir~Yitzhak Gadre}, \bibinfo{person}{Eli VanderBilt}, \bibinfo{person}{Aniruddha Kembhavi}, \bibinfo{person}{Carl Vondrick}, \bibinfo{person}{Georgia Gkioxari}, \bibinfo{person}{Kiana Ehsani}, \bibinfo{person}{Ludwig Schmidt}, {and} \bibinfo{person}{Ali Farhadi}.} \bibinfo{year}{2023}\natexlab{a}.
\newblock \showarticletitle{Objaverse-XL: a universe of 10M+ 3D objects}. In \bibinfo{booktitle}{\emph{Proceedings of the 37th International Conference on Neural Information Processing Systems}} (New Orleans, LA, USA) \emph{(\bibinfo{series}{NIPS '23})}. \bibinfo{publisher}{Curran Associates Inc.}, \bibinfo{address}{Red Hook, NY, USA}, Article \bibinfo{articleno}{1554}, \bibinfo{numpages}{15}~pages.
\newblock


\bibitem[Deitke et~al\mbox{.}(2023b)]%
        {objaverse}
\bibfield{author}{\bibinfo{person}{Matt Deitke}, \bibinfo{person}{Dustin Schwenk}, \bibinfo{person}{Jordi Salvador}, \bibinfo{person}{Luca Weihs}, \bibinfo{person}{Oscar Michel}, \bibinfo{person}{Eli VanderBilt}, \bibinfo{person}{Ludwig Schmidt}, \bibinfo{person}{Kiana Ehsani}, \bibinfo{person}{Aniruddha Kembhavi}, {and} \bibinfo{person}{Ali Farhadi}.} \bibinfo{year}{2023}\natexlab{b}.
\newblock \showarticletitle{Objaverse: A Universe of Annotated 3D Objects}. In \bibinfo{booktitle}{\emph{Proceedings of the IEEE/CVF Conference on Computer Vision and Pattern Recognition (CVPR)}}. \bibinfo{pages}{13142--13153}.
\newblock


\bibitem[Edelstein et~al\mbox{.}(2025)]%
        {edelstein2024sharpitmultiviewmultiviewdiffusion}
\bibfield{author}{\bibinfo{person}{Yiftach Edelstein}, \bibinfo{person}{Or Patashnik}, \bibinfo{person}{Dana Cohen-Bar}, {and} \bibinfo{person}{Lihi Zelnik-Manor}.} \bibinfo{year}{2025}\natexlab{}.
\newblock \showarticletitle{Sharp-It: A Multi-view to Multi-view Diffusion Model for 3D Synthesis and Manipulation}. In \bibinfo{booktitle}{\emph{Proceedings of the Computer Vision and Pattern Recognition Conference (CVPR)}}. \bibinfo{pages}{21458--21468}.
\newblock


\bibitem[Erko{\c{c}} et~al\mbox{.}(2025)]%
        {erkoc2024preditor3d}
\bibfield{author}{\bibinfo{person}{Ziya Erko{\c{c}}}, \bibinfo{person}{Can G{\"u}meli}, \bibinfo{person}{Chaoyang Wang}, \bibinfo{person}{Matthias Nie{\ss}ner}, \bibinfo{person}{Angela Dai}, \bibinfo{person}{Peter Wonka}, \bibinfo{person}{Hsin-Ying Lee}, {and} \bibinfo{person}{Peiye Zhuang}.} \bibinfo{year}{2025}\natexlab{}.
\newblock \showarticletitle{PrEditor3D: Fast and Precise 3D Shape Editing}. In \bibinfo{booktitle}{\emph{Proceedings of the Computer Vision and Pattern Recognition Conference}}. \bibinfo{pages}{640--649}.
\newblock
\urldef\tempurl%
\url{https://ziyaerkoc.com/preditor3d}
\showURL{%
\tempurl}


\bibitem[Gal et~al\mbox{.}(2022)]%
        {gal2021stylegannadaclipguideddomainadaptation}
\bibfield{author}{\bibinfo{person}{Rinon Gal}, \bibinfo{person}{Or Patashnik}, \bibinfo{person}{Haggai Maron}, \bibinfo{person}{Amit~H. Bermano}, \bibinfo{person}{Gal Chechik}, {and} \bibinfo{person}{Daniel Cohen-Or}.} \bibinfo{year}{2022}\natexlab{}.
\newblock \showarticletitle{StyleGAN-NADA: CLIP-guided domain adaptation of image generators}.
\newblock \bibinfo{journal}{\emph{ACM Trans. Graph.}} \bibinfo{volume}{41}, \bibinfo{number}{4}, Article \bibinfo{articleno}{141} (\bibinfo{date}{July} \bibinfo{year}{2022}), \bibinfo{numpages}{13}~pages.
\newblock
\showISSN{0730-0301}
\urldef\tempurl%
\url{https://doi.org/10.1145/3528223.3530164}
\showDOI{\tempurl}


\bibitem[Haque et~al\mbox{.}(2023)]%
        {instructnerf2nerf}
\bibfield{author}{\bibinfo{person}{Ayaan Haque}, \bibinfo{person}{Matthew Tancik}, \bibinfo{person}{Alexei~A Efros}, \bibinfo{person}{Aleksander Holynski}, {and} \bibinfo{person}{Angjoo Kanazawa}.} \bibinfo{year}{2023}\natexlab{}.
\newblock \showarticletitle{Instruct-nerf2nerf: Editing 3d scenes with instructions}. In \bibinfo{booktitle}{\emph{ICCV}}.
\newblock


\bibitem[Hertz et~al\mbox{.}(2023)]%
        {deltadenoisingscore}
\bibfield{author}{\bibinfo{person}{Amir Hertz}, \bibinfo{person}{Kfir Aberman}, {and} \bibinfo{person}{Daniel Cohen-Or}.} \bibinfo{year}{2023}\natexlab{}.
\newblock \showarticletitle{Delta Denoising Score}. In \bibinfo{booktitle}{\emph{Proceedings of the IEEE/CVF International Conference on Computer Vision (ICCV)}}. \bibinfo{pages}{2328--2337}.
\newblock


\bibitem[Huberman-Spiegelglas et~al\mbox{.}(2024)]%
        {editfriendlyddpm}
\bibfield{author}{\bibinfo{person}{Inbar Huberman-Spiegelglas}, \bibinfo{person}{Vladimir Kulikov}, {and} \bibinfo{person}{Tomer Michaeli}.} \bibinfo{year}{2024}\natexlab{}.
\newblock \showarticletitle{An edit friendly ddpm noise space: Inversion and manipulations}. In \bibinfo{booktitle}{\emph{Proceedings of the IEEE/CVF Conference on Computer Vision and Pattern Recognition}}. \bibinfo{pages}{12469--12478}.
\newblock


\bibitem[Jun and Nichol(2023)]%
        {jun2023shapegeneratingconditional3d}
\bibfield{author}{\bibinfo{person}{Heewoo Jun} {and} \bibinfo{person}{Alex Nichol}.} \bibinfo{year}{2023}\natexlab{}.
\newblock \bibinfo{title}{Shap-E: Generating Conditional 3D Implicit Functions}.
\newblock
\newblock
\showeprint[arxiv]{2305.02463}~[cs.CV]
\urldef\tempurl%
\url{https://arxiv.org/abs/2305.02463}
\showURL{%
\tempurl}


\bibitem[Kawar et~al\mbox{.}(2023)]%
        {kawar2023imagic}
\bibfield{author}{\bibinfo{person}{Bahjat Kawar}, \bibinfo{person}{Shiran Zada}, \bibinfo{person}{Oran Lang}, \bibinfo{person}{Omer Tov}, \bibinfo{person}{Huiwen Chang}, \bibinfo{person}{Tali Dekel}, \bibinfo{person}{Inbar Mosseri}, {and} \bibinfo{person}{Michal Irani}.} \bibinfo{year}{2023}\natexlab{}.
\newblock \showarticletitle{Imagic: Text-based real image editing with diffusion models}. In \bibinfo{booktitle}{\emph{Proceedings of the IEEE/CVF conference on computer vision and pattern recognition}}. \bibinfo{pages}{6007--6017}.
\newblock


\bibitem[Kerbl et~al\mbox{.}(2023)]%
        {gs}
\bibfield{author}{\bibinfo{person}{Bernhard Kerbl}, \bibinfo{person}{Georgios Kopanas}, \bibinfo{person}{Thomas Leimk{\"u}hler}, {and} \bibinfo{person}{George Drettakis}.} \bibinfo{year}{2023}\natexlab{}.
\newblock \showarticletitle{3D Gaussian Splatting for Real-Time Radiance Field Rendering}.
\newblock \bibinfo{journal}{\emph{ACM Transactions on Graphics}} \bibinfo{volume}{42}, \bibinfo{number}{4} (\bibinfo{date}{July} \bibinfo{year}{2023}).
\newblock
\urldef\tempurl%
\url{https://repo-sam.inria.fr/fungraph/3d-gaussian-splatting/}
\showURL{%
\tempurl}


\bibitem[Kulikov et~al\mbox{.}(2024)]%
        {flowedit}
\bibfield{author}{\bibinfo{person}{Vladimir Kulikov}, \bibinfo{person}{Matan Kleiner}, \bibinfo{person}{Inbar Huberman-Spiegelglas}, {and} \bibinfo{person}{Tomer Michaeli}.} \bibinfo{year}{2024}\natexlab{}.
\newblock \showarticletitle{FlowEdit: Inversion-Free Text-Based Editing Using Pre-Trained Flow Models}.
\newblock \bibinfo{journal}{\emph{arXiv preprint arXiv:2412.08629}} (\bibinfo{year}{2024}).
\newblock


\bibitem[Labs(2024)]%
        {flux2024}
\bibfield{author}{\bibinfo{person}{Black~Forest Labs}.} \bibinfo{year}{2024}\natexlab{}.
\newblock \bibinfo{title}{FLUX}.
\newblock \bibinfo{howpublished}{\url{https://github.com/black-forest-labs/flux}}.
\newblock


\bibitem[Liu et~al\mbox{.}(2024)]%
        {syncdreamer}
\bibfield{author}{\bibinfo{person}{Yuan Liu}, \bibinfo{person}{Cheng Lin}, \bibinfo{person}{Zijiao Zeng}, \bibinfo{person}{Xiaoxiao Long}, \bibinfo{person}{Lingjie Liu}, \bibinfo{person}{Taku Komura}, {and} \bibinfo{person}{Wenping Wang}.} \bibinfo{year}{2024}\natexlab{}.
\newblock \showarticletitle{Syncdreamer: Generating multiview-consistent images from a single-view image}. In \bibinfo{booktitle}{\emph{ICLR}}.
\newblock


\bibitem[Long et~al\mbox{.}(2024)]%
        {wonder3d}
\bibfield{author}{\bibinfo{person}{Xiaoxiao Long}, \bibinfo{person}{Yuan-Chen Guo}, \bibinfo{person}{Cheng Lin}, \bibinfo{person}{Yuan Liu}, \bibinfo{person}{Zhiyang Dou}, \bibinfo{person}{Lingjie Liu}, \bibinfo{person}{Yuexin Ma}, \bibinfo{person}{Song-Hai Zhang}, \bibinfo{person}{Marc Habermann}, \bibinfo{person}{Christian Theobalt}, {et~al\mbox{.}}} \bibinfo{year}{2024}\natexlab{}.
\newblock \showarticletitle{Wonder3D: Single Image to 3D using Cross-Domain Diffusion}. In \bibinfo{booktitle}{\emph{Proceedings of the IEEE/CVF conference on computer vision and pattern recognition}}. \bibinfo{pages}{9970--9980}.
\newblock


\bibitem[Meng et~al\mbox{.}(2022)]%
        {sdedit}
\bibfield{author}{\bibinfo{person}{Chenlin Meng}, \bibinfo{person}{Yutong He}, \bibinfo{person}{Yang Song}, \bibinfo{person}{Jiaming Song}, \bibinfo{person}{Jiajun Wu}, \bibinfo{person}{Jun-Yan Zhu}, {and} \bibinfo{person}{Stefano Ermon}.} \bibinfo{year}{2022}\natexlab{}.
\newblock \showarticletitle{Sdedit: Guided image synthesis and editing with stochastic differential equations}. In \bibinfo{booktitle}{\emph{ICLR}}.
\newblock


\bibitem[Mikaeili et~al\mbox{.}(2023)]%
        {mikaeili2023sked}
\bibfield{author}{\bibinfo{person}{Aryan Mikaeili}, \bibinfo{person}{Or Perel}, \bibinfo{person}{Mehdi Safaee}, \bibinfo{person}{Daniel Cohen-Or}, {and} \bibinfo{person}{Ali Mahdavi-Amiri}.} \bibinfo{year}{2023}\natexlab{}.
\newblock \showarticletitle{SKED: Sketch-guided Text-based 3D Editing}. In \bibinfo{booktitle}{\emph{Proceedings of the IEEE/CVF International Conference on Computer Vision (ICCV)}}. \bibinfo{pages}{14607--14619}.
\newblock
\urldef\tempurl%
\url{https://arxiv.org/abs/2303.10735}
\showURL{%
\tempurl}


\bibitem[Mildenhall et~al\mbox{.}(2020)]%
        {nerf}
\bibfield{author}{\bibinfo{person}{Ben Mildenhall}, \bibinfo{person}{Pratul~P. Srinivasan}, \bibinfo{person}{Matthew Tancik}, \bibinfo{person}{Jonathan~T. Barron}, \bibinfo{person}{Ravi Ramamoorthi}, {and} \bibinfo{person}{Ren Ng}.} \bibinfo{year}{2020}\natexlab{}.
\newblock \showarticletitle{NeRF: Representing Scenes as Neural Radiance Fields for View Synthesis}. In \bibinfo{booktitle}{\emph{ECCV}}.
\newblock


\bibitem[Mokady et~al\mbox{.}(2023)]%
        {nulltext}
\bibfield{author}{\bibinfo{person}{Ron Mokady}, \bibinfo{person}{Amir Hertz}, \bibinfo{person}{Kfir Aberman}, \bibinfo{person}{Yael Pritch}, {and} \bibinfo{person}{Daniel Cohen-Or}.} \bibinfo{year}{2023}\natexlab{}.
\newblock \showarticletitle{Null-Text Inversion for Editing Real Images Using Guided Diffusion Models}. In \bibinfo{booktitle}{\emph{Proceedings of the IEEE/CVF conference on computer vision and pattern recognition}}. \bibinfo{pages}{6038--6047}.
\newblock


\bibitem[Nichol et~al\mbox{.}(2022)]%
        {glide}
\bibfield{author}{\bibinfo{person}{Alexander~Quinn Nichol}, \bibinfo{person}{Prafulla Dhariwal}, \bibinfo{person}{Aditya Ramesh}, \bibinfo{person}{Pranav Shyam}, \bibinfo{person}{Pamela Mishkin}, \bibinfo{person}{Bob Mcgrew}, \bibinfo{person}{Ilya Sutskever}, {and} \bibinfo{person}{Mark Chen}.} \bibinfo{year}{2022}\natexlab{}.
\newblock \showarticletitle{{GLIDE}: Towards Photorealistic Image Generation and Editing with Text-Guided Diffusion Models}. In \bibinfo{booktitle}{\emph{Proceedings of the 39th International Conference on Machine Learning}} \emph{(\bibinfo{series}{Proceedings of Machine Learning Research}, Vol.~\bibinfo{volume}{162})}, \bibfield{editor}{\bibinfo{person}{Kamalika Chaudhuri}, \bibinfo{person}{Stefanie Jegelka}, \bibinfo{person}{Le~Song}, \bibinfo{person}{Csaba Szepesvari}, \bibinfo{person}{Gang Niu}, {and} \bibinfo{person}{Sivan Sabato}} (Eds.). \bibinfo{publisher}{PMLR}, \bibinfo{pages}{16784--16804}.
\newblock
\urldef\tempurl%
\url{https://proceedings.mlr.press/v162/nichol22a.html}
\showURL{%
\tempurl}


\bibitem[Parmar et~al\mbox{.}(2023)]%
        {pix2pixzero2023}
\bibfield{author}{\bibinfo{person}{Gaurav Parmar}, \bibinfo{person}{Krishna Kumar~Singh}, \bibinfo{person}{Richard Zhang}, \bibinfo{person}{Yijun Li}, \bibinfo{person}{Jingwan Lu}, {and} \bibinfo{person}{Jun-Yan Zhu}.} \bibinfo{year}{2023}\natexlab{}.
\newblock \showarticletitle{Zero-shot image-to-image translation}. In \bibinfo{booktitle}{\emph{ACM SIGGRAPH 2023 conference proceedings}}. \bibinfo{pages}{1--11}.
\newblock


\bibitem[Patashnik et~al\mbox{.}(2024)]%
        {qnerf}
\bibfield{author}{\bibinfo{person}{Or Patashnik}, \bibinfo{person}{Rinon Gal}, \bibinfo{person}{Daniel Cohen-Or}, \bibinfo{person}{Jun-Yan Zhu}, {and} \bibinfo{person}{Fernando De~La~Torre}.} \bibinfo{year}{2024}\natexlab{}.
\newblock \showarticletitle{Consolidating Attention Features for Multi-view Image Editing}. In \bibinfo{booktitle}{\emph{SIGGRAPH Asia 2024 Conference Papers}} (Tokyo, Japan) \emph{(\bibinfo{series}{SA '24})}. \bibinfo{publisher}{Association for Computing Machinery}, \bibinfo{address}{New York, NY, USA}, Article \bibinfo{articleno}{40}, \bibinfo{numpages}{12}~pages.
\newblock
\showISBNx{9798400711312}
\urldef\tempurl%
\url{https://doi.org/10.1145/3680528.3687611}
\showDOI{\tempurl}


\bibitem[Poole et~al\mbox{.}(2023)]%
        {dreamfusion}
\bibfield{author}{\bibinfo{person}{Ben Poole}, \bibinfo{person}{Ajay Jain}, \bibinfo{person}{Jonathan~T Barron}, {and} \bibinfo{person}{Ben Mildenhall}.} \bibinfo{year}{2023}\natexlab{}.
\newblock \showarticletitle{DreamFusion: Text-to-3D using 2D Diffusion}. In \bibinfo{booktitle}{\emph{ICLR}}.
\newblock


\bibitem[Salimans and Ho(2022)]%
        {salimans2022progressive}
\bibfield{author}{\bibinfo{person}{Tim Salimans} {and} \bibinfo{person}{Jonathan Ho}.} \bibinfo{year}{2022}\natexlab{}.
\newblock \bibinfo{title}{Progressive Distillation for Fast Sampling of Diffusion Models}.
\newblock
\newblock
\showeprint[arxiv]{2202.00512}~[cs.LG]
\urldef\tempurl%
\url{https://arxiv.org/abs/2202.00512}
\showURL{%
\tempurl}


\bibitem[Sella et~al\mbox{.}(2023)]%
        {voxe}
\bibfield{author}{\bibinfo{person}{Etai Sella}, \bibinfo{person}{Gal Fiebelman}, \bibinfo{person}{Peter Hedman}, {and} \bibinfo{person}{Hadar Averbuch-Elor}.} \bibinfo{year}{2023}\natexlab{}.
\newblock \showarticletitle{Vox-e: Text-guided voxel editing of 3d objects}. In \bibinfo{booktitle}{\emph{Proceedings of the IEEE/CVF international conference on computer vision}}. \bibinfo{pages}{430--440}.
\newblock


\bibitem[Shi et~al\mbox{.}(2023)]%
        {zero123++}
\bibfield{author}{\bibinfo{person}{Ruoxi Shi}, \bibinfo{person}{Hansheng Chen}, \bibinfo{person}{Zhuoyang Zhang}, \bibinfo{person}{Minghua Liu}, \bibinfo{person}{Chao Xu}, \bibinfo{person}{Xinyue Wei}, \bibinfo{person}{Linghao Chen}, \bibinfo{person}{Chong Zeng}, {and} \bibinfo{person}{Hao Su}.} \bibinfo{year}{2023}\natexlab{}.
\newblock \bibinfo{title}{Zero123++: a Single Image to Consistent Multi-view Diffusion Base Model}.
\newblock
\newblock
\showeprint[arxiv]{2310.15110}~[cs.CV]
\urldef\tempurl%
\url{https://arxiv.org/abs/2310.15110}
\showURL{%
\tempurl}


\bibitem[Shi et~al\mbox{.}(2024)]%
        {mvdream}
\bibfield{author}{\bibinfo{person}{Yichun Shi}, \bibinfo{person}{Peng Wang}, \bibinfo{person}{Jianglong Ye}, \bibinfo{person}{Mai Long}, \bibinfo{person}{Kejie Li}, {and} \bibinfo{person}{Xiao Yang}.} \bibinfo{year}{2024}\natexlab{}.
\newblock \showarticletitle{Mvdream: Multi-view diffusion for 3d generation}. In \bibinfo{booktitle}{\emph{ICLR}}.
\newblock


\bibitem[Tang et~al\mbox{.}(2024)]%
        {lgm}
\bibfield{author}{\bibinfo{person}{Jiaxiang Tang}, \bibinfo{person}{Zhaoxi Chen}, \bibinfo{person}{Xiaokang Chen}, \bibinfo{person}{Tengfei Wang}, \bibinfo{person}{Gang Zeng}, {and} \bibinfo{person}{Ziwei Liu}.} \bibinfo{year}{2024}\natexlab{}.
\newblock \showarticletitle{LGM: Large Multi-view Gaussian Model for~High-Resolution 3D Content Creation}. In \bibinfo{booktitle}{\emph{Computer Vision – ECCV 2024: 18th European Conference, Milan, Italy, September 29–October 4, 2024, Proceedings, Part IV}} (Milan, Italy). \bibinfo{publisher}{Springer-Verlag}, \bibinfo{address}{Berlin, Heidelberg}, \bibinfo{pages}{1–18}.
\newblock
\showISBNx{978-3-031-73234-8}
\urldef\tempurl%
\url{https://doi.org/10.1007/978-3-031-73235-5_1}
\showDOI{\tempurl}


\bibitem[Tumanyan et~al\mbox{.}(2023)]%
        {tumanyan2023plug}
\bibfield{author}{\bibinfo{person}{Narek Tumanyan}, \bibinfo{person}{Michal Geyer}, \bibinfo{person}{Shai Bagon}, {and} \bibinfo{person}{Tali Dekel}.} \bibinfo{year}{2023}\natexlab{}.
\newblock \showarticletitle{Plug-and-Play Diffusion Features for Text-Driven Image-to-Image Translation}. In \bibinfo{booktitle}{\emph{Proceedings of the IEEE/CVF Conference on Computer Vision and Pattern Recognition}}. \bibinfo{pages}{1921--1930}.
\newblock


\bibitem[Wang et~al\mbox{.}(2024)]%
        {crm}
\bibfield{author}{\bibinfo{person}{Zhengyi Wang}, \bibinfo{person}{Yikai Wang}, \bibinfo{person}{Yifei Chen}, \bibinfo{person}{Chendong Xiang}, \bibinfo{person}{Shuo Chen}, \bibinfo{person}{Dajiang Yu}, \bibinfo{person}{Chongxuan Li}, \bibinfo{person}{Hang Su}, {and} \bibinfo{person}{Jun Zhu}.} \bibinfo{year}{2024}\natexlab{}.
\newblock \showarticletitle{CRM: Single Image to 3D Textured Mesh with Convolutional Reconstruction Model}. In \bibinfo{booktitle}{\emph{Computer Vision – ECCV 2024: 18th European Conference, Milan, Italy, September 29–October 4, 2024, Proceedings, Part XXXI}} (Milan, Italy). \bibinfo{publisher}{Springer-Verlag}, \bibinfo{address}{Berlin, Heidelberg}, \bibinfo{pages}{57–74}.
\newblock
\showISBNx{978-3-031-72750-4}
\urldef\tempurl%
\url{https://doi.org/10.1007/978-3-031-72751-1_4}
\showDOI{\tempurl}


\bibitem[Weber et~al\mbox{.}(2024)]%
        {nerfiller}
\bibfield{author}{\bibinfo{person}{Ethan Weber}, \bibinfo{person}{Aleksander Holynski}, \bibinfo{person}{Varun Jampani}, \bibinfo{person}{Saurabh Saxena}, \bibinfo{person}{Noah Snavely}, \bibinfo{person}{Abhishek Kar}, {and} \bibinfo{person}{Angjoo Kanazawa}.} \bibinfo{year}{2024}\natexlab{}.
\newblock \showarticletitle{NeRFiller: Completing Scenes via Generative 3D Inpainting}. In \bibinfo{booktitle}{\emph{Proceedings of the IEEE/CVF Conference on Computer Vision and Pattern Recognition}}. \bibinfo{pages}{20731--20741}.
\newblock


\bibitem[Xiang et~al\mbox{.}(2024)]%
        {trellis}
\bibfield{author}{\bibinfo{person}{Jianfeng Xiang}, \bibinfo{person}{Zelong Lv}, \bibinfo{person}{Sicheng Xu}, \bibinfo{person}{Yu Deng}, \bibinfo{person}{Ruicheng Wang}, \bibinfo{person}{Bowen Zhang}, \bibinfo{person}{Dong Chen}, \bibinfo{person}{Xin Tong}, {and} \bibinfo{person}{Jiaolong Yang}.} \bibinfo{year}{2024}\natexlab{}.
\newblock \showarticletitle{Structured 3D Latents for Scalable and Versatile 3D Generation}.
\newblock \bibinfo{journal}{\emph{arXiv preprint arXiv:2412.01506}} (\bibinfo{year}{2024}).
\newblock


\bibitem[Xu et~al\mbox{.}(2024)]%
        {instantmesh}
\bibfield{author}{\bibinfo{person}{Jiale Xu}, \bibinfo{person}{Weihao Cheng}, \bibinfo{person}{Yiming Gao}, \bibinfo{person}{Xintao Wang}, \bibinfo{person}{Shenghua Gao}, {and} \bibinfo{person}{Ying Shan}.} \bibinfo{year}{2024}\natexlab{}.
\newblock \showarticletitle{InstantMesh: Efficient 3D Mesh Generation from a Single Image with Sparse-view Large Reconstruction Models}.
\newblock \bibinfo{journal}{\emph{arXiv preprint arXiv:2404.07191}} (\bibinfo{year}{2024}).
\newblock


\bibitem[Zhuang et~al\mbox{.}(2024)]%
        {zhuang2024tipeditoraccurate3deditor}
\bibfield{author}{\bibinfo{person}{Jingyu Zhuang}, \bibinfo{person}{Di Kang}, \bibinfo{person}{Yan-Pei Cao}, \bibinfo{person}{Guanbin Li}, \bibinfo{person}{Liang Lin}, {and} \bibinfo{person}{Ying Shan}.} \bibinfo{year}{2024}\natexlab{}.
\newblock \showarticletitle{TIP-Editor: An Accurate 3D Editor Following Both Text-Prompts And Image-Prompts}.
\newblock \bibinfo{journal}{\emph{ACM Transactions on Graphics (TOG)}} \bibinfo{volume}{43}, \bibinfo{number}{4} (\bibinfo{year}{2024}), \bibinfo{pages}{1--12}.
\newblock


\bibitem[Zhuang et~al\mbox{.}(2023)]%
        {zhuang2023dreameditortextdriven3dscene}
\bibfield{author}{\bibinfo{person}{Jingyu Zhuang}, \bibinfo{person}{Chen Wang}, \bibinfo{person}{Liang Lin}, \bibinfo{person}{Lingjie Liu}, {and} \bibinfo{person}{Guanbin Li}.} \bibinfo{year}{2023}\natexlab{}.
\newblock \showarticletitle{DreamEditor: Text-Driven 3D Scene Editing with Neural Fields}. In \bibinfo{booktitle}{\emph{SIGGRAPH Asia 2023 Conference Papers}}. \bibinfo{pages}{1--10}.
\newblock


\end{thebibliography}
